\documentclass[aps,prb,twocolumn,superscriptaddress]{revtex4-2}
\usepackage{graphicx}
\usepackage{siunitx}
\usepackage{bm}
\usepackage[caption=false]{subfig}
\usepackage{mathtools}
\usepackage{multirow}
\usepackage{amssymb}
\usepackage{braket}
\usepackage{amsmath}
\usepackage{setspace}
\usepackage{rotating}
\usepackage{placeins}
\usepackage{cancel}
\usepackage{xcolor}
\usepackage{adjustbox}
\usepackage{afterpage}
\usepackage{spreadtab}
\usepackage{tabularx}
\usepackage{pax}

\newcommand\clearrow{\global\let\rowmac\relax}

\usepackage{hyperref}
\hypersetup{
hyperindex=true,
breaklinks=true,   
colorlinks=true,   
pdfusetitle=true,  
linkcolor=blue,
citecolor=blue,
urlcolor=blue,
}
\setcitestyle{super,open={},close={}}
\renewcommand*{\onlinecite}[1]{%
  \begingroup
    \romannumeral-`\x 
    \setcitestyle{numbers}%
    \cite{#1}%
  \endgroup   
}
\DeclareRobustCommand*{\onlinecite}[1]{%
  \begingroup
    \romannumeral-`\x 
    \setcitestyle{numbers}%
    \cite{#1}%
  \endgroup
}

\usepackage{pdfpages} 
\usepackage{pgffor} 

\makeatletter
\AtBeginDocument{\let\LS@rot\@undefined}
\makeatother

\def\supplementfilename{si.pdf}

\pdfximage{\supplementfilename}
\def\numbersupplementpages{\the\pdflastximagepages}

\newif\ifarXiv
\arXivtrue 

\begin{document}

\title{General invariance and equilibrium conditions for lattice dynamics in 1D, 2D, and 3D materials}

\author{Changpeng Lin}
\email{changpeng.lin@epfl.ch}
\affiliation{%
Theory and Simulation of Materials (THEOS), \'Ecole Polytechnique F\'ed\'erale de Lausanne, CH-1015 Lausanne, Switzerland
}%
\affiliation{%
National Centre for Computational Design and Discovery of Novel Materials (MARVEL), \'Ecole Polytechnique F\'ed\'erale de Lausanne, CH-1015 Lausanne, Switzerland
}%
\author{Samuel Ponc\'e}
\affiliation{%
European Theoretical Spectroscopy Facility, Institute of Condensed Matter and Nanosciences, Universit\'e catholique de Louvain, Chemin des \'Etoiles 8, B-1348 Louvain-la-Neuve, Belgium
}%
\affiliation{%
Theory and Simulation of Materials (THEOS), \'Ecole Polytechnique F\'ed\'erale de Lausanne, CH-1015 Lausanne, Switzerland
}%
\author{Nicola Marzari}
\affiliation{%
Theory and Simulation of Materials (THEOS), \'Ecole Polytechnique F\'ed\'erale de Lausanne, CH-1015 Lausanne, Switzerland
}%
\affiliation{%
National Centre for Computational Design and Discovery of Novel Materials (MARVEL), \'Ecole Polytechnique F\'ed\'erale de Lausanne, CH-1015 Lausanne, Switzerland
}%

\date{\today}

\begin{abstract}
The long-wavelength behavior of vibrational modes plays a central role in carrier transport, phonon-assisted optical properties, superconductivity, and thermomechanical and thermoelectric properties of materials. 
Here, we present general invariance and equilibrium conditions of the lattice potential; these allow to recover the quadratic dispersions of flexural phonons in low-dimensional materials, in agreement with the phenomenological model for long-wavelength bending modes.
We also prove that for any low-dimensional material the bending modes can have a purely out-of-plane polarization in the vacuum direction and a quadratic dispersion in the long-wavelength limit.
In addition, we propose an effective approach to treat invariance conditions in crystals with non-vanishing Born effective charges where the long-range dipole-dipole interactions induce a contribution
to the lattice potential and stress tensor.
Our approach is successfully applied to the phonon dispersions of 158 two-dimensional materials, highlighting its critical relevance in the study of phonon-mediated properties of low-dimensional materials.
\end{abstract}

\maketitle

\section*{INTRODUCTION}

The vibrations of atoms inside crystalline solids are fundamental to many aspects of condensed matter physics and materials science. 
The lattice vibrations play a crucial role in a wealth of phenomena including phase transitions~\cite{dove1993}, phonon-mediated superconductivity~\cite{grimvall1981,giustino2017}, and electronic~\cite{Ponce2020} and thermal transport~\cite{shinde2014,qian2021}.
The lattice-dynamical theory of solids was established at the beginning of the twentieth century with seminal works from Einstein, Born, von K\'arm\'an and Debye~\cite{einstein1907,born1912,debye1912,born1954}. 
With advances in density-functional theory (DFT) and density-functional perturbation theory (DFPT)~\cite{baroni2001,gonze1997},  the parameter-free calculations of phonon dispersions from first principles have become a routine task implemented in many software packages~\cite{alfe2009,togo2015,giannozzi2017,gonze2016,wang2016,Petretto2018,eriksson2019}.
However,  there still exist some controversies in the calculated phonon dispersions of low-dimensional (LD) materials, where the long-wavelength behavior of flexural phonons are not fully understood. 
First-principles calculations performed by different groups showed either a linear or quadratic dispersion relation for the flexural acoustic (ZA) modes in two-dimensional (2D) materials and two bending modes in one-dimensional (1D) materials (see the detailed discussion in Ref.~\onlinecite{carrete2016}).
It has been recently argued~\cite{carrete2016,eriksson2019,croy2020} that the lack of rotational invariance in the harmonic lattice Hamiltonian with non-vanishing external stress from DFT calculations is responsible for the linear dispersion of the ZA mode; it becomes quadratic after including rotational invariance and vanishing stress condition.
Like translational invariance, there are various reasons for the broken rotational invariance in the interatomic force constants (IFCs) obtained from DFT and DFPT; these include incomplete basis sets, insufficient Brillouin zone sampling or numerical stability of algorithms.
In contrast, IFCs obtained from semi-empirical force fields or models do not typically suffer from such issues, since they can satisfy translational and rotational invariances by construction~\cite{dresselhaus2000,mahan2004,mingo2008,libbi2020}.

Translational and rotational invariances drive the conservation of total momentum and angular momentum in crystals~\cite{kosevich2006,leibfried1961} (Noether's theorem~\cite{noether1971}); these are known as Born-Huang invariance conditions~\cite{born1954,begbie1947}.
In the 1960s, several works from Keating~\cite{keating1966,keating1966-2}, Gazis and Wallis~\cite{gazis1966} questioned the 
scope of such conditions and derived additional conditions for a harmonic lattice to be rotationally invariant.
From the requirement of unchanged potential energy under any arbitrary rotation operation, Gaizs and Wallis~\cite{gazis1966} derived additional constraints for the rotational invariance of a harmonic lattice.  
In a similar way, through the potential energy expressed in terms of the rotation parameter up to the second order, Keating~\cite{keating1966,keating1966-2} also formulated a new set of equations for harmonic IFCs of the simple cubic lattice.
However, Sarkar and Sengupta~\cite{sarkar1977} demonstrated that these are both equivalent to the Born-Huang invariance conditions which are complete.
Finally, Pick, Cohen, and Martin~\cite{pick1970} derived the connection between the microscopic theory and the phenomenological approach, and the microscopic acoustic sum rule for bulk crystals.

In this work, we investigate important aspects of the Born-Huang invariance conditions as well as the so-called equilibrium conditions (i.e. Huang conditions) in determining the phonon dispersions of materials and their consequence for LD materials.
We introduce the polar Born-Huang invariances and polar Huang conditions for harmonic IFCs in infrared-active solids where the existing long-range forces from the dipole-dipole interactions need to be treated separately.
Indeed, Huang had already demonstrated this in the case of bulk ionic crystals~\cite{huang1949,huang1950}, in which the long-range Coulomb interactions need a special treatment.
However, this condition seems to have been omitted in recent investigations on the rotational invariance of 2D materials~\cite{carrete2016,eriksson2019,croy2020}.
The detailed implementation of these conditions for correcting the IFCs either from real-space small displacements or reciprocal-space DFPT calculations are then presented.
Last, based on the secular equation for the long-wavelength vibrations, we demonstrate that the bending modes in any LD material should have a purely out-of-plane polarization in the vacuum direction and a quadratic dispersion in the long-wavelength limit.
We apply the present approach to silicon, triclinic calcium triphosphide (CaP$_3$), graphene, 2D molybdenum disulfide (MoS$_2$), 1D single-wall (8,0) carbon nanotube (CNT) and (4,4) boron nitride nanotube (BNNT), and find that the fulfillment of the rotational invariance and vanishing stress condition is crucial in LD materials.
We also verify that these conditions have negligible effects in 3D materials, as expected. 
Finally, these invariance conditions are applied to correct the phonon dispersions of 245 candidate 2D materials from the high-throughput database of Ref.~\onlinecite{mounet2018}.
In that study, the authors used a simple sum rule, resulting in 187 phonon dispersions presenting soft modes or an incorrect linear dispersion of the ZA mode. 
With the rotational invariance and equilibrium conditions correctly applied, the phonon dispersions of 158 materials become real and display a quadratic ZA branch in the long-wavelength limit.
The remaining 87 materials are either dynamically unstable (54) or require tighter numerical convergence (33).

\section*{RESULTS AND DISCUSSION}\label{sec:results_discuss}

\paragraph*{\normalfont\textbf{Born-Huang invariance conditions}}\label{sec:theory_BH}

The vibrational properties of solids can be determined by the nuclear Hamiltonian constructed based on the second-order IFCs.
Neglecting higher-order anharmonic contributions, a Taylor expansion of the Born-Oppenheimer potential energy $E$ in the atomic displacements up to the second order gives:
\begin{equation}\label{eq:expand_potential}
E=E_0+\sum_{\varkappa\alpha}\Phi_{\varkappa\alpha} u_{\varkappa\alpha}+\frac{1}{2}\!\!\sum_{\varkappa\alpha,\varkappa'\beta}\Phi_{\varkappa\alpha,\varkappa'\beta}u_{\varkappa\alpha} u_{\varkappa'\beta},
\end{equation}
where $E_0$ is the total potential energy of the chosen equilibrium reference structure, and $\Phi_{\varkappa\alpha} = \partial E / \partial u_{\varkappa\alpha}$ and $\Phi_{\varkappa\alpha,\varkappa'\beta} = \partial^2 E/ \partial u_{\varkappa\alpha}\partial u_{\varkappa'\beta}$ are the corresponding first- and second-order IFCs, respectively, with atomic indices $\varkappa$ and $\varkappa'$ in the Born-von K\'arm\'an supercell and the Cartesian directions $\alpha$ and $\beta$; the atom index $\varkappa\equiv\{l, \kappa\}$ denotes the atomic index $\kappa$ in the unit cell $l$, and $u_{\varkappa\alpha}$ is the corresponding atomic displacement from the equilibrium position.
Apart from the space group symmetry of the given crystal type and permutation symmetry, these IFCs are also required to satisfy both the global translational and rotational invariances due to the conservation of total crystal and angular momenta.

The Born-Huang invariance conditions provide the acoustic sum rules for both the translational and rotational invariances of a harmonic lattice.
These constraint equations on the first- and second-order IFCs read~\cite{leibfried1961}:
\begin{align}
\sum_\varkappa \Phi_{\varkappa\alpha} &= 0, \label{eq:ti_1st} \\
\sum_\varkappa \Phi_{\varkappa\alpha} \tau_{\varkappa\beta} &= \sum_\varkappa \Phi_{\varkappa\beta} \tau_{\varkappa\alpha}, \label{eq:ri_1st}\\ 
\sum_{\varkappa'} \Phi_{\varkappa\alpha,\varkappa'\beta} &= 0, \label{eq:ti_2nd} \\
\sum_{\varkappa'} \Phi_{\varkappa\alpha,\varkappa'\beta} \tau_{\varkappa'\gamma} + \delta_{\alpha\gamma} \Phi_{\varkappa\beta} &= \sum_{\varkappa'} \Phi_{\varkappa\alpha,\varkappa'\gamma} \tau_{\varkappa'\beta} + \delta_{\alpha\beta} \Phi_{\varkappa\gamma},\label{eq:ri_2nd}
\end{align}
where $\tau_{\varkappa\alpha}$ is the equilibrium position of atom $\varkappa$ in the Cartesian direction $\alpha$ and $\delta_{\alpha\beta}$ represents the Kronecker delta.
In these expressions, Equations~\eqref{eq:ti_1st} and \eqref{eq:ri_1st} are the acoustic sum rules for the translational and rotational invariances of the first-order IFCs, while Equations~\eqref{eq:ti_2nd} and \eqref{eq:ri_2nd} are their counterparts for the second-order IFCs.
When atoms are at their equilibrium positions, this set of constraint equations can be derived by adding a global translation or an infinitesimal rotation to all atoms in the system, which results in an unchanged lattice potential and zero forces acting on atoms.
These acoustic sum rules correspond to the conservation of total momentum and angular momentum in the crystal.
In general, the rotational sum rules link the $(n+1)$-th order to the $n$-th order of the Taylor-expanded potential~\cite{leibfried1961}.
For instance, Equation~\eqref{eq:ri_2nd} links the second-order IFCs to the first-order.

\paragraph*{\normalfont\textbf{Equilibrium conditions}}\label{sec:theory_EC}
For an infinite crystal, the lattice sites are perfectly periodic throughout the whole bulk region, and this requirement has two important consequences.
One is the total force on each atom must be zero, which means that the first-order IFCs in Equation~\eqref{eq:expand_potential} vanish at the equilibrium. 
This condition gives rise to the Born-Huang invariances imposed on the second-order IFCs as being rotationally invariant:
\begin{equation}\label{eq:B-H}
\sum_{\varkappa'} \Phi_{\varkappa\alpha,\varkappa'\beta} \tau_{\varkappa'\gamma} = \sum_{\varkappa'} \Phi_{\varkappa\alpha,\varkappa'\gamma} \tau_{\varkappa'\beta}.
\end{equation}
The other consequence is the vanishing of the stress tensor that governs the equilibrium volume of the unit cell, eliminating any surface effect as an infinite lattice.
The stress tensor for a finite lattice is given by~\cite{leibfried1961,sarkar1977}
\begin{equation}\label{eq:stress_1st}
\sigma_{\alpha\beta}=\frac{1}{\Omega} \sum_\varkappa \Phi_{\varkappa\alpha} \tau_{\varkappa\beta},
\end{equation}
where $\Omega$ is the volume of a finite lattice, and the first-order IFCs $\Phi_{\varkappa\alpha}$ account for the surface forces and vanish in the bulk region~\cite{leibfried1961}.
In order to derive the stress tensor for an infinite lattice, we multiply the rotational invariance, Equation~\eqref{eq:ri_2nd}, by the atomic position $\tau_{\varkappa\delta}$ and sum over $\varkappa$. 
Then, using Equation~\eqref{eq:stress_1st} we arrive at
\begin{equation}\label{eq:stress_2nd}
T_{\alpha\delta,\beta\gamma} - \Omega\sigma_{\gamma\delta} \delta_{\alpha\beta} = T_{\alpha\delta,\gamma\beta} -  \Omega \sigma_{\beta\delta} \delta_{\alpha\gamma} ,
\end{equation}
where we have introduced
\begin{equation}\label{eq:notation_T}
T_{\alpha\delta,\beta\gamma}  \equiv \sum_{\varkappa\varkappa'} \Phi_{\varkappa\alpha\varkappa'\beta} \tau_{\varkappa\delta} \tau_{\varkappa'\gamma}= T_{\beta\gamma,\alpha\delta}.
\end{equation}
Since $T_{\alpha\delta,\beta\gamma}$ might not be symmetric near the surface, we define
\begin{align}
T^{\rm sym}_{\alpha\beta,\gamma\delta} & \equiv \frac{T_{\alpha\gamma,\beta\delta}+T_{\alpha\delta,\beta\gamma}}{2} \label{eq:H_sym} \\
& = \frac{1}{2} \sum_{\varkappa\varkappa'}\Phi_{\varkappa\alpha,\varkappa'\beta}(\tau_{\varkappa\gamma} \tau_{\varkappa'\delta} + \tau_{\varkappa\delta} \tau_{\varkappa'\gamma}) \label{eq:H_sym2}.
\end{align}
With the help of the translational invariance, Equation~\eqref{eq:ti_2nd}, we can further express Equation~\eqref{eq:H_sym2} as
\begin{align}\label{eq:H_inf}
T^{\rm sym}_{\alpha\beta,\gamma\delta} =& -\frac{1}{2}\sum_{\varkappa\varkappa'} \Phi_{\varkappa\alpha,\varkappa'\beta}(\tau_{\varkappa\gamma} - \tau_{\varkappa'\gamma})(\tau_{\varkappa\delta} - \tau_{\varkappa'\delta}) \nonumber \\
=& -\frac{1}{2}\sum_{\varkappa\varkappa'} \Phi_{\varkappa\alpha,\varkappa'\beta} \tau_{\varkappa\varkappa'\gamma}\tau_{\varkappa\varkappa'\delta},
\end{align}
where $\tau_{\varkappa\varkappa'\gamma} \equiv \tau_{\varkappa'\gamma} - \tau_{\varkappa\gamma}$ is the distance between two atoms along the Cartesian direction $\gamma$.
By construction, the symmetrized tensor has the following symmetry relations
\begin{equation}\label{eq:sym_relation}
T^{\rm sym}_{\alpha\beta,\gamma\delta} = T^{\rm sym}_{\beta\alpha,\gamma\delta} = T^{\rm sym}_{\alpha\beta,\delta\gamma}.
\end{equation}
To derive the stress tensor for an infinite crystal, we further subtract both sides of Equation~\eqref{eq:H_sym} by the quantity $\Omega \sigma_{\gamma\delta}\delta_{\alpha\beta}$ to obtain
\begin{multline}\label{eq:stress_inf1}
T^{\rm sym}_{\alpha\beta,\gamma\delta}- \Omega\sigma_{\gamma\delta} \delta_{\alpha\beta} =\\
 \frac{1}{2} \Big[ T_{\alpha\gamma,\beta\delta} - \Omega \sigma_{\gamma\delta}\delta_{\alpha\beta} + T_{\alpha\delta,\beta\gamma} - \Omega \sigma_{\delta\gamma}\delta_{\alpha\beta} \Big],
\end{multline}
where the symmetry of the stress tensor $\sigma_{\gamma\delta}=\sigma_{\delta\gamma}$ is used.
From Equations~\eqref{eq:stress_2nd} and \eqref{eq:notation_T}, it can be noted that the right-hand side of Equation~\eqref{eq:stress_inf1} is invariant by interchanging $\alpha$ with $\delta$ and also $\beta$ with $\gamma$.
Therefore, the left-hand side of Equation~\eqref{eq:stress_inf1} has the following symmetry
\begin{equation}\label{eq:stress_inf2}
T^{\rm sym}_{\alpha\beta,\gamma\delta}-\Omega\sigma_{\gamma\delta}\delta_{\alpha\beta} = T^{\rm sym}_{\gamma\delta,\alpha\beta}-\Omega\sigma_{\alpha\beta}\delta_{\gamma\delta},
\end{equation}
which is the expression of the stress tensor for infinite crystals~\cite{leibfried1961}.
At equilibrium, the stress tensor vanishes, which gives the Huang conditions $T^{\rm sym}_{\alpha\beta,\gamma\delta} = T^{\rm sym}_{\gamma\delta,\alpha\beta}$:
\begin{equation}\label{eq:H-I}
\sum_{\varkappa\varkappa'} \Phi_{\varkappa\alpha,\varkappa'\beta} \tau_{\varkappa\varkappa'\gamma} \tau_{\varkappa\varkappa'\delta} = \sum_{\varkappa\varkappa'} \Phi_{\varkappa\gamma,\varkappa'\delta} \tau_{\varkappa\varkappa'\alpha} \tau_{\varkappa\varkappa'\beta}.
\end{equation}
In general, there are 36 Huang conditions due to the fourth-rank tensor $T^{\rm sym}_{\alpha\beta,\gamma\delta}$, which can be reduced to 15 by exploiting the symmetry relations in Equation~\eqref{eq:sym_relation}.
For the anisotropic components of stress tensor, they become
\begin{align}\label{eq:stress_aniso1}
\sigma_{\alpha\beta} =& \frac{T^{\rm sym}_{\beta\beta,\beta\alpha}-T^{\rm sym}_{\beta\alpha,\beta\beta}}{\Omega}\\
\sigma_{\alpha\alpha}-\sigma_{\beta\beta} =& \frac{T^{\rm sym}_{\beta\beta,\alpha\alpha}-T^{\rm sym}_{\alpha\alpha,\beta\beta}}{\Omega}, \label{eq:strerss_aniso2}
\end{align}
for $\alpha\neq\beta$.
We emphasize here that the Huang conditions are not the constraints for a potential to be rotationally invariant but the ones for a vanishing stress tensor. 
The set of Equations~\eqref{eq:ti_2nd}, \eqref{eq:B-H} and \eqref{eq:H-I} constitute the complete description of invariance conditions for the lattice dynamics of crystalline solids at the equilibrium. 
Leaving the translational invariance aside, we focus on the rotational invariance, Equation~\eqref{eq:B-H}, and the equilibrium conditions, Equation~\eqref{eq:H-I}, which we collectively refers to as invariance conditions.

\paragraph*{\normalfont\textbf{Invariance conditions to interatomic force constants}}\label{sec:implement}

In practice, the real-space IFCs can be obtained by fitting the DFT forces using the small displacement method~\cite{alfe2009} or 
 a Fourier transformation within DFPT~\cite{gonze1997,baroni2001}. 
The IFCs are then transformed into reciprocal space on dense momentum grids, giving rise to the dynamical matrices. 
This can be achieved by a Fourier interpolation of the IFCs.

The small displacement method fits the IFCs in the Taylor-expanded interatomic forces:
\begin{equation}\label{eq:F_compent}
F_{\varkappa\alpha} = -\sum_{\varkappa'\beta}\Phi_{\varkappa\alpha,\varkappa'\beta} u_{\varkappa'\beta}+\cdots.
\end{equation}
This can be recast in matrix form as a linear regression problem~\cite{zhou2019}
\begin{equation}\label{eq:F_matrix}
\bold{F} = \mathbb{A}\cdot\bold{\Phi} = \mathbb{A}\cdot\mathbb{C}\cdot \boldmath{\phi},
\end{equation}
where $\mathbb{A}$ is the displacement matrix, $\bold{\Phi}$ is an IFC vector, and $\mathbb{C}$ is null space constructed from all symmetry constraints on the IFCs. 
The constraints contain those from space group symmetry,  permutation symmetry and the invariance conditions discussed so far. 
Through the null space $\mathbb{C}$, the IFCs $\bold{\Phi}$ can be projected into the independent IFC parameters $\boldsymbol{\phi}$.
Then, a least-square or compressive sensing technique~\cite{zhou2019} can be applied to determine the unknown parameters $\boldsymbol{\phi}$, and the IFCs $\bold{\Phi}$ are further obtained via $\bold{\Phi}=\mathbb{C}\cdot\boldsymbol{\phi}$.

In DFPT calculations, the IFCs are corrected to make them fulfill the invariance conditions and then Fourier interpolated. 
Two approaches exist to correct the IFCs. 
The first one consists in projecting the IFCs onto a subspace, spanned by the vector basis of the invariance conditions and obtaining the solution by nearest distance minimization~\cite{mounet2005}.
In order to get the projection $\bold{\Phi}^{\rm p}$ of the IFCs in such a subspace, the orthogonal basis set $\{g_i\}$ of the subspace needs to be built from the constraint matrix of invariance conditions and yields
\begin{equation}\label{eq:projection}
\bold{\Phi}^{\rm p}=\sum_i(\bold{\Phi} \cdot g_i)g_i,
\end{equation}
which is the nearest distance between the original and the corrected IFCs.
The corrected IFCs $\bold{\Phi}^{\rm c}$ that satisfy the invariance conditions are obtained via $\bold{\Phi}^{\rm c}= \bold{\Phi}-\bold{\Phi}^{\rm p}$.
We have implemented the invariance conditions through this optimal projection of IFCs in the \textsc{Quantum ESPRESSO} distribution~\cite{giannozzi2009,giannozzi2017}.

The second approach adds a correction to the IFCs and then minimizes the correction to remain as close as possible to the original IFCs.
We define the deviation from the invariance conditions as
\begin{equation}\label{eq:deviation_d}
\bold{d} \equiv \mathbb{I}\cdot\mathbb{C^\prime}\cdot\boldmath{\phi},
\end{equation}
where $\mathbb{I}$ is the constraint matrix of the invariance conditions. 
The null space $\mathbb{C^\prime}$ introduced in Equation~\eqref{eq:deviation_d} ensures that any correction made will not break other symmetries, such as space group symmetry. 
The prime on null space indicates that $\mathbb{C^\prime}$ is evaluated excluding the invariance conditions that we want to correct.
We apply a small correction $\Delta\boldsymbol{\phi}$ such that
\begin{equation}\label{eq:correction_0}
\mathbb{I}\cdot\mathbb{C^\prime}\cdot(\boldsymbol{\phi}+\Delta \boldsymbol{\phi})=\bold{0},
\end{equation} 
which means that $\mathbb{I} \cdot \mathbb{C^\prime} \cdot \Delta \boldsymbol{\phi} = \bold{-d}$.
We can solve this problem efficiently by utilizing ridge regression~\cite{hilt1977} with the help of a penalty term $\mu\Vert \Delta \boldsymbol{\phi}\Vert^2_2$ where $\Vert \dots \Vert_2$ denotes the $\ell_2$-norm and  $\mu$ is a hyperparameter with values usually in the range 10$^{-2}$ to 10$^{-6}$, which ensures that the correction $\Delta \boldsymbol{\phi}$ on the IFCs is as small as possible:
\begin{equation}\label{eq:ridge}
\Delta\boldmath{\phi}=\arg \min_{\Delta\boldmath{\phi}}\Vert\bold{d}+\mathbb{I}\cdot\mathbb{C^\prime}\cdot\Delta\boldmath{\phi}\Vert^2_2+\mu\Vert\Delta\boldmath{\phi}\Vert^2_2.
\end{equation}
A similar implementation that applies ridge regression to impose the invariance conditions on the IFCs can be found in the \textsc{hiPhive} software~\cite{eriksson2019}.

In the case of crystals, the periodic-boundary conditions need to be taken into account in order to impose the invariance conditions properly, and this can be tackled easily by constructing the Wigner-Seitz supercell.
Thanks to the periodicity of the lattice and translational invariance, all the interactions are considered between the atoms in the chosen central unit cell and the other atoms in the supercell. 
To construct the Wigner-Seitz supercell, their relative distances are recalculated as the nearest periodic image. 
In the case of equal distances among their periodic images, a weight inversely proportional to its cardinality is given.  
As a result, the sum of the weights is unity because all of the atom pairs in the supercell contribute uniquely to the potential energy. 
This approach is similar to the one implemented in the \textsc{EPW} code~\cite{Ponce2016,Ponce2021} to improve the accuracy of the interpolated  dynamical matrix, which optimizes atomic pairs to ensure that the distance between each two of atomic positions are within a cutoff radius.

\paragraph*{\normalfont\textbf{Fourier interpolation of interatomic force constants}}

The dynamical matrix $D_{\kappa\alpha,\kappa'\beta}(\mathbf{q})$ is computed via a Fourier interpolation of the real-space IFCs as:
\begin{equation}\label{eq:dy2fc}
D_{\kappa\alpha,\kappa'\beta}(\mathbf{q}) = \sum_{\mathbf{R}} \Phi_{\kappa\alpha,\kappa'\beta}(\mathbf{0},\mathbf{R}) {\rm e}^{-{\rm i}\mathbf{q}\cdot\mathbf{R}},
\end{equation}
where $\bold{q}$ is a phonon wavevector defined on an arbitrary grid in reciprocal space.
Due to the transitional invariance, the IFCs $\Phi_{\varkappa\alpha,\varkappa'\beta}$ in crystalline solids depend only on the relative position between the two atoms $\varkappa$ and $\varkappa'$.
Therefore, it is convenient to consider only the IFCs $\Phi_{\kappa\alpha,\kappa'\beta}(\mathbf{R})$ between the atom labelled by $\kappa$ within the reference unit cell $\mathbf{0}$ and the atom labelled by $\kappa'$ within another unit cell $\mathbf{R}$.

However, in the case of infrared-active materials, the atomic vibrations generate long-range Coulomb interactions that can deteriorate the quality of the Fourier interpolations~\cite{baroni2001,gonze1997,sohier2017}. 
A common strategy~\cite{gonze1997} to overcome this problem is to split the dynamical matrix into a short-range ($\mathcal{S}$) and long-range ($\mathcal{L}$) component as:
\begin{equation}
D_{\kappa\alpha,\kappa'\beta}(\mathbf{q}) = D_{\kappa\alpha,\kappa'\beta}^{\mathcal{S}}(\mathbf{q}) + D_{\kappa\alpha,\kappa'\beta}^{\mathcal{L}}(\mathbf{q}),
\end{equation}
such that the short-range part is a smooth function of $\mathbf{q}$.
If we consider only dipole-dipole interactions~\cite{gonze1997,Royo2020}, the long-range part takes the following form:
\begin{equation}\label{eq:nac_D}
D_{\kappa\alpha,\kappa'\beta}^{\mathcal{L}}(\mathbf{q\rightarrow0}) = \frac{W_{\rm c}(\bold{q})e^2}{\Omega_0}\Big[\big(\bold{q}\cdot\bold{Z}^*_{\kappa}\big)_\alpha \big(\bold{q}\cdot\bold{Z}^*_{\kappa^\prime}\big)_\beta\Big],
\end{equation}
which is non-analytic at $\mathbf{q}=\mathbf{0}$ and induces the direction-dependent LO-TO splitting.
To deal with the $\mathbf{q} = \mathbf{0}$ term exactly,  the Ewald summation technique~\cite{gonze1997,baroni2001} is adopted in practice with a sum over reciprocal $\mathbf{G}$ vectors. 
In this expression, $e$ is the elementary charge unit, $\Omega_0$ is the volume of unit cell, $\bold{Z}^*_{\kappa}$ is the Born effective charge tensor of the atom $\kappa$ within the unit cell, and $W_{\rm c}(\bold{q})$ is the screened Coulomb potential which depends on the dimensionality of the system as~\cite{sohier2017,royo2021,rivano2022}
\begin{align}
&W_{\rm c}^{\rm 3D}(\bold{q})=\frac{4\pi}{\bold{q}\cdot \boldsymbol{\epsilon} \cdot\bold{q}}, \label{eq:W_3D} \\
&W_{\rm c}^{\rm 2D}(\bold{q})=\frac{2\pi}{|\bold{q}|\left(1+\frac{\bold{q}\cdot \mathbf{r}_{\rm eff}\cdot\bold{q}}{|\bold{q}|^2}|\bold{q}|\right)}, \label{eq:W_2D} \\
&W_{\rm c}^{\rm 1D}(\bold{q})=\frac{ 4 \pi}{\epsilon_\mathrm{1D}|\bold{q}|^2 \pi t^2} \Bigg\{ 1-2I_1 (|\bold{q}| t) K_1 (|\bold{q}| t)\bigg[1- \nonumber \\
&\frac{2\epsilon_\mathrm{1D}\sqrt\pi |\bold{q}| tI_1(|\bold{q}| t)K_0(|\bold{q}| t)-G^{2 2}_{2 4}(|\bold{q}|^2 t^2)} {2\sqrt \pi |\bold{q}| t\big[\epsilon_\mathrm{1D}I_1(|\bold{q}| t)K_0(|\bold{q}| t)+I_0(|\bold{q}| t)K_1(|\bold{q}| t)\big]}\bigg]\Bigg\}, \label{eq:W_1D}
\end{align}
where $\boldsymbol{\epsilon}$ is the ion-clamped dielectric tensor of 3D system, $\bold{r}_{\rm eff}\approx \boldsymbol{\epsilon} b/2$ is the in-plane matrix of effective screening length, $b$ is the thickness of the 2D material, $\epsilon_{\rm 1D}$ is the 1D dielectric constant, $t$ is the effective screening radius, $I_n(x)$ and $K_n(x)$ are the $n$th-order modified cylindrical Bessel functions, and $G^{mn}_{pq}$ is the Meijer G-function~\cite{meijer1936}.
The 1D long-range formula, Equation~\eqref{eq:W_1D}, has been recently derived and more details can be found in Ref.~\onlinecite{rivano2022}.
The long-range part is substracted from the dynamical matrix on the coarse grid, the remaining short-range part is Fourier transformed into real space to get the IFCs.

Inspired by this approach, we formulate the polar Born-Huang invariances and polar Huang conditions by separating the total IFCs into short-range and long-range IFCs,
\begin{multline}\label{eq:polar_BH}
\sum_{\varkappa'}\Big(\Phi^{\mathcal{S}}_{\varkappa\alpha,\varkappa'\beta}  + 
 \Phi^{\mathcal{L}}_{\varkappa\alpha,\varkappa'\beta}\Big )\tau_{\varkappa'\gamma}
= \\
 \sum_{\varkappa'}\Big(\Phi^{\mathcal{S}}_{\varkappa\alpha,\varkappa'\gamma} +  \Phi^{\mathcal{L}}_{\varkappa\alpha,\varkappa'\gamma}\Big) \tau_{\varkappa'\beta},
\end{multline}
and
\begin{multline}\label{eq:polar_H}
\sum_{\varkappa\varkappa'}\Big( \Phi^{\mathcal{S}}_{\varkappa\alpha,\varkappa'\beta} +
 \Phi^{\mathcal{L}}_{\varkappa\alpha,\varkappa'\beta} \Big) \tau_{\varkappa\varkappa'\gamma} \tau_{\varkappa\varkappa'\delta} = \\
\sum_{\varkappa\varkappa'} \Big( \Phi^{\mathcal{S}}_{\varkappa\gamma,\varkappa'\delta} 
+
 \Phi^{\mathcal{L}}_{\varkappa\gamma,\varkappa'\delta} \Big) \tau_{\varkappa\varkappa'\alpha} \tau_{\varkappa\varkappa'\beta}.
\end{multline}
Since the $\Phi^{\mathcal{L}}$ are analytically defined, we apply the corrections due to the invariance conditions only on $\Phi^{\mathcal{S}}$. 
Due to the presence of the long-range terms, we have to solve an inhomogeneous set of linear equations which is here achieved by using a ridge regression method~\citep{hilt1977}.
We then Fourier interpolate the resulting corrected short-range IFCs at arbitrary momentum points and add back the analytic long-range part.   
This approach is an alternative to applying the Born-Huang invariances, Equation~\eqref{eq:B-H}, and Huang conditions, Equation~\eqref{eq:H-I}, 
on the \textit{total} real-space IFCs.
We have verified in Fig.~\ref{fig:MoS2_DFPT} that both methods give the same interpolated 2D phonon band structures.
We also reiterate that we have neglected dynamical quadrupoles in the long-range treatments which have been recently shown to be important in some cases~\cite{Brunin2020,Royo2020,royo2021,Ponce2021} and have been left for future works.

Finally, in the case of low-symmetry infrared-active solids, we have observed that the total dynamical matrices are not guaranteed to be Hermitian.
As observed in Ref.~\onlinecite{zhou2019-2}, the dipole-dipole IFC ansatz (e.g. Equation~(4) of Ref.~\onlinecite{Gonze1994} or Equation~(70) of Ref.~\onlinecite{gonze1997}) which is used in the analytic form of the long-range dynamical matrix does not guarantee Hermiticity for a general crystal. 
Zhou \textit{et al.}~\cite{zhou2019-2} addressed this issue by adding a real-space short-range correction to the IFC matrix.
Here, for simplicity, we directly symmetrize the long-range part of the on-site dynamical matrix by its conjugate transpose and Fourier transform 
both the short-range and long-range parts into real space. 
We then apply the invariance conditions on the total IFCs and subsequently substract the symmetrized long-range in real space. 
Last, we Fourier interpolate the short-range IFCs to arbitrary momentum and add the long-range part symmetrized in the same way.

\paragraph*{\normalfont\textbf{Bending modes in 2D materials}}\label{sec:long-wavelength}

The lattice dynamics of crystalline solids is governed by the secular equation~\cite{dove1993,born1954}
\begin{equation}\label{eq:eigeneq}
[\omega_\nu(\mathbf{q})]^2e_{\nu,\kappa\alpha}(\mathbf{q})=\sum_{\kappa'\beta}\frac{D_{\kappa\alpha,\kappa'\beta}(\mathbf{q})}{\sqrt{m_\kappa m_{\kappa'}}}e_{\nu,\kappa'\beta}(\mathbf{q}),
\end{equation}
where $m_\kappa$ is the mass of the $\kappa$-th atom within the unit cell.
The eigenvector $e_{\nu,\kappa\alpha}(\mathbf{q})$ is obtained by diagonalizing the mass-scaled dynamical matrix $D_{\kappa\alpha,\kappa'\beta}(\mathbf{q})/\sqrt{m_\kappa m_{\kappa'}}$ with the corresponding eigenfrequency $\omega_\nu(\mathbf{q})$ for the phonon mode $\nu$ at momentum  $\mathbf{q}$, and $\nu$ ranges from $1$ to $3n$ with $n$ atoms in the unit cell.
To analyze the dispersion relation of bending modes in the long-wavelength limit ($\mathbf{q}\rightarrow0$), we 
perform a Taylor expansion of Equation~\eqref{eq:eigeneq} in terms of $\mathbf{q}$.
The resulting long-wavelength equation for lattice vibrations is~\cite{croy2020,born1954}:
\begin{multline}\label{eq:acou_wave}
\sum_\kappa m_\kappa[\omega^{(1)}_\nu(\mathbf{q})]^2u_{\nu\alpha}=\\
\sum_{\beta\gamma\delta} \Big[ T^{\rm sym}_{\alpha\beta,\gamma\delta}q_\gamma q_\delta + T^{\rm int}_{\alpha\gamma,\beta\delta}q_\gamma q_\delta\Big]u_{\nu\beta},
\end{multline}
where $\omega^{(1)}_\nu(\mathbf{q}) \equiv \frac{\partial \omega_\nu(\mathbf{q})}{\partial \mathbf{q}}\Big|_{\mathbf{q}=0} \cdot\mathbf{q}$. 
In this equation, $u_{\nu\alpha}$ is an arbitrary polarization vector which is introduced via the solution for the zero-order equation of the Taylor expansion, $e^{(0)}_{\nu,\kappa\alpha}(\mathbf{q})=\sqrt{m_\kappa}u_{\nu\alpha}$.
The $T^{\rm sym}_{\alpha\beta,\gamma\delta}$ defined in Equation~\eqref{eq:H_inf} 
can be rewritten in term of a sum over primitive cells as
\begin{equation}\label{eq:bracket}
T^{\rm sym}_{\alpha\beta,\gamma\delta}=-\frac{1}{2}\sum_{\mathbf{R}}\sum_{\kappa\kappa'}\Phi_{\kappa\alpha,\kappa'\beta}(\mathbf{R}) \tau_{\kappa\kappa'\gamma}(\mathbf{R}) \tau_{\kappa\kappa'\delta}(\mathbf{R}),
\end{equation}
and $T^{\rm int}_{\alpha\gamma,\beta\delta}$ is the contribution associated with the relaxation of internal coordinates in response to external stress fields~\cite{born1954}:
\begin{align}\label{eq:curly}
T^{\rm int}_{\alpha\gamma,\beta\delta}=&-\sum_{\kappa\kappa'}\sum_{\lambda\mu}\frac{\Gamma_{\kappa\lambda,\kappa'\mu}}{\sqrt{m_\kappa m_{\kappa'}}} \nonumber \\
&\times \Bigg[\sum_{\mathbf{R}}\sum_{\kappa''}\Phi_{\kappa\lambda,\kappa''\alpha}(\mathbf{R})\tau_{\kappa\kappa''\gamma}(\mathbf{R})\Bigg] \nonumber \\
&\times \Bigg[\sum_{\mathbf{R'}}\sum_{\kappa'''}\Phi_{\kappa'\mu,\kappa'''\beta}(\mathbf{R'})\tau_{\kappa'\kappa'''\delta}(\mathbf{R'})\Bigg],
\end{align}
with the $\nu\times \nu$ matrix $\Gamma_{\kappa\lambda,\kappa'\mu}$ denoting the inverse of the mass-scaled IFC matrix $\sum_{\mathbf{R}}\Phi_{\kappa\lambda,\kappa'\mu}(\mathbf{R})/\sqrt{m_{\kappa}m_{\kappa'}}$.
Unfortunately, the IFC matrix is singular because of the acoustic sum rules for translational invariance and the $\Gamma_{\kappa\lambda,\kappa'\mu}$ is thus introduced as
\begin{equation}\label{eq:Gamma}
\Gamma_{\kappa\lambda,\kappa'\mu}=
\begin{cases}
\hat{\Gamma}_{\kappa\lambda,\kappa'\mu} &\kappa,\kappa'\neq1\\
0 & \text{otherwise}
\end{cases},
\end{equation}
where the pseudoinverse $\hat{\Gamma}_{\kappa\lambda,\kappa'\mu}$ is calculated and satisfies the relation~\cite{born1954}
\begin{equation}\label{eq:inverse}
\sum_\mu\sum_{\kappa'=2}\frac{\hat{\Gamma}_{\kappa\lambda,\kappa'\mu}}{\sqrt{m_{\kappa'}m_{\kappa''}}}\sum_{\mathbf{R}}\Phi_{\kappa'\mu,\kappa''\beta}(\mathbf{R})=\delta_{\lambda\beta}\delta_{\kappa\kappa''}.
\end{equation}
Furthermore, the tensor $T^{\rm int}_{\alpha\gamma,\beta\delta}$ satisfies the following symmetry relations~\cite{huang1950,born1954}:
\begin{equation}\label{eq:H'_sym}
T^{\rm int}_{\alpha\gamma,\beta\delta}=T^{\rm int}_{\gamma\alpha,\beta\delta}=T^{\rm int}_{\beta\delta,\alpha\gamma},
\end{equation}
which result from the rotational invariance, Equation~\eqref{eq:B-H}, and the fact that $\Gamma_{\kappa\lambda,\kappa'\mu}=\Gamma_{\kappa'\mu,\kappa\lambda}$.

For the ZA bending modes of 2D materials in the long-wavelength limit, if they were not purely polarized along the vacuum direction $z$ and coupled with the in-plane vibrations, the leading linear term in the ZA dispersion would read:
\begin{multline}\label{eq:ZA_mode}
[\omega^{(1)}_{\rm ZA}(\mathbf{q})]^2 =\\
 \frac{1}{\sum_\kappa m_\kappa}\sum_\beta\frac{u_{\beta} }{u_z}\sum_{\gamma,\delta\in\{x,y\}}\Big[T^{\rm sym}_{z\beta,\gamma\delta}+T^{\rm int}_{z \gamma,\beta\delta}\Big]q_\gamma q_\delta,
\end{multline}
where the phonon band index $\nu \equiv {\rm ZA}$ is omitted for clarity. 
By using the Huang conditions and the symmetry of $T^{\rm int}_{\alpha\gamma,\beta\delta}$,  the identity $T^{\rm sym}_{z\beta,\gamma\delta}+T^{\rm int}_{z \gamma,\beta\delta}=T^{\rm sym}_{\gamma\delta,z\beta}+T^{\rm int}_{\gamma z,\delta\beta}$ holds.
As demonstrated in Supplementary Note~1~\cite{si}, we have $T^{\rm sym}_{\gamma\delta,z\beta}+T^{\rm int}_{\gamma z,\delta\beta} = 0$
even in case of non-purely out-of-plane modes \textit{if} the rotational invariance of the lattice potential is fullfilled \textit{and} the system is free of stress, leading to the vanishing linear dispersion term. 
In the long-wavelength limit,  the dispersion of ZA modes thus becomes quadratic with a purely out-of-plane polarization.
Having proven this, we can therefore define the bending conditions as:
\begin{equation}
T^{\rm sym}_{\gamma\delta,z\beta}=-T^{\rm int}_{\gamma z,\delta\beta},
\end{equation}
which can be imposed in case of numerical noise in the calculations to ensure a quadratic dispersion of the flexural modes.

More generally, these conditions for recovering quadratic flexural modes are also applicable to 1D materials, where there are two acoustic bending modes polarizing in two vacuum directions.
In 1D materials with periodicity in the $z$ direction, the bending conditions thus become $T^{\rm sym}_{\gamma\delta,x\beta}=-T^{\rm int}_{\gamma x,\delta\beta}$ and $T^{\rm sym}_{\gamma\delta,y\beta}=-T^{\rm int}_{\gamma y,\delta\beta}$ for polarizations in the $x$ and $y$ directions, respectively.
We note that if LD materials are subjected to external stress fields, the linear term will then dominate the phonon dispersion in the long-wavelength limit. 
In bulk materials, no such bending conditions exist due to the periodicity in all three directions.

\paragraph*{\normalfont\textbf{Application to bulk materials}}\label{sec:bulk}

\begin{figure}[t]
  \centering
  \includegraphics[width=0.99\linewidth]{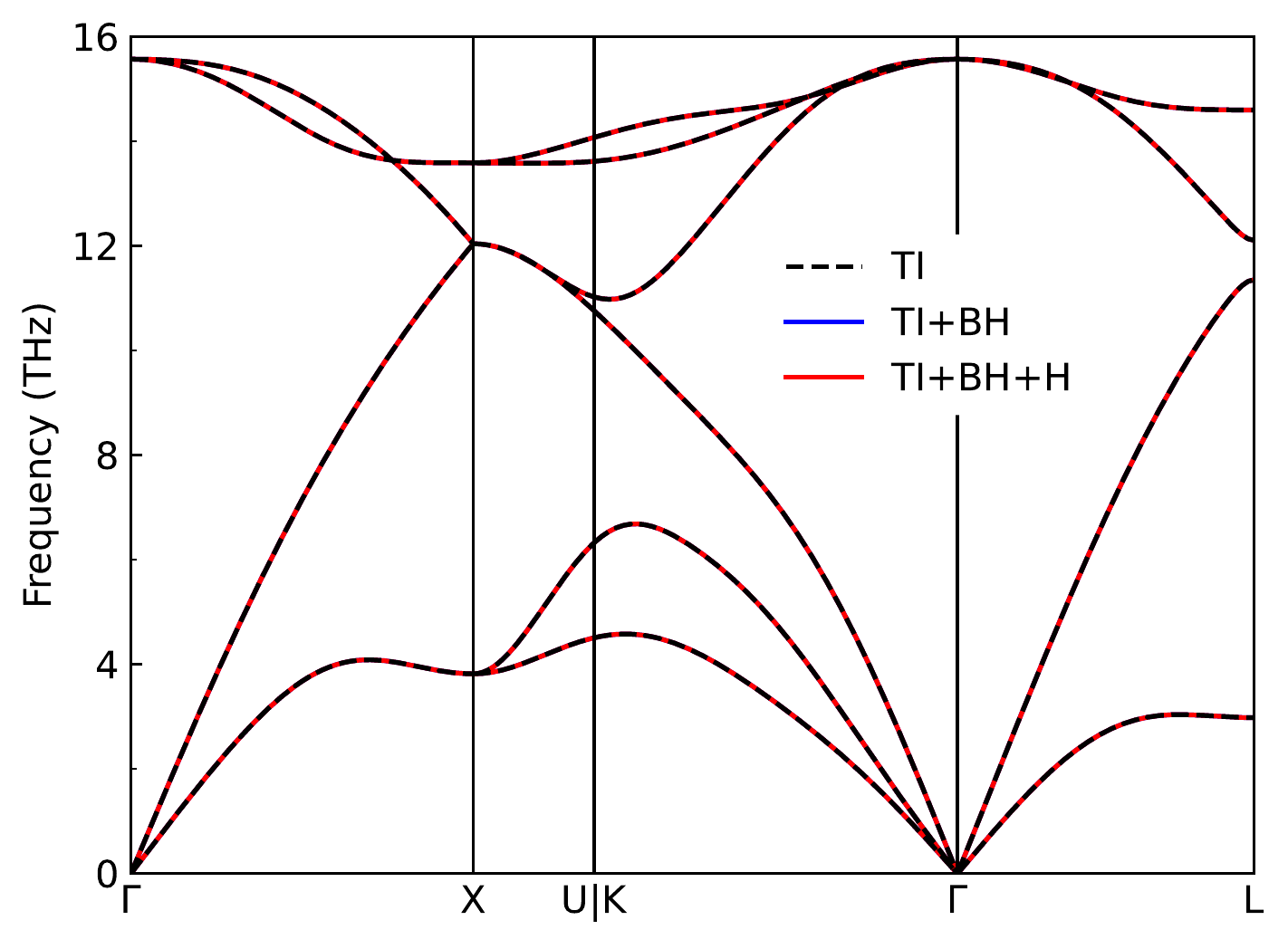}
  \caption{\label{fig:Si}
Effects of the invariance conditions on the phonon dispersion of bulk silicon. 
TI denotes the correction of translational invariance on the IFCs,  while BH and H represent the corrections by Born-Huang rotational invariance and Huang conditions, respectively.
}
\end{figure}

The influence of the rotational invariance on the lattice dynamics of bulk materials is almost unexplored.
We show the phonon dispersion of silicon in Fig.~\ref{fig:Si} and find that both the Born-Huang invariances (red line) alone and the combination of Born-Huang invariances and Huang conditions (blue line) have little effect, compared with the phonon dispersion obtained from the IFCs without the aforementioned conditions imposed (black dashed line).  
This result is not surprising given that the IFCs of high-symmetry bulk crystals with space inversion automatically fulfill rotational invariance~\cite{sluiter1998}.

\begin{figure}
  \centering
  \includegraphics[width=0.99\linewidth]{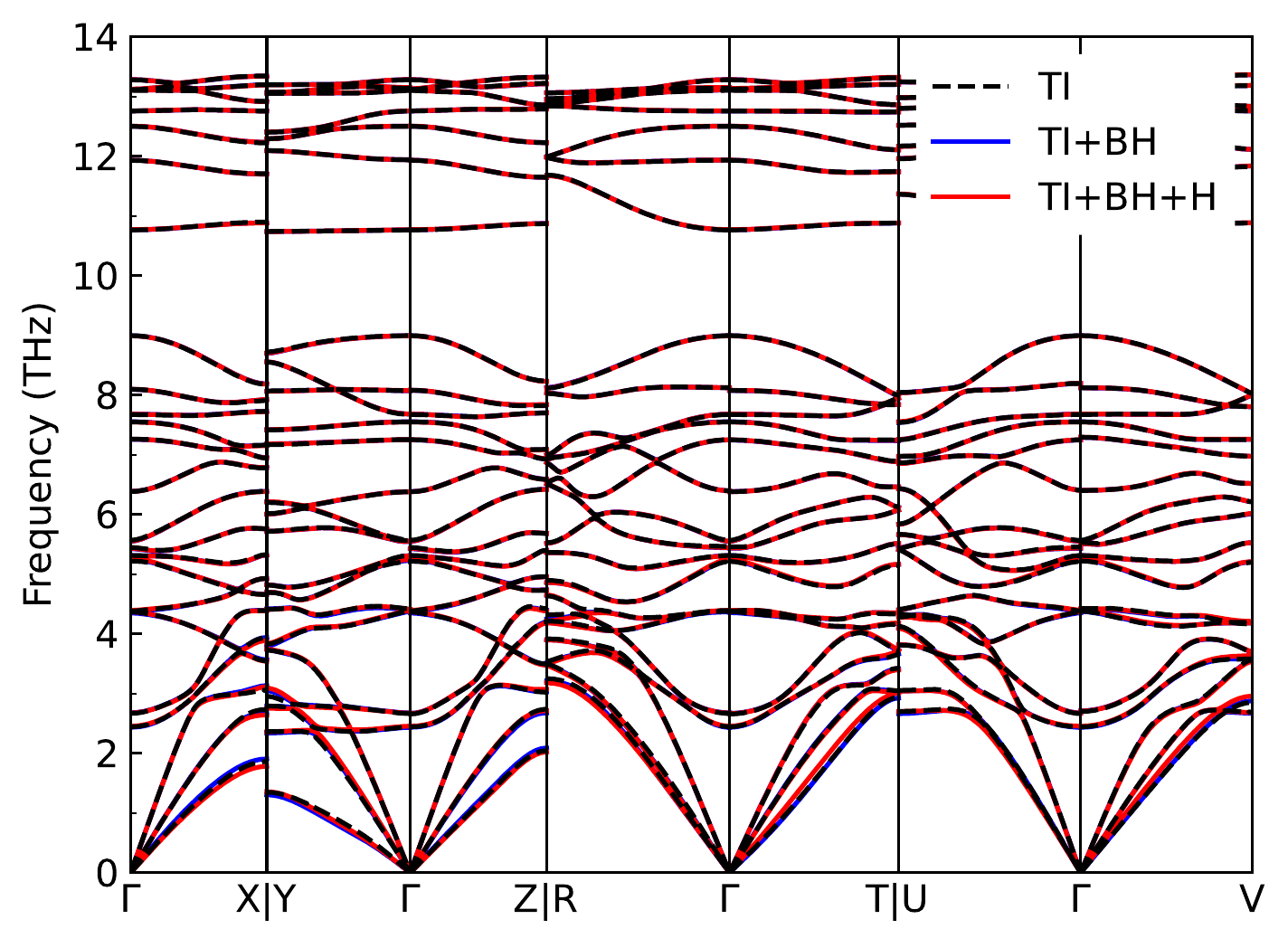}
  \caption{\label{fig:CaP3}
Influence of invariance conditions on the phonon dispersion of the infrared-active bulk CaP$_3$. 
TI denotes the correction of translational invariance on the IFCs,  while BH and H represent the corrections by Born-Huang rotational invariance and Huang conditions, respectively.
}
\end{figure}

As an example of infrared-active bulk materials, we calculate the phonon dispersion of triclinic CaP$_3$ shown in Fig.~\ref{fig:CaP3}, with the LO-TO splitting included.
The crystal symmetry of CaP$_3$ with the space group of $\rm {P\bar{1}}$ is low: only the identity and the spatial inversion operations. 
For a $2\times2\times2$ supercell with 64 atoms, there are 160 unique atomic pairs which means 1440 IFC components.
The total number of independent IFC parameters in CaP$_3$ is 1215 after considering the space group symmetry, the permutation symmetry and the translational invariance. 
When we add the Born-Huang invariances on the IFCs, the number of independent IFCs is reduced to 1179.
The number of independent IFC parameters can further decrease to 1164 if the Huang conditions are included as symmetry constraints.
Besides, there is a large deviation from these two invariance conditions reflected by the Frobenius norm of $\mathbf{d}$ in Equation~\eqref{eq:deviation_d}; it vanishes with a magnitude of $10^{-6}$ after the invariance conditions are enforced, which indicates that the Born-Huang invariances and Huang conditions are now satisfied. 
However, the phonon dispersions of CaP$_3$ in Fig.~\ref{fig:CaP3} are almost unchanged.
There is only a small modification of the lowest-lying transverse acoustic (TA) branch.
Hence, we conclude that the violation of the rotational invariance does not affect the phonon dispersion of bulk materials at equilibrium, and the corrections of the invariance conditions on the IFCs of bulk materials are not important, even for low-symmetry crystals. 
This can be rationalized from the fact that bulk materials are periodic along all directions and thus do not have any rigid rotation.
Nonetheless, for the elastic tensor $c_{ij}$ (in Voigt notation) determined directly from the second-order IFCs, the broken symmetry relation  (i.e. $c_{ij}\neq c_{ji}$) of low-symmetry solids can be observed due to the violation of the rotational invariance and equilibrium conditions~\cite{sluiter1998}.

\paragraph*{\normalfont\textbf{Application to 2D materials}}\label{sec:2D}

\begin{figure}[t]
  \centering
  \includegraphics[width=0.99\linewidth]{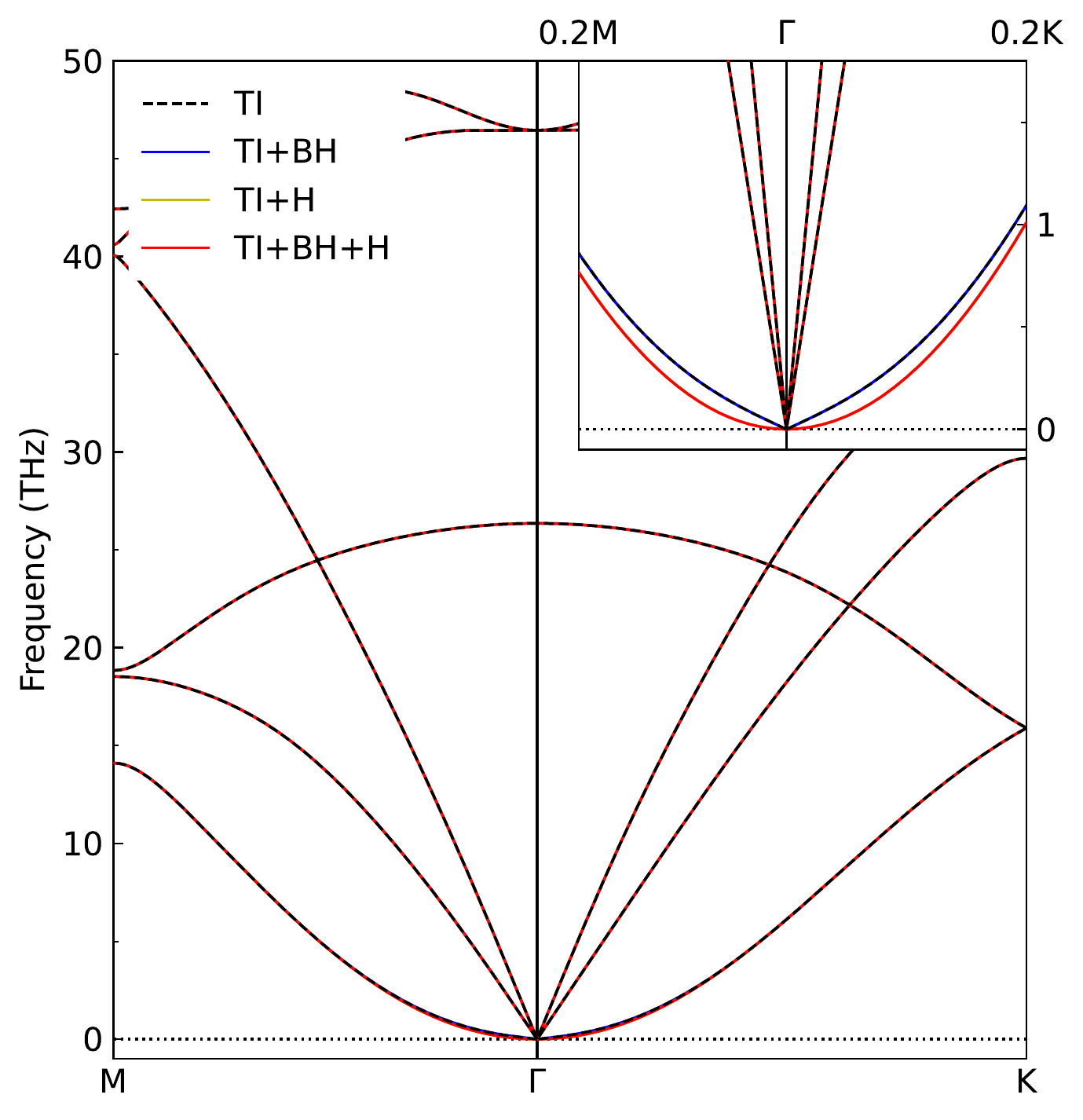}
  \caption{\label{fig:Gr}
Invariance conditions applied on the phonon dispersion of graphene. 
TI denotes the correction of translational invariance on the IFCs,  while BH and H represent the corrections by Born-Huang rotational invariance and Huang conditions, respectively.
The inset shows a zoom-in result around the $\Gamma$ point.
The yellow line coincides with the red one as the rotational invariance is automatically satisfied in graphene.
}
\end{figure}

As revealed by a few recent studies~\cite{carrete2016,eriksson2019,croy2020}, the violation of the rotational invariance and vanishing stress condition are expected to have a remarkable effect on the lattice-dynamical properties of 2D materials, which can lead to an unphysical dispersion relation of the out-of-plane ZA branch.
The influence of such two invariance conditions on the phonon dispersions of infrared-inactive 2D materials is illustrated in Fig.~\ref{fig:Gr}, where we take graphene as a representative material.
In general, the lack of these two invariance conditions can result in a ZA mode either displaying imaginary frequencies or a non-quadratic dispersion relation in the long-wavelength limit. 
We confirm these findings by fitting the IFCs obtained from randomly displaced configurations. 
As can be seen in the inset of Fig.~\ref{fig:Gr}, the pristine phonon dispersion (black dashed line) of graphene without the Born-Huang invariances and Huang conditions exhibits a linear ZA branch close to the zone center. 
When we add the Born-Huang invariances to make the IFCs satisfy the global rotational invariance, the result (blue line) remains unchanged with respect to the pristine one.
This happens because the crystal symmetry of graphene is high and the rotational invariance of its lattice potential is automatically satisfied.
In contrast, only the red and yellow lines, where the additional Huang conditions are fulfilled, show the expected parabolic dispersion of the ZA branch around the $\Gamma$ point.
More generally, both the Born-Huang invariances and Huang conditions must be satisfied to observe the physically correct ZA phonons with quadratic dispersions in the long-wavelength limit.
To illustrate this, we consider an MoS$_2$ monolayer (see Supplementary Fig.~1~\cite{si}), which lacks the additional inversion symmetry.
We see that fulfilling the vanishing stress condition alone is not sufficient for obtaining a parabolic ZA branch, and that the rotational invariance is required as well.
We therefore deduce that residual stress remains after structural optimization and that this equilibrium condition enforces strict null stress to recover the quadratic dispersion relation of the ZA phonons.
In particular, there is still a non-negligible stress in the vacuum direction even after a high-quality structural optimization with a Coulomb cutoff for 2D materials~\citep{sohier2017,sohier2017-2}, owing to the existing weak layer interactions from the periodic images.
This observation is consistent with the dispersion model of long-wavelength bending modes in Equation~\eqref{eq:ZA_mode} which indicates that any external stress will give rise to a non-vanishing linear dispersion term and hence suppresses the next-leading quadratic dispersion.
Overall, both the Born-Huang rotational invariance and Huang conditions (vanishing stress condition) are of great importance for 2D materials to have the expected behavior of the ZA branch in the long-wavelength limit.

\begin{figure}[t]
  \centering
  \includegraphics[width=0.99\linewidth]{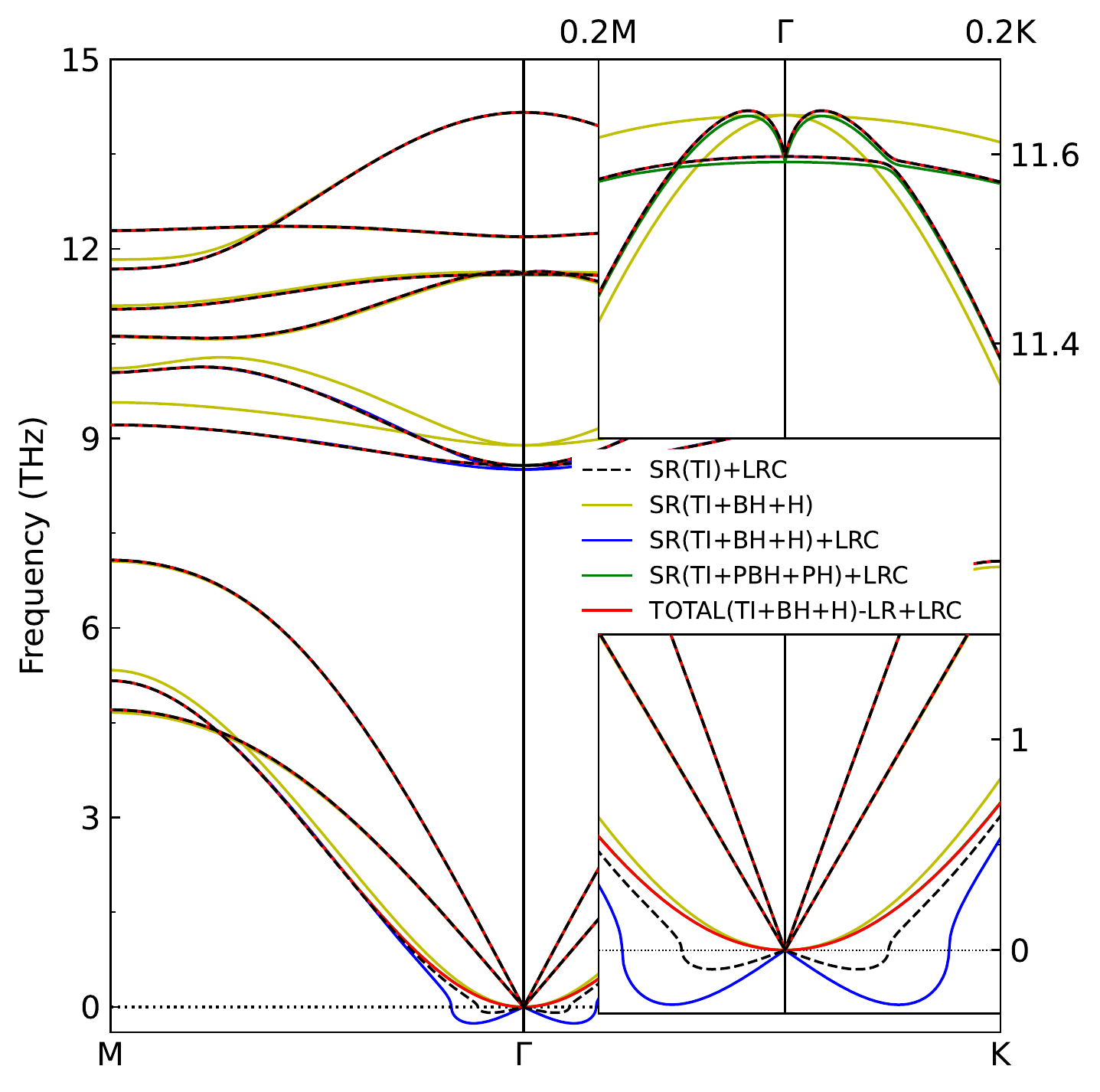}
  \caption{\label{fig:MoS2_DFPT}
Comparison between the phonon dispersions of an infrared-active 2D MoS$_2$ with the corrections of Born-Huang (BH) rotational invariance and Huang (H) conditions on the short-range (SR) IFCs and on the total IFCs.
The polar Born-Huang (PBH) rotational invariance and polar Huang (PH) conditions are also considered. 
The TI denotes the correction of translational invariance.
When correcting the total IFCs, the contributions from the long-range (LR) IFCs are further removed in real space and an analytic long-range correction (LRC) is then added back in reciprocal space to recover the LO-TO splitting. 
The inset show zoom-in results around the $\Gamma$ point and the LO-TO splitting at the $\rm E'$ mode.
}
\end{figure}

Most 2D insulators and semiconductors are infrared-active with non-vanishing Born effective charges and long-range interactions.
In such cases, the long-range Coulomb interactions provide additional contributions to the IFCs~\cite{huang1949,huang1950}.
The enforcement of the invariance conditions for infrared-active materials should therefore be performed with caution.
We take the MoS$_2$ monolayer as an example to showcase this special treatment. 
In 2D materials, the LO-TO splitting vanishes at the zone center, and the LO dispersions become linear in $\mathbf{q}$ with a finite slope~\cite{sohier2017,sohier2017-2}.
As shown in Fig.~\ref{fig:MoS2_DFPT}, the phonon dispersions of MoS$_2$ monolayer calculated directly from the pristine IFCs from DFPT (black dashed line) exhibit small imaginary frequencies in the ZA branch close to the zone center, as a consequence of the broken invariance conditions. 
If we add the corrections according to the invariance conditions from Equations~\eqref{eq:B-H} and \eqref{eq:H-I} on the short-range IFCs only (blue line), there still exist some imaginary frequencies around the $\Gamma$ point, belonging to the ZA phonons. 
This can be understood by realizing that the long-range IFCs indeed contribute to the stress tensor and the rotational invariance of the lattice potential.
When the long-range interactions are added back analytically, the resulting total dynamical matrices in reciprocal space will not fulfill the Born-Huang invariances and Huang conditions. 
This is further confirmed by looking the yellow line in Fig.~\ref{fig:MoS2_DFPT}, where the non-analytic correction corresponding to the long-range interactions are removed, and the result presents a stable ZA branch in the parabolic shape near the $\Gamma$ point (i.e. the short-range IFCs satisfying the invariance conditions).
To deal with the invariance conditions properly in infrared-active solids, we need to consider Equation~\eqref{eq:polar_BH} for the polar Born-Huang invariances and Equation~\eqref{eq:polar_H} for the polar Huang conditions to make the total IFCs meet the invariance conditions (green line).
The long-range IFCs contributed from the dipole-dipole interactions are considered when imposing the invariance conditions on the short-range IFCs only, and an analytic long-range correction is further introduced back in reciprocal space to recover the LO-TO splitting.
As indicated by the red line in Fig.~\ref{fig:MoS2_DFPT}, an equivalent approach is to apply the normal invariance conditions on the total IFCs, substract the long-range part obtained by Fourier transformation, and Fourier interpolate the resulting short-range IFCs with an analytic long-range contribution added back.
Both approaches yield the correct ZA branch and the 2D LO-TO splitting shown in the inset of Fig.~\ref{fig:MoS2_DFPT}.
Apart from the effects on the LO-TO splitting of optical phonons, we observe that the long-range dipole-dipole interactions also modify the ZA modes by lowering their energies when comparing the results in yellow and red lines.
In the rest of the manuscript, we always use the last approach (red line) in infrared-active materials.

\paragraph*{\normalfont\textbf{Application to 1D materials}}\label{sec:1D}

\begin{figure}[t]
  \centering
  \includegraphics[width=0.99\linewidth]{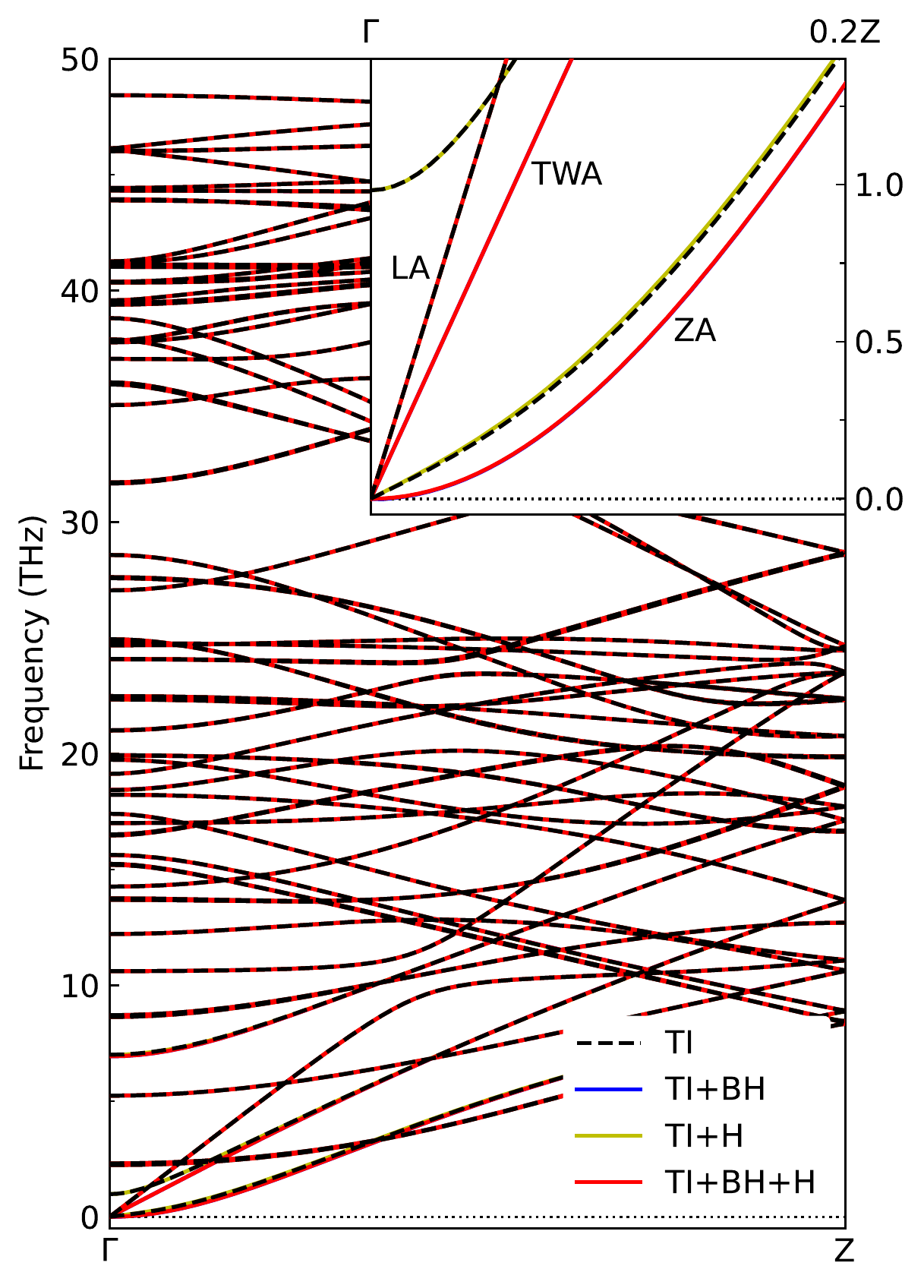}
  \caption{\label{fig:CNT}
Corrections due to the invariance conditions on the phonon dispersion of an infrared-inactive 1D single-wall (8,0) carbon nanotube. 
TI denotes the correction of translational invariance on the IFCs,  while BH and H represent the corrections by Born-Huang rotational invariance and Huang conditions, respectively. 
The inset shows a zoom-in result around the $\Gamma$ point, with four acoustic branches labelled as longitudinal acoustic (LA), twisting acoustic (TWA) and flexural acoustic (ZA, doubly degenerate) modes, respectively.
The blue line coincides with the red one, which indicates the Huang conditions have no effect and the external stress is already vanishing after structural optimization.}
\end{figure}

\begin{figure}[t]
  \centering
  \includegraphics[width=0.99\linewidth]{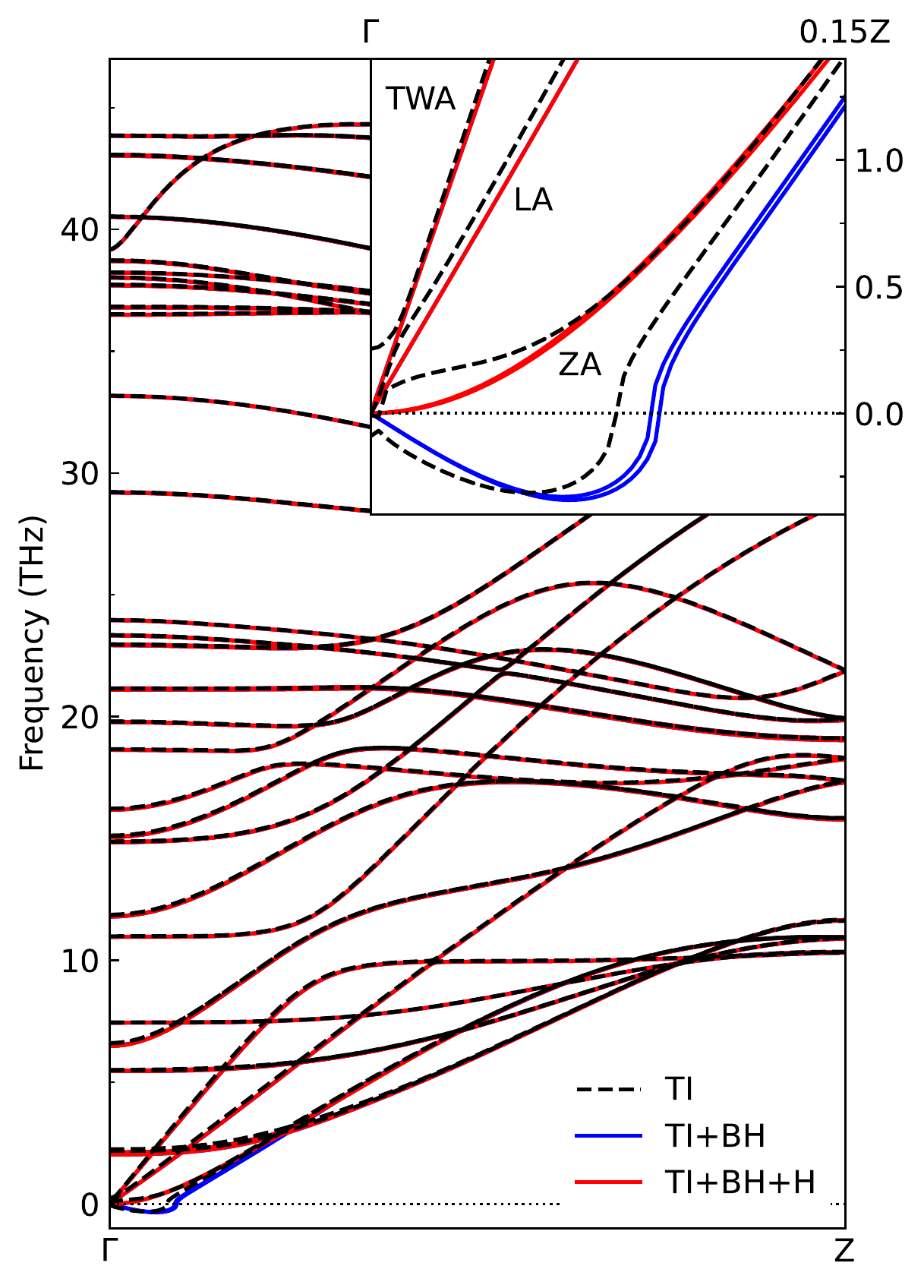}
  \caption{\label{fig:BNNT}
Application of the invariance conditions to the phonon dispersion of an infrared-active 1D single-wall (4,4) boron nitride nanotube. 
TI denotes the correction of translational invariance on the IFCs,  while BH and H represent the corrections by Born-Huang rotational invariance and Huang conditions, respectively. 
The inset shows a zoom-in result around the $\Gamma$ point, with four acoustic branches labelled as longitudinal acoustic (LA), twisting acoustic (TWA) and flexural acoustic (ZA, doubly degenerate) modes, respectively.
}
\end{figure}

Another important class of LD materials are 1D materials.
The effects of the invariance conditions on their lattice-dynamical properties have not yet received much attention.
We choose the single-wall (8,0) CNT and (4,4) BNNT as representative infrared-inactive and -active 1D cases, respectively.
We illustrate in Fig.~\ref{fig:CNT} the corrections from the invariance conditions on the phonon dispersion of (8,8) CNT, where we can see that the results with only the translational invariance applied yield only three vanishing acoustic frequencies at the Brillouin zone center.
In 1D systems with more than one atom in the unit cell, four acoustic branches are expected and correspond to the three global translational degrees of freedom in the three cartesian directions and one rotational degree of freedom along the 1D axis. 
This is correctly recovered when applying the Born-Huang invariances as shown with a red line in Fig.~\ref{fig:CNT}.
As shown previously~\cite{mounet2005,mounet2004,ye2004,libbi2020}, the frequency of the twisting acoustic mode due to the global rotation along the nanotube axis can become finite at the $\Gamma$ point (around 1.0~THz in our case) if the Born-Huang rotational invariance is not enforced.
Also, the dispersion relation of the two acoustic branches corresponding to the bending motions becomes parabolic in the long-wavelength limit after adding the corrections from the Born-Huang rotational invariance.
The Huang conditions for vanishing external stress seem to have no influence on the case of single-wall (8,8) CNT, i.e. the blue line coinciding with the red one.
The conditions of vanishing stress have been already satisfied after a tight structural optimization.
In general, both the Born-Huang invariances and Huang conditions should be fulfilled in order to find the correct dispersions of two bending modes in 1D materials.
It is the case for the infrared-active (4,4) BNNT shown in Fig.~\ref{fig:BNNT}, where the two bending modes with quadratic dispersions are only recovered when both invariance conditions are imposed.
Furthermore, there are long-range Coulomb interactions due to the non-vanishing Born effective charges in BNNT.
The quadratic phonon dispersions of two bending modes and the emergence of twisting acoustic modes with vanishing frequency at the zone center can only be observed with the polar invariance conditions introduced in Equations~\eqref{eq:polar_BH} and~\eqref{eq:polar_H}.
Overall, our proposed invariance scheme can be applied to any material ranging from bulk to 2D and to 1D systems to recover exact lattice-dynamical properties.

\begin{figure*}[ht]
  \centering
  \includegraphics[width=0.99\linewidth]{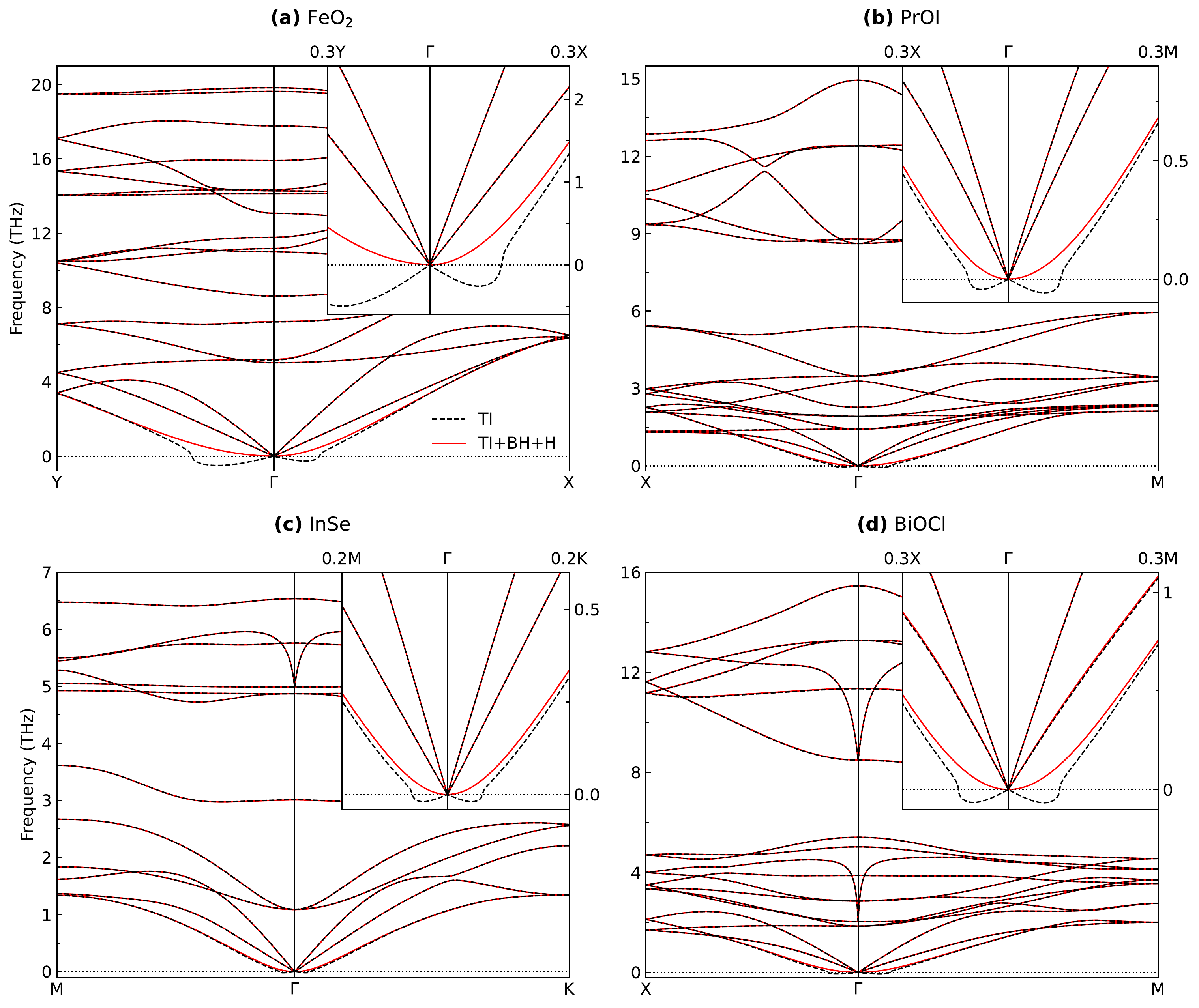}
  \caption{\label{fig:2D_database}
Corrected phonon dispersions of four representative 2D materials from the 2D material database of Ref.~\onlinecite{mounet2020}, (a) FeO$_2$, (b) PrOI, (c) InSe and (d) BiOCl. 
The inset shows a zoom-in result around the $\Gamma$ point.
The black dashed lines are the results computed from the DFPT calculations with the translational invariance (TI) fulfilled, while the red lines denote the results where the Born-Huang (BH) invariances and Huang (H) conditions have also been enforced.
}
\end{figure*}

\paragraph*{\normalfont\textbf{2D materials database}}\label{sec:2D_database}

A recent high-throughput screening by Mounet \emph{et al.}~\cite{mounet2018} on experimentally known layered compounds has identified 1825 materials that can be easily or potentially exfoliated. 
The vibrational properties of 245 of these 2D materials were calculated and explored, with all the data available on the Materials Cloud~\cite{mounet2020}. 
However, a large proportion of the calculated phonon dispersions showed imaginary frequencies, which suggests that the predicted 2D materials would be dynamically unstable. 
Since most of the unstable phonon branches are ZA modes in the zone center region, we apply here our invariance scheme to see if these 
can be stabilized.

We select four representative 2D materials from the database, namely FeO$_2$, PrOI, InSe and BiOCl, and report their phonon dispersions in Fig.~\ref{fig:2D_database}. 
As can be seen, after the invariance conditions are enforced, all of the four phonon dispersions become stable with a quadratic ZA mode in the long-wavelength limit. 
Besides, FeO$_2$ and PrOI are metals, while InSe and BiOCl are inrared-active semiconductors with non-zero Born effective charges. 
The results in Fig.~\ref{fig:2D_database} highlight the effectiveness of the invariance conditions described in this work. 
Finally, our simulations shed a new light on the origin of imaginary frequencies existing in the 2D material database
and shows that in most cases this was due to a violation of some invariance conditions. 
We therefore expect that most 2D materials reported are stable and will be synthesized in the future.

The updated database is available on the Materials Could archive~\cite{lin2022}, and 
the phonon dispersions for all of 245 materials can be found in Supplementary Information~\cite{si} of this study.
After investigation, there are 54 materials that are not dynamically stable, with pronounced imaginary frequencies not located around the $\Gamma$ point in the database (see Supplementary Table~1~\cite{si}).
After these truly unstable 2D structures are excluded, we still have 33 candidates that exhibit an unstable ZA branch with small imaginary frequencies near the $\Gamma$ point after the invariance conditions are enforced (see the detailed information of these 2D materials in Supplementary Table~2~\cite{si}).
There are mainly two reasons why the correction of the invariance conditions fails to recover a stable and quadratic ZA mode at the zone center.
One of them is the numerical accuracy of the structure optimization and DFPT calculation, which might not be sufficient to produce accurate real-space IFCs.
The other is more subtle: one might need to expand Equation~\eqref{eq:expand_potential} also in terms of strain-electric field~\cite{Wu2005} and include the treatment of higher-order multipolar interactions~\cite{Ponce2021,Royo2020,Brunin2020,royo2021}, which we neglect here for simplicity.
To reveal the first problem, we have selected a few of cases including HfS$_2$, NbF$_4$, CoO$_2$, TaS$_2$ and C where the structure optimization and subsequent DFPT were repeated with more stringent calculation parameters.
The corresponding phonon dispersions based on these new IFCs with the invariance conditions applied become stable and display a quadratic ZA mode in the long-wavelength limit (see the results with the legend \textit{this work} in Supplementary Note 4~\cite{si}).
In a few cases involving magnetic 2D materials, the self-consistent solution found might not be the lowest-energy ground state.
We did not investigate these cases further. 
For the second case, we observe that black phosphorus and arsenene have a non-vanishing $yz$ components of the Born effective charge tensor, resulting in small soft modes due to the neglect of electric ﬁeld in calculating IFCs and higher-order long-range electrostatic fields in the interpolation of IFCs, which could be important in the presence of piezoelectricity or flexoelectricity~\cite{Wu2005,Ma2020,Royo2020,royo2022}.

In conclusion, we have presented a systematic study on the invariance and equilibrium conditions in the lattice dynamics of crystalline solids, as well as its implementation in the framework of modern first-principles simulations. 
The results highlight the importance of enforcing both the Born-Huang invariances and Huang conditions in order to recover the quadratic dispersion relation of the ZA phonons in LD materials around the Brillouin zone center. 
This is in agreement with the theoretical model discussed in Ref.~\onlinecite{croy2020} and this work for bending waves, where the first-order dispersion vanishes as the bending conditions regardless of the crystal type of LD solids.
As expected, such two invariance conditions are found to have negligible influence on the lattice dynamics of bulk materials, even for low-symmetry crystals. 
Besides, we have introduced polar Born-Huang invariances and polar Huang conditions, which allow to separately treat the short-range and long-range IFCs for infrared-active solids. 
This scheme can work effectively in infrared-active 2D materials where the long-range dipole-dipole interactions contribute to the fulfillment of the invariance conditions. 
To demonstrate the effectiveness of the method developed, we apply this invariance scheme to over two hundred 2D material candidates from a high-throughput database, most of which originally displayed an unstable phonon dispersion with the imaginary ZA modes close to the $\Gamma$ point. 
With four types of representative 2D materials from the database illustrated here, we demonstrate that most of their phonon dispersions become stable with a physically quadratic ZA branch in the long-wavelength limit, after the invariance conditions are imposed.
We believe the proposed rotational invariance and equilibrium conditions will have wide applications in the field of LD materials, benefiting future lattice-dynamical studies of novel materials, and the calculations of their mechanical, thermodynamic, vibrational, and spectroscopic properties.

\section*{METHODS}
The DFT and DFPT calculations are performed with \textsc{Quantum ESPRESSO}~\cite{giannozzi2009,giannozzi2017} and the v1.1 \textsc{SSSP} efficiency pseudopotentials' library~\cite{prandini2018}. 
To describe the exchange and correlation of electrons, we employ the Perdew, Burke, and Ernzerhof’s parametrization for solids (PBEsol)~\cite{perdew2008} of the generalized gradient approximation (GGA). 
In the case of 2D materials, the Coulomb cutoff for 2D system implemented by Sohier~\emph{et al. }~\cite{sohier2017-2} is adopted to remove the long-range interactions in the vacuum direction, which is important to obtain the correct electronic screening and LO-TO splitting in 2D materials~\cite{sohier2017}. 
To compute the IFCs directly in real space from the small displacement method, we use an in-house code which is interfaced with \textsc{Quantum ESPRESSO}.
The subsequent Fourier interpolation to reciprocal space is obtained with \textsc{Phonopy}~\cite{togo2015} to compute the phonon dispersions.

\paragraph*{\normalfont\textbf{Silicon}} The optimized lattice parameter is 5.43 Å, with convergence thresholds for pressure and forces smaller than 10$^{-2}$~kbar and 10$^{-4}$~Ry/Bohr, respectively.
The same criteria for other materials are used unless stated.
To calculate the real-space IFCs, we use a plane-wave kinetic energy cutoff of 60~Ry with a $\Gamma$-point grid computed on a 
 $4\times4\times4$ supercell and perform a random displacement with the magnitude of 0.01~Å for all atoms.
A least-square fit of the Taylor-expanded forces to the DFT forces is carried out to find the best solution for the IFCs.

\paragraph*{\normalfont\textbf{Triclinic CaP$_3$}} The structure is relaxed with a plane-wave kinetic energy cutoff of 30~Ry and a $10\times10\times10$ electronic grid.
Similar to the case of silicon,  we build a $2\times2\times2$ supercell, and the real-space DFT calculations based on a $2\times2\times2$ grid are carried out for 30 displaced structures to get the entire IFCs.
In addition, a DFPT calculation at $\bold{q=0}$ is performed to get the dielectric tensor and Born effective charges.

\paragraph*{\normalfont\textbf{Graphene}} To simulate graphene as a 2D material, a vacuum of 20 Å is set in the out-of-plane direction to remove the interaction between periodic images. 
The structure is relaxed using a plane-wave kinetic energy cutoff of 80~Ry and a $20\times20\times20$ electronic grid. 
The optimized lattice constant of graphene is 2.46~Å. 
To calculate the phonon dispersions, the same real-space procedure is adopted. 
We create a $6\times6\times1$ supercell and the corresponding DFT calculations for 10 displaced supercells, with an electronic sampling of $3\times3\times3$.

\paragraph*{\normalfont\textbf{2D MoS$_2$}} We choose a vacuum of 25 Å to avoid the spurious interactions in the out-of-plane direction. 
With a plane-wave kinetic energy cutoff of 55~Ry and a $14\times14\times1$ electronic grid,  we obtain the optimized lattice constant as 3.14~Å based on tighter relaxation criteria for pressure (smaller than 10$^{-3}$~kbar) and forces (smaller than 10$^{-5}$~Ry/Bohr).
For the lattice-dynamical calculations, we perform a DFPT calculation on a $6\times6\times1$ phonon grid, and the IFCs are obtained through the inverse Fourier transform of the dynamical matrices.

\paragraph*{\normalfont\textbf{1D materials}} For the single-wall (8,0) CNT, the periodic direction is chosen along the $z$ direction with a vacuum of 20~Å in both the $x$ and $y$ directions, and the lattice constant of the optimized structure is 4.26~Å based on the relaxation criteria of pressure (smaller than 10$^{-2}$~kbar) and force (smaller than 10$^{-5}$~Ry/Bohr).
We create a $1\times1\times4$ supercell and generated 60 displaced configurations using the same procedure as before to obtain the real-space IFCs.
The DFT calculations for these displaced supercells are performed with a plane-wave kinetic energy cutoff of 80~Ry and a $\Gamma$-point electronic grid.
For the infrared-active single-wall (4,4) BNNT, the IFCs are obtained by DFPT calculations as in Ref.~\onlinecite{rivano2022}, with the appropriate Coulomb cutoff for 1D systems.

\vspace*{0.5cm}

\section*{DATA AVAILABILITY}
The material structures, pseudopotentials, interatomic force constants and related data in order to reproduce this study can be found on the Materials Cloud Archive~\cite{lin2022}.

\section*{CODE AVAILABILITY}
The code developed and used in this work is integrated into the \textsc{q2r.x} and \textsc{matdyn.x} executables of the \textsc{Quantum ESPRESSO} distribution, which will be released in the next version of \textsc{Quantum ESPRESSO}.

\section*{ACKNOWLEDGEMENTS}
We would like to thank Norma Rivano and Thibault Sohier for sharing with us the results of Ref.~\onlinecite{rivano2022} prior to publication, and for the calculations of the interatomic force constants of the single-wall (4,4) boron nitride nanotube.
We are also grateful to Alexander Croy for providing us with additional information on Ref.~\onlinecite{croy2020}.
We also thank Junfeng Qiao, Elsa Passaro, Giovanni Pizzi, Yanbo Li, Jian Han, Xavier Gonze, Jiongzhi Zheng, Raffaele Resta, Massimiliano Stengel and Miquel Royo for useful discussions.
This work is supported by the Sinergia project of the Swiss National Science Foundation (No. CRSII5\_189924).
S.P. acknowledges financial support from the Belgian F.R.S.-FNRS. 

\vspace*{0.5cm}

\section*{AUTHOR CONTRIBUTIONS}
C.L. implemented the invariance conditions and performed the calculations.
S.P. and N.M. supervised the project.
All authors analyzed the results and contributed to writing the manuscript.

\section*{COMPETING INTERESTS}
The authors declare no competing interests. 


\bibliography{main_arxiv.bbl}

\newpage

\ifarXiv
    

\section*{Supplementary material}

# Supplemental information: General invariance and equilibrium conditions for lattice dynamics in 1D, 2D, and 3D materials

Changpeng Lin,[1, 2, *] Samuel Poncé,[3, 1] and Nicola Marzari[1, 2]

[1] *Theory and Simulation of Materials (THEOS), École Polytechnique Fédérale de Lausanne, CH-1015 Lausanne, Switzerland*
[2] *National Centre for Computational Design and Discovery of Novel Materials (MARVEL),*
*École Polytechnique Fédérale de Lausanne, CH-1015 Lausanne, Switzerland*
[3] *European Theoretical Spectroscopy Facility, Institute of Condensed Matter and Nanosciences,*
*Université catholique de Louvain, Chemin des Étoiles 8, B-1348 Louvain-la-Neuve, Belgium*

## CONTENTS

———

* changpeng.lin@epfl.ch

## SUPPLEMENTARY NOTE 1: THE BENDING CONDITIONS IN 2D MATERIALS

As discussed in the main text of this work, the first-order linear dispersion of the bending ZA mode in 2D materials reads

$$
\begin{aligned}
[\omega_{\mathrm{ZA}}^{(1)}(\mathbf{q})]^2 &= \frac{1}{\sum_\kappa m_\kappa} \sum_\beta \frac{u_\beta}{u_z} \sum_{\gamma,\delta \in \{x,y\}} \left[ T_{z\beta,\gamma\delta}^{\mathrm{sym}} + T_{z\gamma,\beta\delta}^{\mathrm{int}} \right] q_\gamma q_\delta \\
&= \frac{1}{\sum_\kappa m_\kappa} \sum_\beta \frac{u_\beta}{u_z} \sum_{\gamma,\delta \in \{x,y\}} \left[ T_{\gamma\delta,z\beta}^{\mathrm{sym}} + T_{\gamma z,\delta\beta}^{\mathrm{int}} \right] q_\gamma q_\delta.
\end{aligned}
\tag{1}
$$

In the second step of Equation (1), the Huang conditions and the symmetry of the tensor $T_{\alpha\gamma,\beta\delta}^{\mathrm{int}}$ (see Equations (15) and (38) in the main text) are used. We denote the $z$ axis as the vacuum direction for 2D materials. As a result, the $z$ component of distance between two atoms does not depend on the unit cell position $\mathbf{R}$, and one can have the bending conditions

$$
\begin{aligned}
T_{\gamma\delta,z\beta}^{\mathrm{sym}} + T_{\gamma z,\delta\beta}^{\mathrm{int}} &= -\frac{1}{2} \sum_{\mathbf{R},\kappa\kappa'} \Phi_{\kappa\gamma,\kappa'\delta}^{\mathbf{R}} \tau_{\kappa\kappa'}^{\mathbf{R}} z \tau_{\kappa\kappa'\beta}^{\mathbf{R}} - \sum_{\kappa\lambda,\kappa'\mu} \frac{\Gamma_{\kappa\lambda,\kappa'\mu}}{\sqrt{m_\kappa m_{\kappa'}}} \sum_{\mathbf{R},\kappa''} \Phi_{\kappa\lambda,\kappa''\gamma}^{\mathbf{R}} \tau_{\kappa\kappa''} z \sum_{\mathbf{R}',\kappa'''} \Phi_{\kappa'\mu,\kappa'''\delta}^{\mathbf{R}'} \tau_{\kappa'\kappa'''\beta}^{\mathbf{R}'} \\
&= -\frac{1}{2} \sum_{\mathbf{R},\kappa''\kappa'''} \Phi_{\kappa''\gamma,\kappa'''\delta}^{\mathbf{R}} \tau_{\kappa''\kappa'''} z \tau_{\kappa''\kappa'''\beta}^{\mathbf{R}} - \sum_{\kappa\lambda,\kappa'\mu} \frac{\Gamma_{\kappa'\mu,\kappa\lambda}}{\sqrt{m_\kappa m_{\kappa'}}} \sum_{\mathbf{R},\kappa''} \Phi_{\kappa\lambda,\kappa''\gamma}^{\mathbf{R}} \tau_{\kappa\kappa''} z \sum_{\mathbf{R}',\kappa'''} \Phi_{\kappa'\mu,\kappa'''\delta}^{\mathbf{R}'} \tau_{\kappa'''}^{\mathbf{R}'} \\
&= -\frac{1}{2} \sum_{\mathbf{R},\kappa''\kappa'''} \Phi_{\kappa''\gamma,\kappa'''\delta}^{\mathbf{R}} \tau_{\kappa''\kappa'''} z \tau_{\kappa''\kappa'''\beta}^{\mathbf{R}} - \sum_{\kappa'\kappa''\mu} \tau_{\kappa''z} \delta_{\mu\gamma} \delta_{\kappa'\kappa''} \sum_{\mathbf{R}',\kappa'''} \Phi_{\kappa'\mu,\kappa'''\delta}^{\mathbf{R}'} \tau_{\kappa'''\beta}^{\mathbf{R}'} \\
&= -\frac{1}{2} \sum_{\mathbf{R},\kappa''\kappa'''} \Phi_{\kappa''\gamma,\kappa'''\delta}^{\mathbf{R}} \tau_{\kappa''z} \tau_{\kappa''\kappa'''\beta}^{\mathbf{R}} + \frac{1}{2} \sum_{\mathbf{R},\kappa''\kappa'''} \Phi_{\kappa''\gamma,\kappa'''\delta}^{\mathbf{R}} \tau_{\kappa''z} \tau_{\kappa'''\beta}^{\mathbf{R}} - \sum_{\mathbf{R}',\kappa''\kappa'''} \tau_{\kappa''z} \Phi_{\kappa'\mu,\kappa'''\delta}^{\mathbf{R}'} \tau_{\kappa'''\beta}^{\mathbf{R}'} \\
&= \frac{1}{2} \sum_{-\mathbf{R},\kappa''\kappa'''} \Phi_{\kappa'''\delta,\kappa''\gamma}^{-\mathbf{R}} \tau_{\kappa''z} \tau_{\kappa''\kappa'''\beta}^{-\mathbf{R}} - \frac{1}{2} \sum_{\mathbf{R},\kappa''\kappa'''} \Phi_{\kappa''\gamma,\kappa'''\delta}^{\mathbf{R}} \tau_{\kappa''z} \tau_{\kappa'''\beta}^{\mathbf{R}} \\
&= \frac{1}{2} \sum_{-\mathbf{R},\kappa''\kappa'''} \Phi_{\kappa'''\delta,\kappa''\gamma}^{-\mathbf{R}} \tau_{\kappa''z} \tau_{\kappa'''\beta}^{-\mathbf{R}} - \frac{1}{2} \sum_{\mathbf{R},\kappa''\kappa'''} \Phi_{\kappa''\gamma,\kappa'''\delta}^{\mathbf{R}} \tau_{\kappa''z} \tau_{\kappa'''\beta}^{\mathbf{R}} \\
&= 0,
\end{aligned}
\tag{2}
$$

with $\Phi_{\kappa\gamma,\kappa'\delta}^{\mathbf{R}} \equiv \Phi_{\kappa\gamma,\kappa'\delta}(\mathbf{R})$ and $\tau_{\kappa\kappa'\beta}^{\mathbf{R}} \equiv \tau_{\kappa\kappa'\beta}(\mathbf{R})$. The translational invariance and the properties of $\Gamma_{\kappa\lambda,\kappa'\mu}$ matrix (see Equations (4) and (37) in the main text) are used in the derivation with the rearrangement of summation indices. Therefore, the linear dispersion term $\omega_{\mathrm{ZA}}^{(1)}(\mathbf{q})$ for the ZA mode of 2D materials vanishes, and its dispersion relation becomes quadratic in the long-wavelength limit. These bending conditions $T_{\gamma\delta,z\beta}^{\mathrm{sym}} = -T_{\gamma z,\delta\beta}^{\mathrm{int}}$ also allow the ZA branch to have a purely out-of-plane polarization in the vacuum $z$ direction, fully decoupling from the in-plane vibrational modes.

## SUPPLEMENTARY NOTE 2: SUMMARY OF THE 2D MATERIALS DATABASE

The current 2D materials database available on the Materials Could archive[1] contains 245 entries for the phonon dispersions of 2D materials[2]. However, 54 entries are dynamically unstable; the phonon spectra display remarkable imaginary frequencies not located around the Brillouin zone center, as summarized in Supplementary Table 1. After excluding these cases, we have imposed the corrections of the Born-Huang rotational invariance and Huang conditions on the original IFCs from Materials Could archive[1] and recalculated the corresponding phonon dispersions. We find that the corrections of the two invariance conditions failed for 33 entries of the remaining 191 candidates in the 2D database, which are listed out in Supplementary Table 2. By analyzing these results, we identify two main reasons that result in the failure of the invariance conditions to yield stable quadratic dispersion for the ZA branch. One of them is inaccurate IFCs in real space as obtained from DFPT calculations, due to the choice of convergence parameters and not perfectly relaxed structures. To showcase this issue, we select a few entries (i.e. $HfS_2$, $NbF_4$, $CoO_2$, $TaS_2$ and C) and perform further structural optimizations, and calculate again the IFCs by DFPT. The phonon dispersions from these new DFPT calculations are presented in Supplementary Note 4 with the legend *this work*. In addition, for some magnetic materials the self-consistent solution found might not be the lowest-energy ground state; the neglect of higher-order multipolar interactions in the interpolation of IFCs could also be important for piezoelectric and flexoelectric materials[3,4].

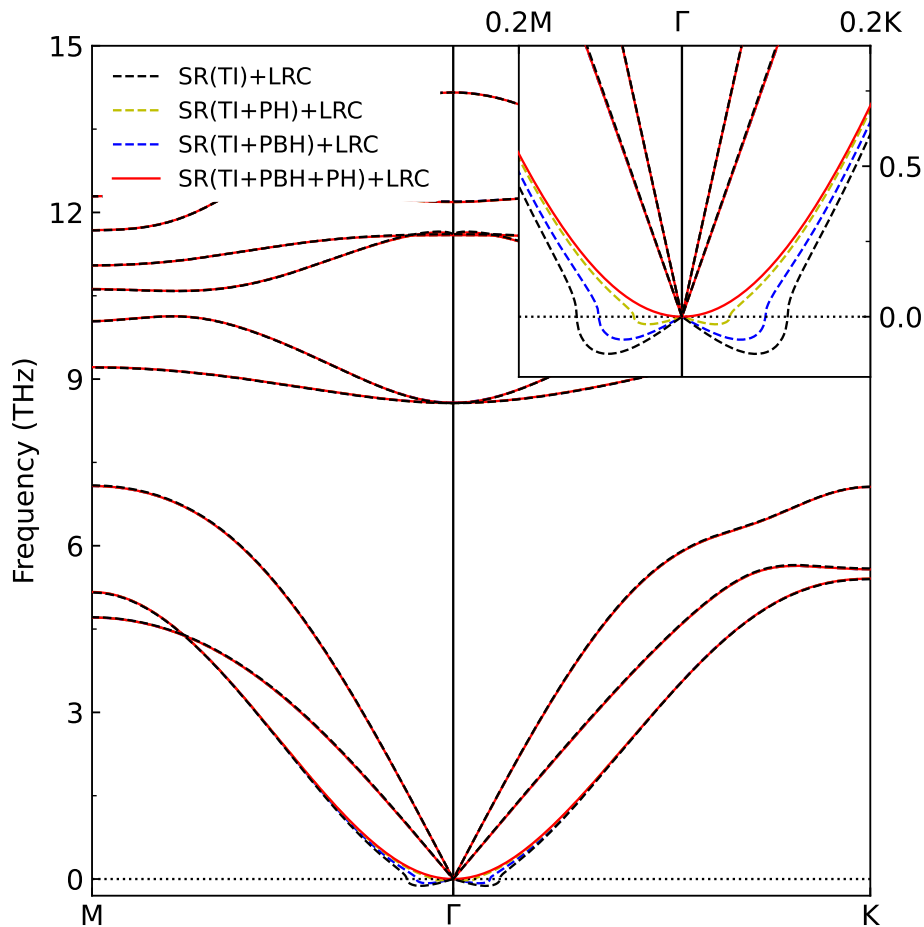

**Supplemental Figure 1.** Effects of each invariance condition on the phonon dispersion of 2D $MoS_2$. TI denotes correction of the translational invariance on the IFCs, while BH and H represent the further corrections by Born-Huang rotational invariance and Huang conditions, respectively. The corrections are made on the short-range (SR) IFCs through the polar Born-Huang (PBH) rotational invariance and the polar Huang (PH) conditions. The long-range correction (LRC) is added analytically at arbitrary **q**-point after Fourier interpolation.

**Supplementary Table 1**. 54 entries of unstable materials in the 2D database.

| No. | Formula | Space group | Point group |
|-----|---------|-------------|-------------|
| 1 | TiOBr | Pmmn | mmm |
| 2 | ReSe$_2$ | P-6m2 | -6m2 |
| 3 | PbF$_4$ | P4/mmm | 4/mmm |
| 4 | LiBH$_4$ | C2 | 2 |
| 5 | HfTe$_2$ | P-3m1 | -3m |
| 6 | SnO | C2/m | 2/m |
| 7 | SnO | P4/mmm | 4/mmm |
| 8 | VTe$_2$ | P-3m1 | -3m1 |
| 9 | MoS$_2$ | P-3m1 | -3m1 |
| 10 | OsOCl$_2$ | Pmmm | mmm |
| 11 | CoI$_2$ | C2/m | 2/m |
| 12 | TiBr$_2$ | P-3m1 | -3m1 |
| 13 | Co(OH)$_2$ | C2/m | 2/m |
| 14 | SbTeI | C2/m | 2/m |
| 15 | CrSe$_2$ | C2/m | 2/m |
| 16 | VOCl | Pmmm | mmm |
| 17 | NaCN | Pmm2 | mm2 |
| 18 | ZrTiSe$_4$ | P2 | 2 |
| 19 | CuAgTe$_2$ | Pm | m |
| 20 | Cu$_2$Te | C2/m | 2/m |
| 21 | Na$_2$PdH$_2$ | P4/mmm | 4/mmm |
| 22 | ZrTiTe$_2$ | P2/m | 2/m |
| 23 | TiOCl | Pmmn | mmm |
| 24 | TlF | P4/mmm | 4/mmm |
| 25 | ZrTe$_2$ | P-3m1 | -3m |
| 26 | KTlCl$_4$ | P-4 | -4 |
| 27 | Ti$_2$PTe$_2$ | P-3m1 | -3m |
| 28 | AgBr | P2_1/m | 2/m |
| 29 | AgF$_2$ | P2_1/c | 2/m |
| 30 | TiSe$_2$ | P-3m1 | -3m |
| 31 | OLuI | P4/mmm | 4/mmm |
| 32 | PbO | P4/mmm | 4/mmm |
| 33 | TaSe$_2$ | P-3m1 | -3m |
| 34 | AgNO$_2$ | P1 | 1 |
| 35 | Hf$_3$Te$_2$ | P4/mmm | 4/mmm |
| 36 | SnF$_4$ | P4/mmm | 4/mmm |
| 37 | CuTe | Pmmn | mmm |
| 38 | FeOCl | Pmmn | mmm |
| 39 | TiS$_2$ | P-3m1 | -3m |
| 40 | GeI$_2$ | P-6m2 | -6m2 |
| 41 | TiNI | Pmmn | mmm |
| 42 | TaSe$_2$ | P-6m2 | -6m2 |
| 43 | TiCl2 | P-3m1 | -3m |
| 44 | RbCl | P4/mmm | 4/mmm |
| 45 | TiTe$_2$ | P-3m3 | -3m |
| 46 | FeS | P4/mmm | 4/mmm |
| 47 | ZrNI | Pmmn | mmm |
| 48 | NbTe$_2$ | P-3m1 | -3m |
| 49 | NbSe$_2$ | P-3m1 | -3m |
| 50 | Tl$_2$S | P-3m1 | -3m |
| 51 | RuOCl$_2$ | Pmmm | mmm |
| 52 | La$_2$GeI$_2$ | P-3m1 | -3m |
| 53 | NbS$_2$ | P-3m1 | -3m |
| 54 | TaS$_2$ | P-3m1 | -3m |

**Supplementary Table 2**. 33 entries of 2D materials in the database where the invariance conditions failed to recover a stable ZA mode.

| No. | Formula | Space group | Point group |
|-----|---------|-------------|-------------|
| 1 | $HfS_2$ | P-3m1 | -3m |
| 2 | $NbF_4$ | P4/mmm | 4/mmm |
| 3 | $MnBr_2$ | C2/m | 2/m |
| 4 | $MnCl_2$ | C2/m | 2/m |
| 5 | $CoO_2$ | P-3m1 | -3m |
| 6 | HoSI | Pmmn | mmm |
| 7 | $LiAuI_4$ | P-1 | -1 |
| 8 | As | Pmna | mmm |
| 9 | $CoCl_2$ | P-3m1 | -3m |
| 10 | FeTe | P4/nmm | 4/mmm |
| 11 | $PdCl_2$ | $P2_1/c$ | 2/m |
| 12 | $VOCl_2$ | Pmm2 | mm2 |
| 13 | $TaS_2$ | P-6m2 | -6m2 |
| 14 | $PdS_2$ | $P2_1/c$ | 2/m |
| 15 | $MnI_2$ | C2/m | 2/m |
| 16 | GeSe | $Pmn2_1$ | mm2 |
| 17 | $In_2Se_3$ | C2 | 2 |
| 18 | $CuCl_2$ | C2/m | 2/m |
| 19 | $AgClO_4$ | P-42m | -42m |
| 20 | YbOCl | P-3m1 | -3m |
| 21 | $ZrS_2$ | P-3m1 | -3m |
| 22 | $CoBr_2$ | P-3m1 | -3m |
| 23 | $NbS_2$ | P-6m2 | -6m2 |
| 24 | C | P6/mmm | 6/mmm |
| 25 | $CrBr_2$ | P-1 | -1 |
| 26 | P | Pmna | mmm |
| 27 | $SrI_2$ | Pmmn | mmm |
| 28 | GaTeCl | $Pmn2_1$ | mm2 |
| 29 | $NiI_2$ | P-3m1 | -3m |
| 30 | KTlO | $P2_1/m$ | 2/m |
| 31 | FeSe | P4/nmm | 4/mmm |
| 32 | $CrI_2$ | $P2_1/m$ | 2/m |
| 33 | SnTe | $Pmn2_1$ | mm2 |

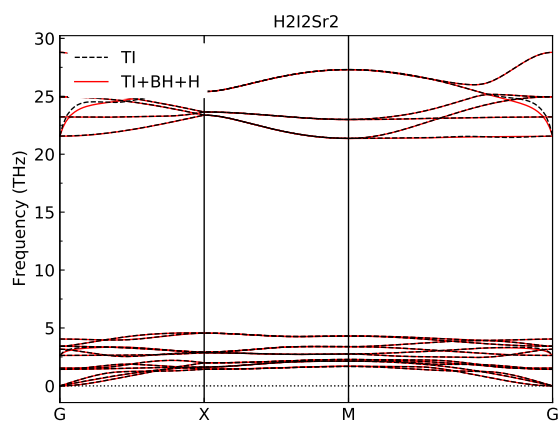

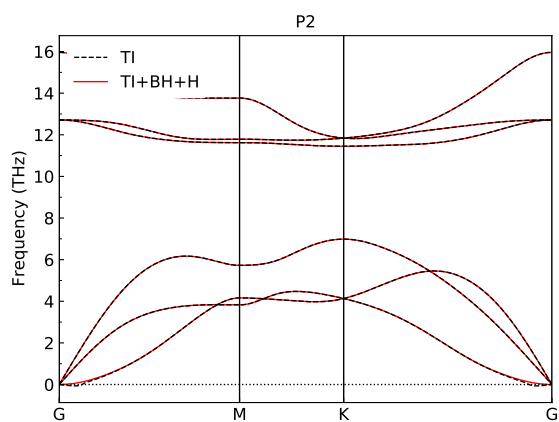

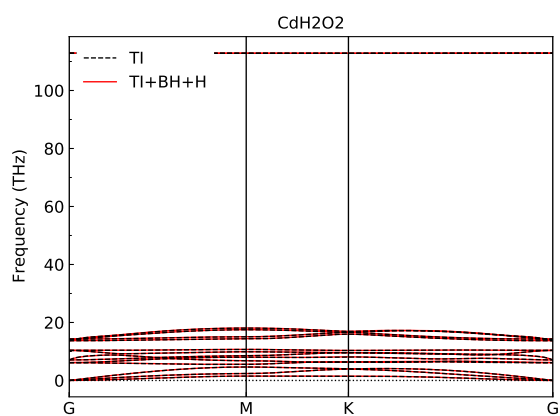

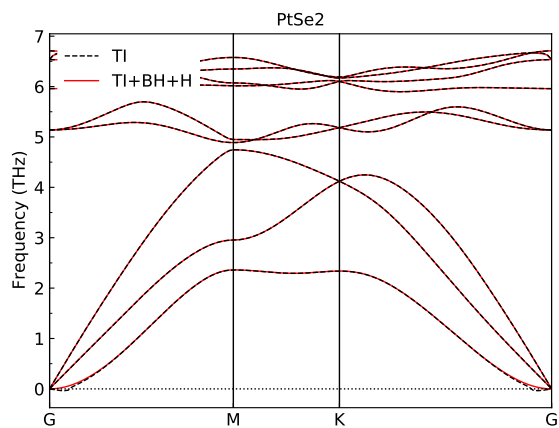

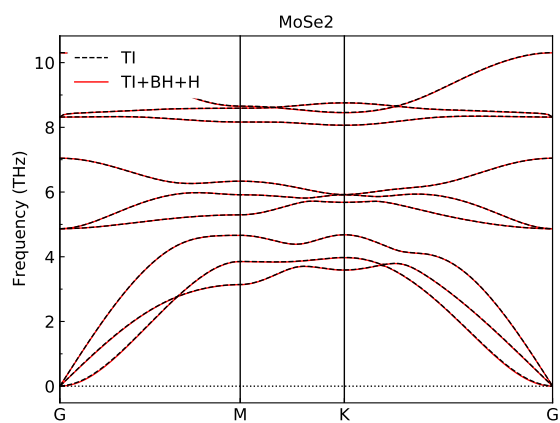

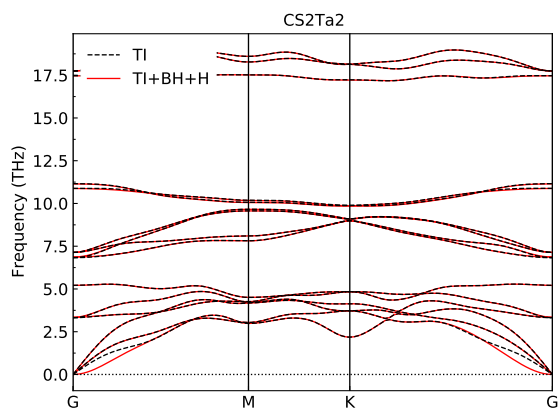

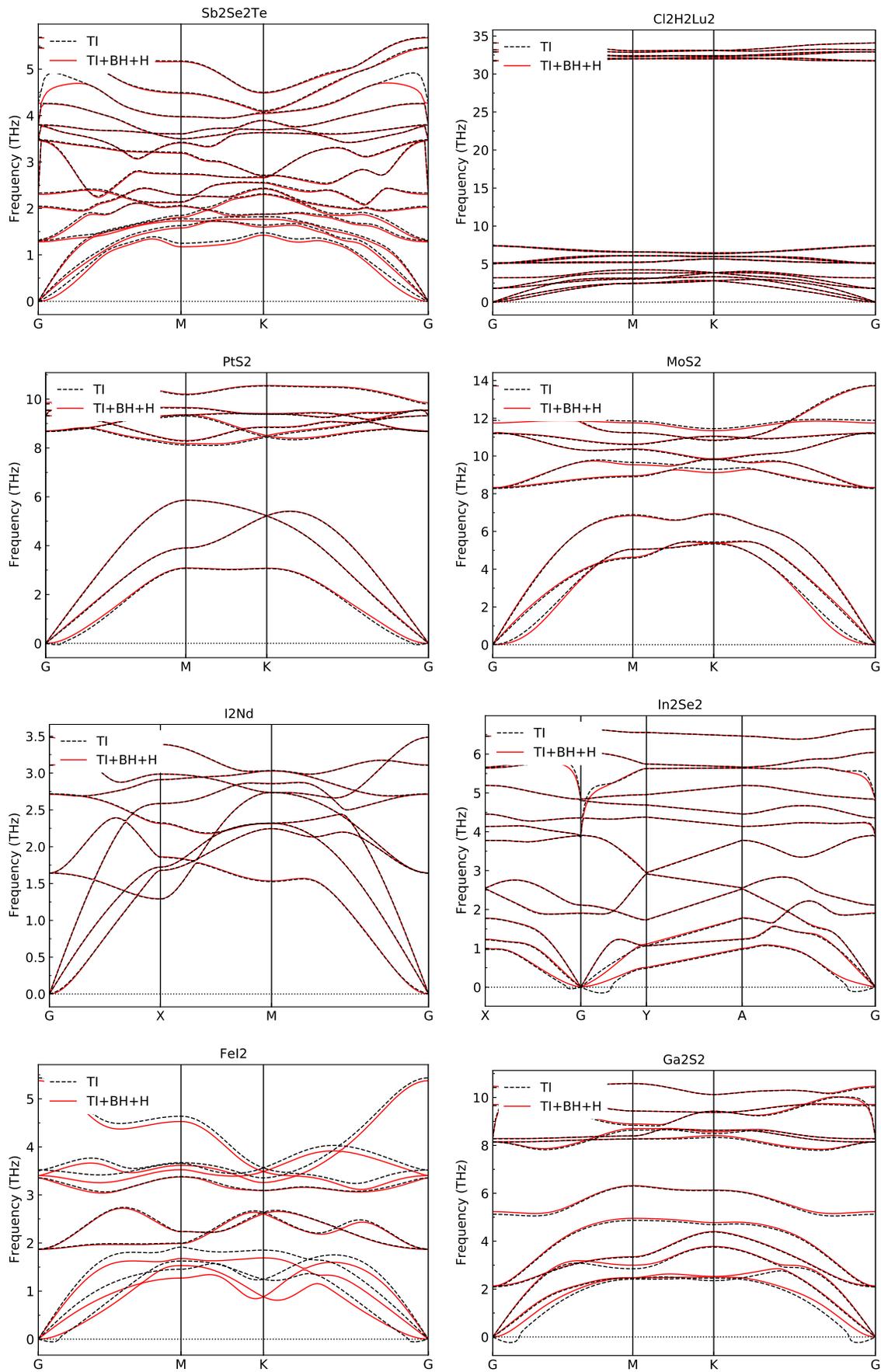

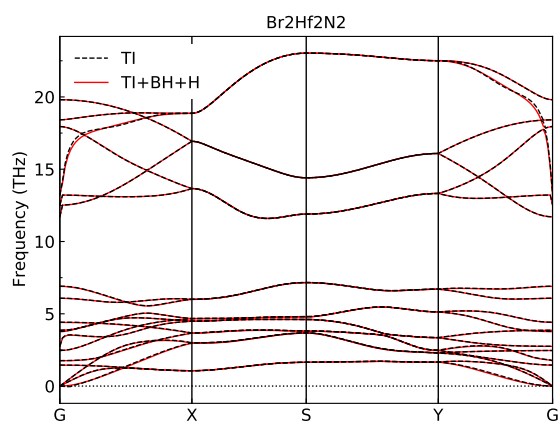

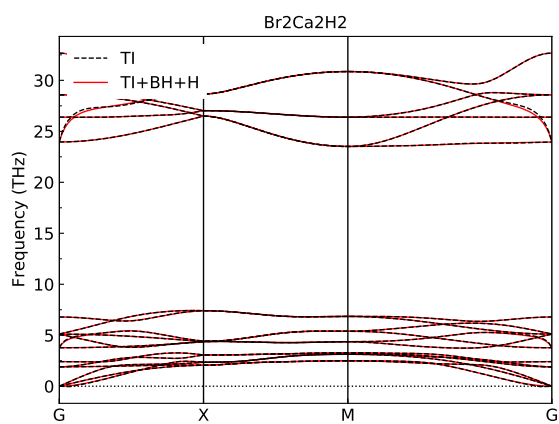

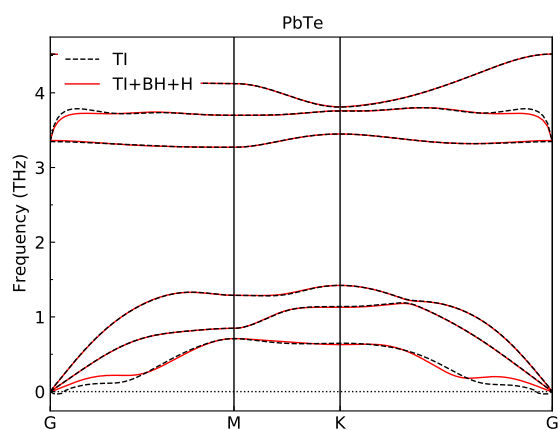

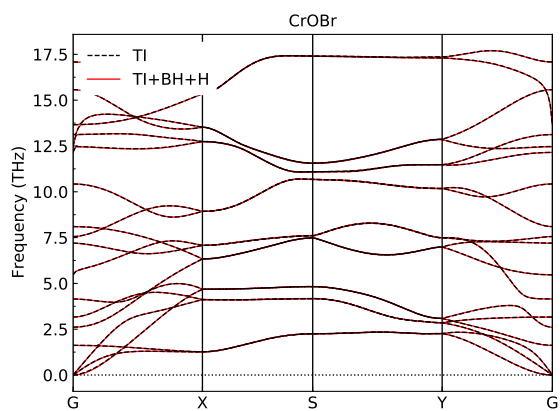

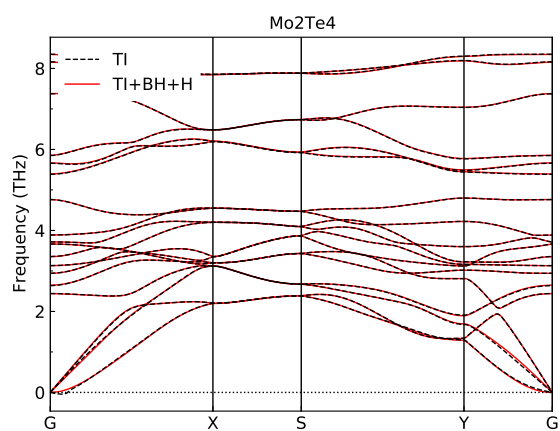

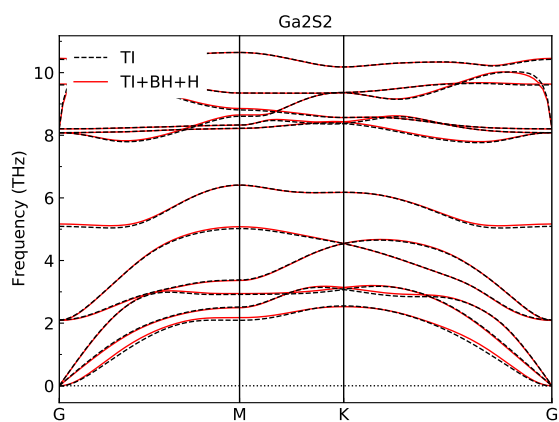

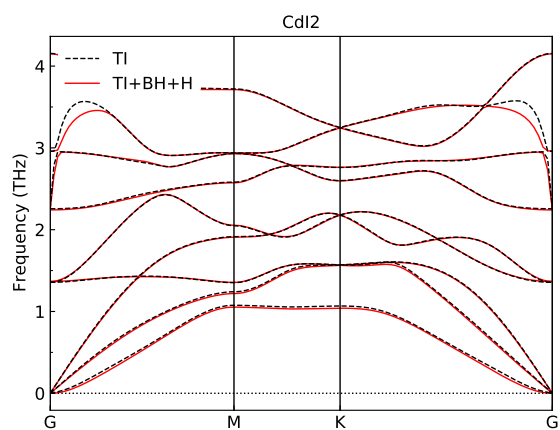

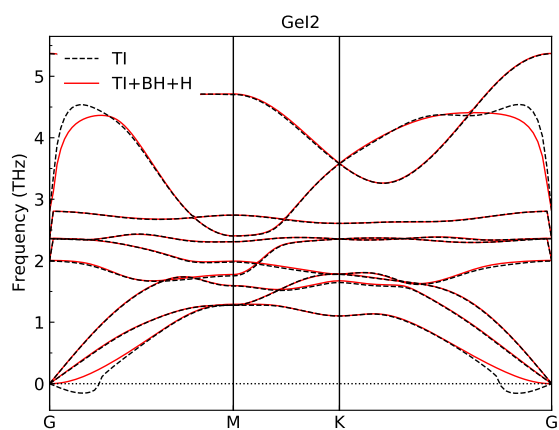

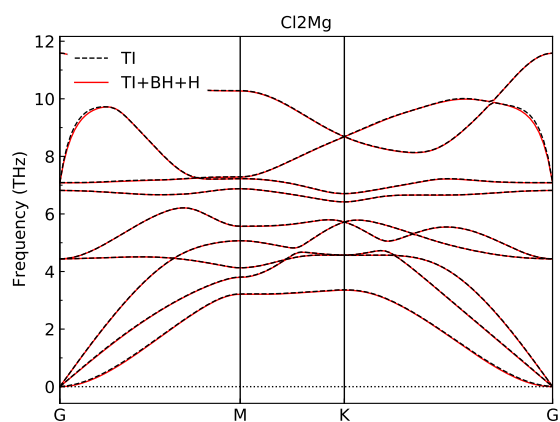

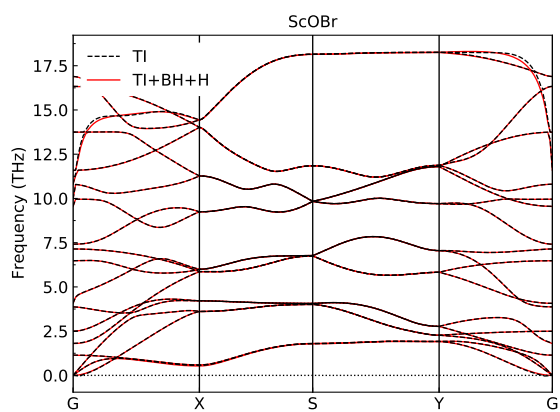

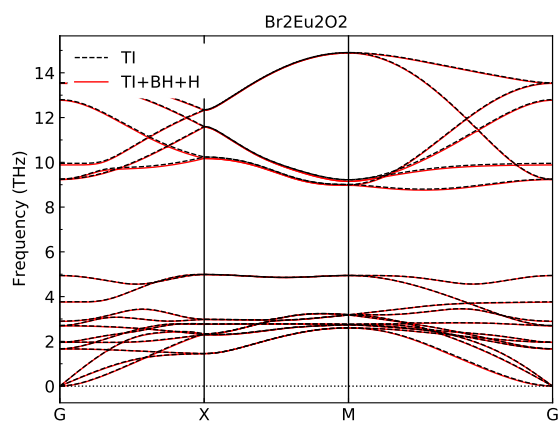

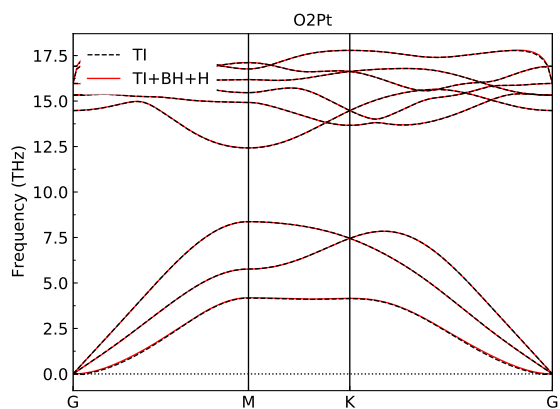

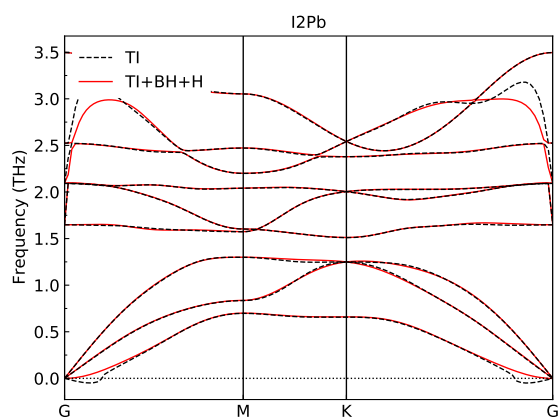

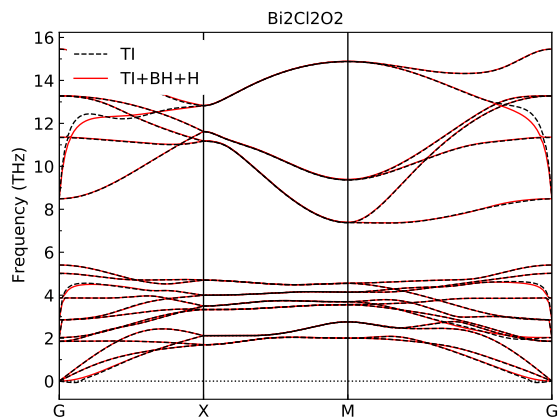

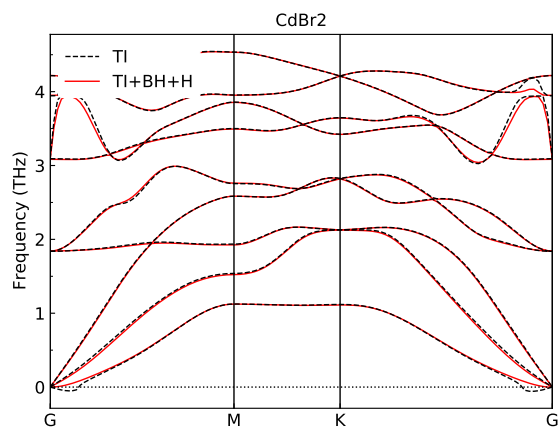

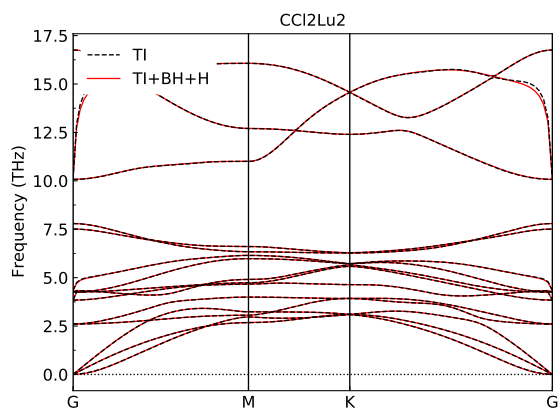

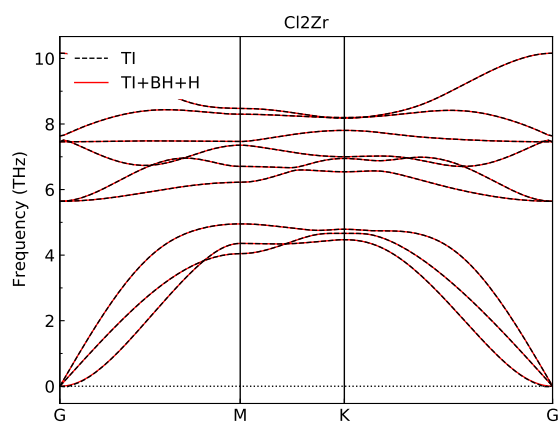

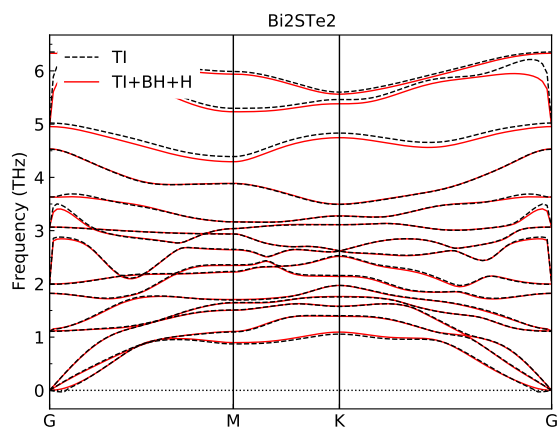

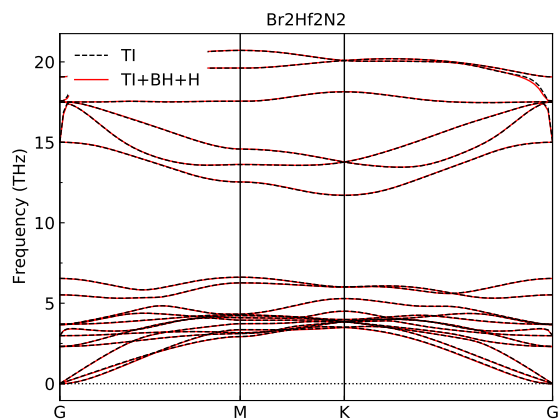

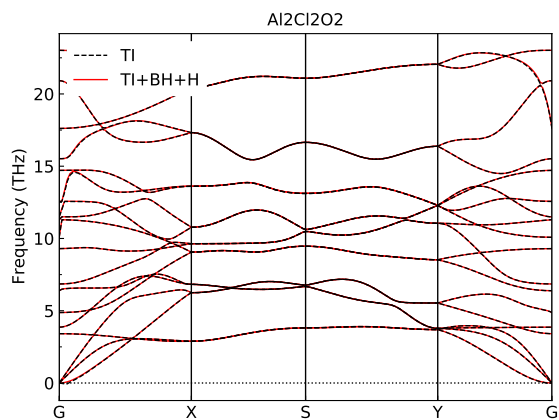

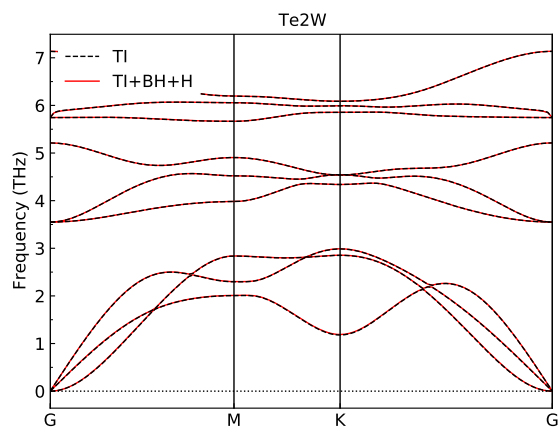

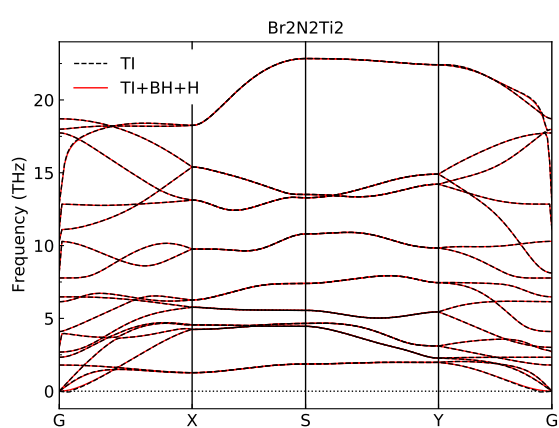

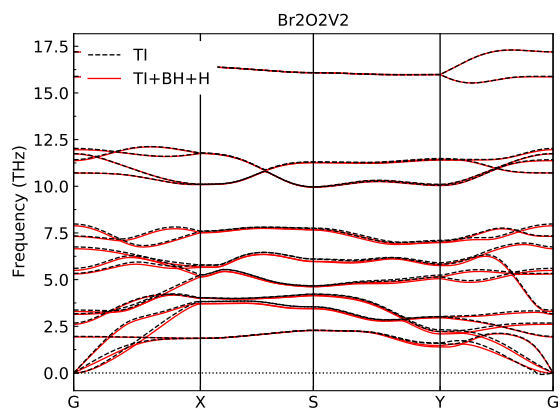

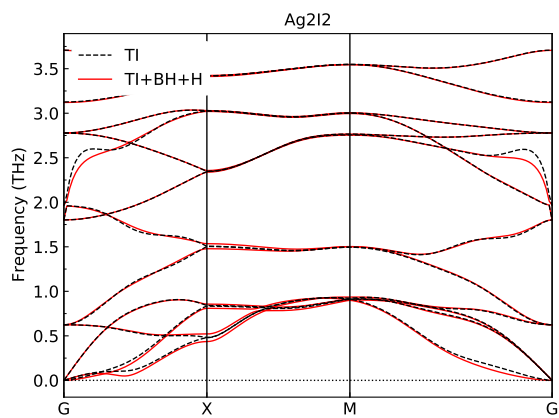

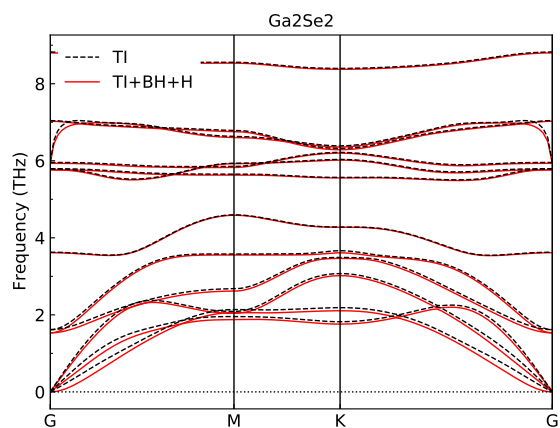

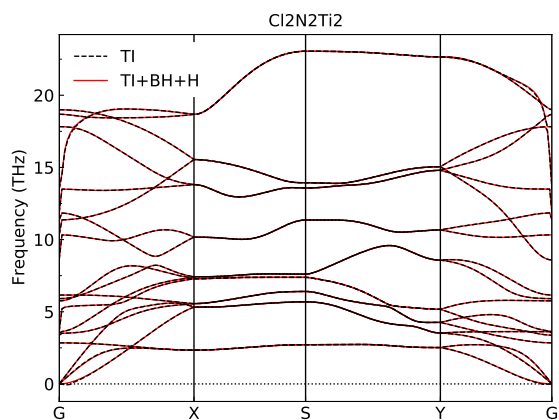

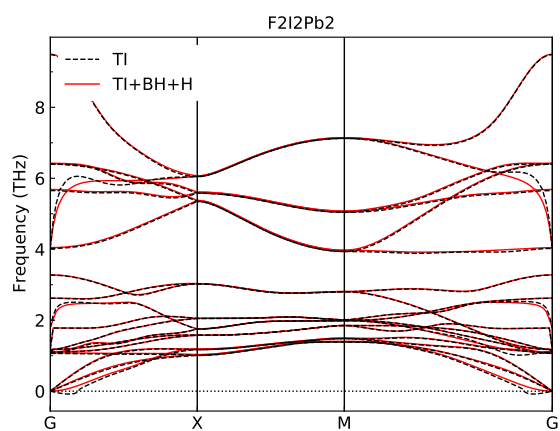

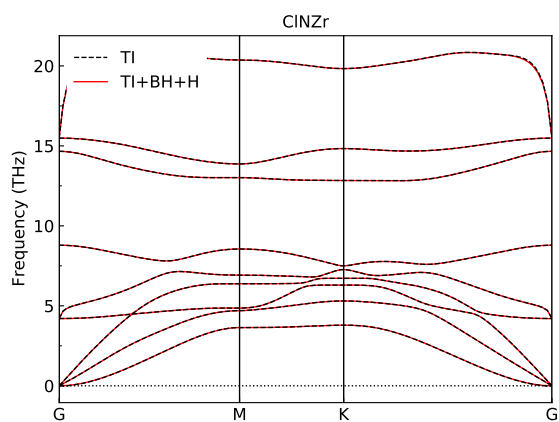

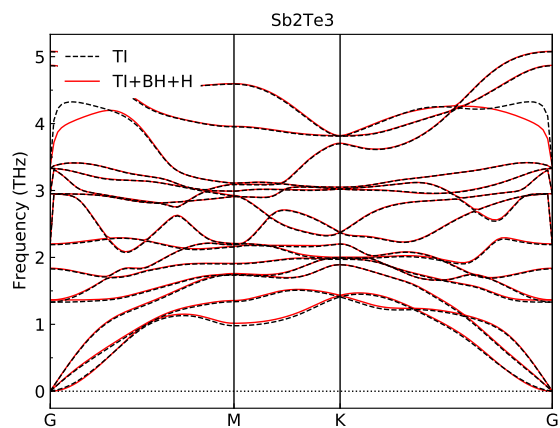

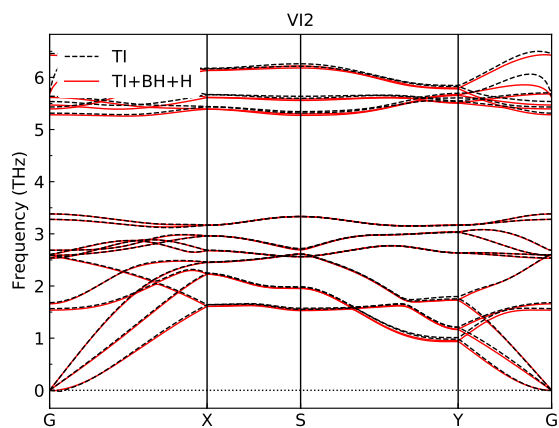

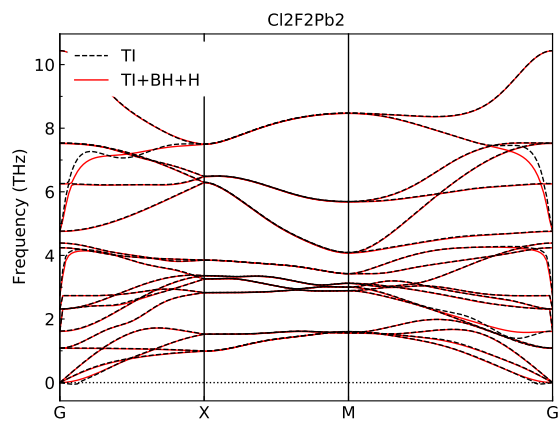

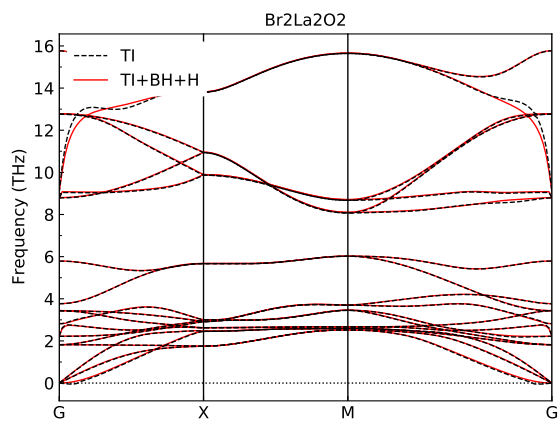

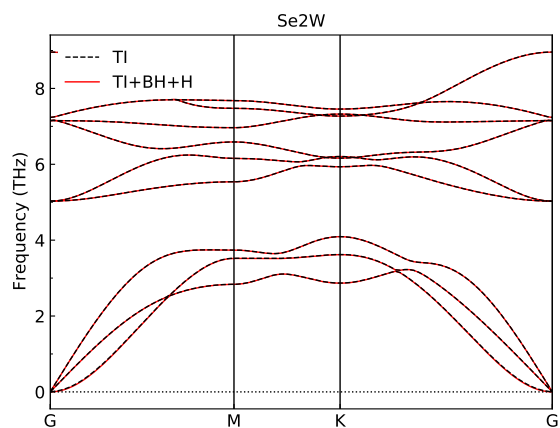

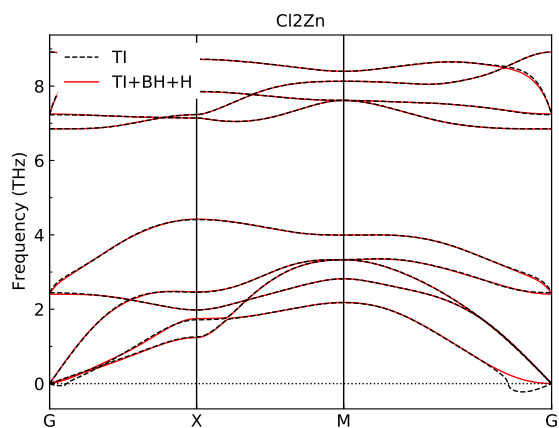

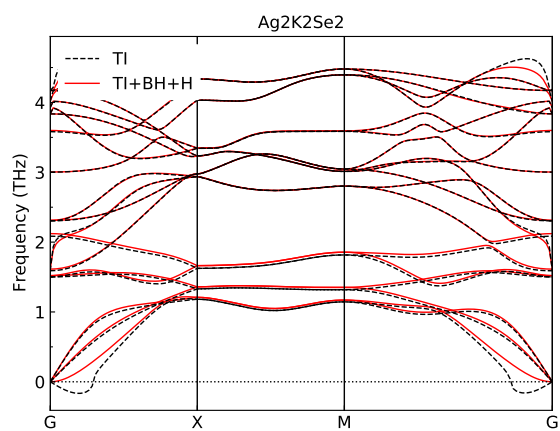

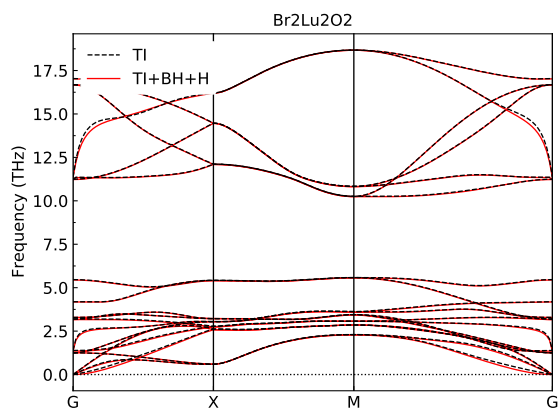

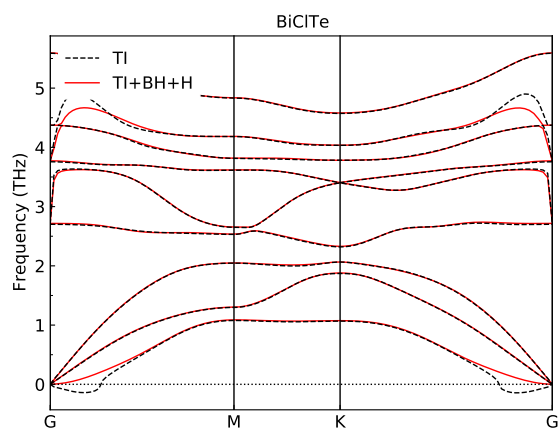

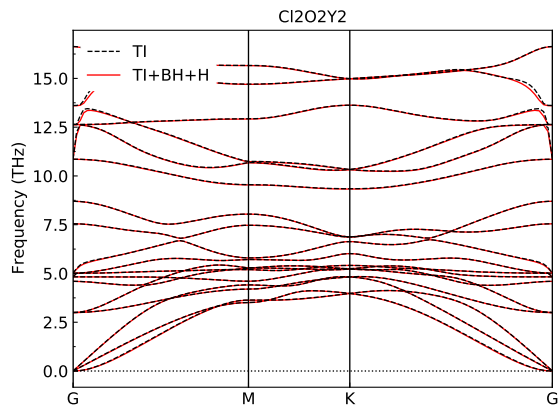

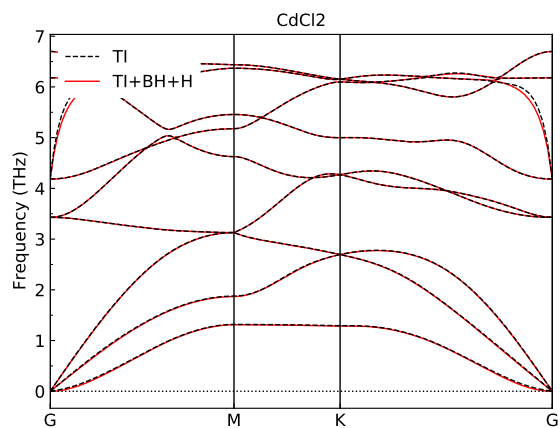

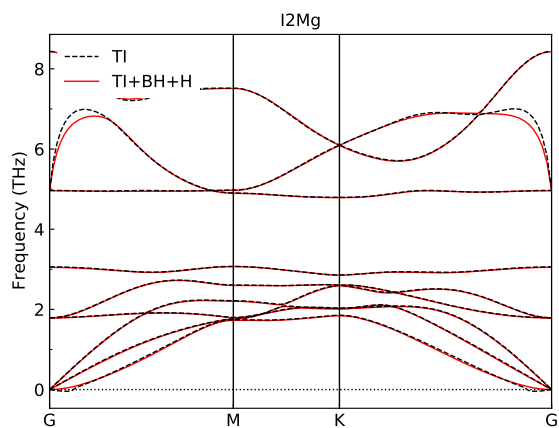

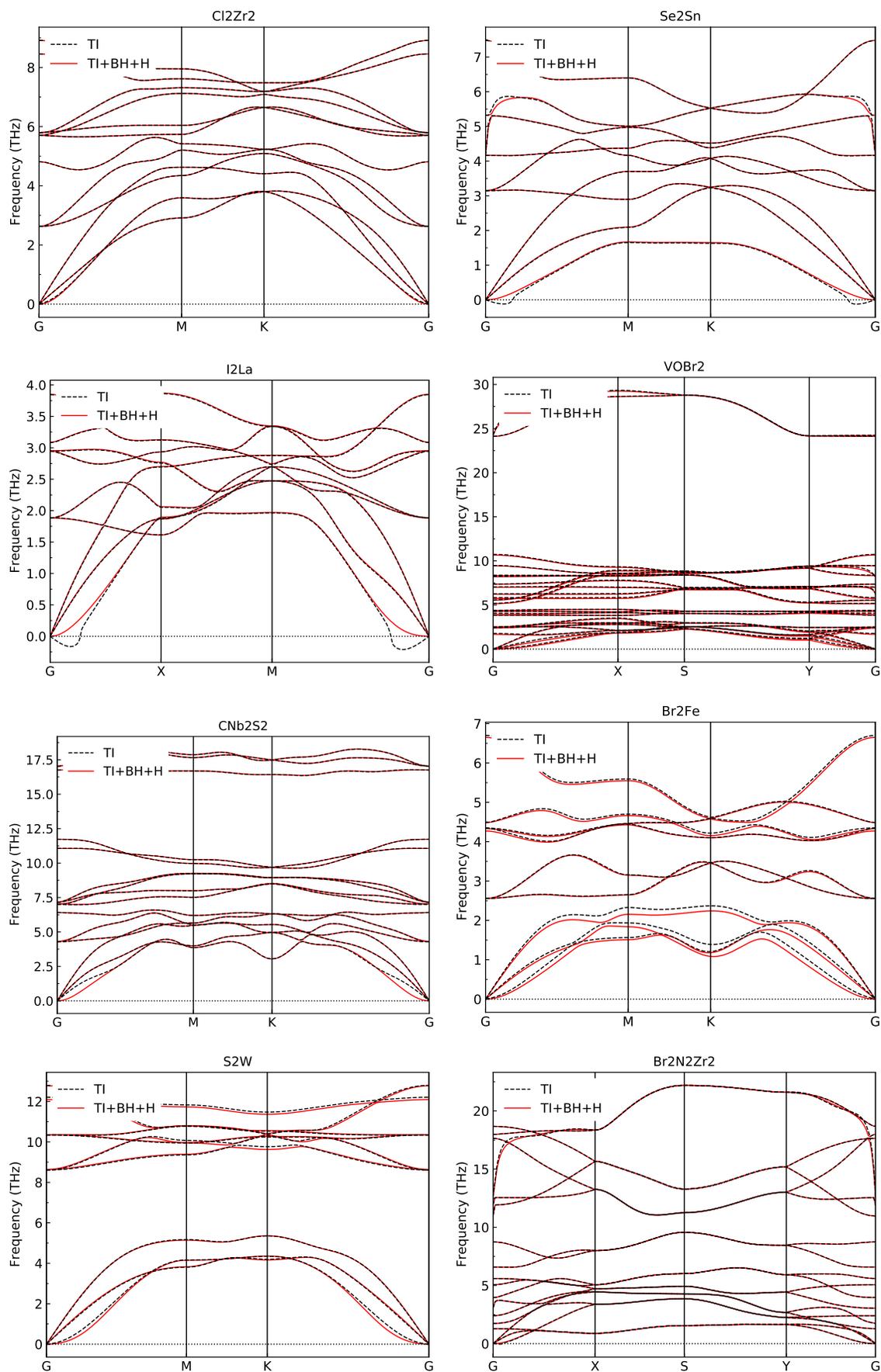

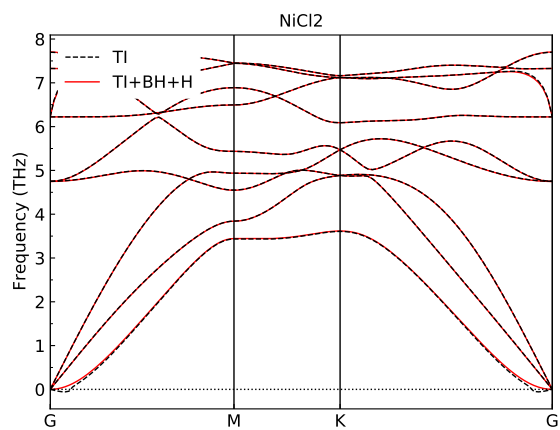
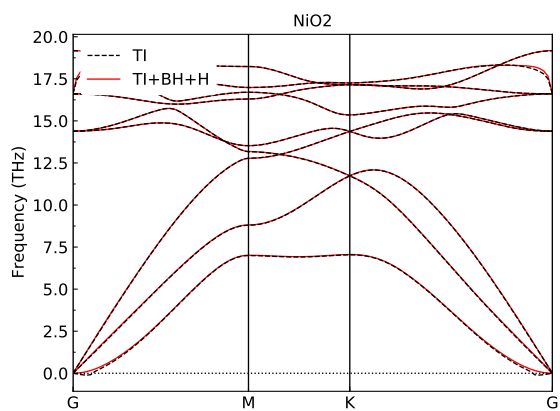
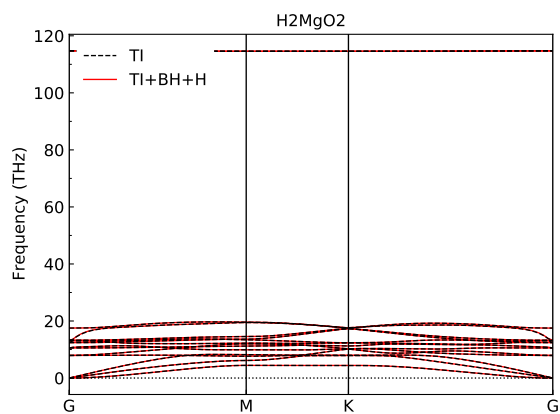
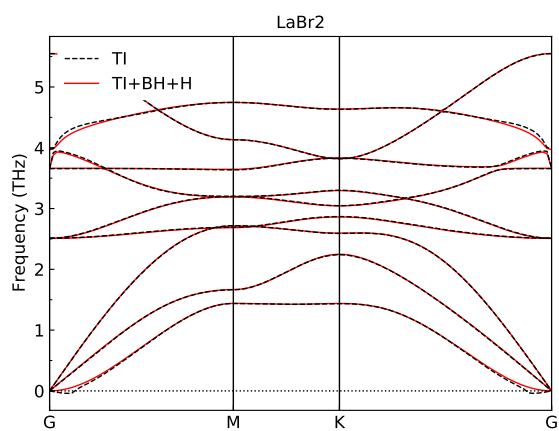
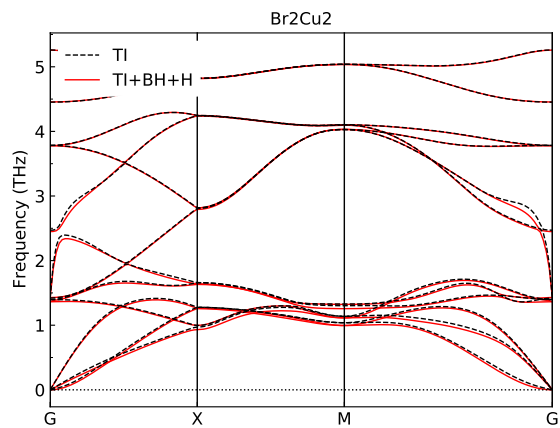
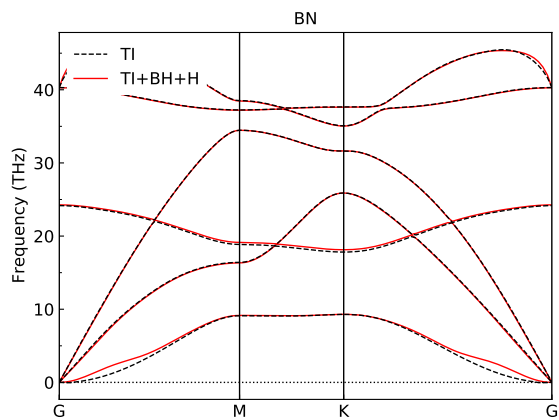
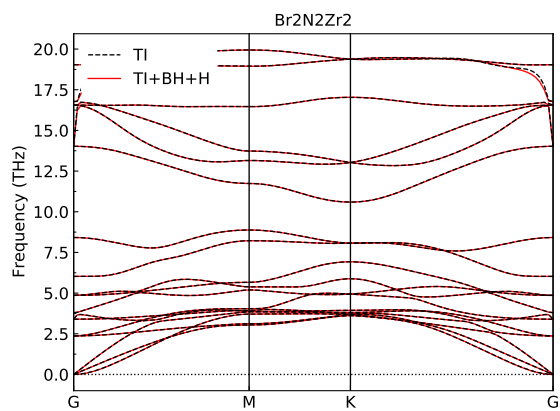
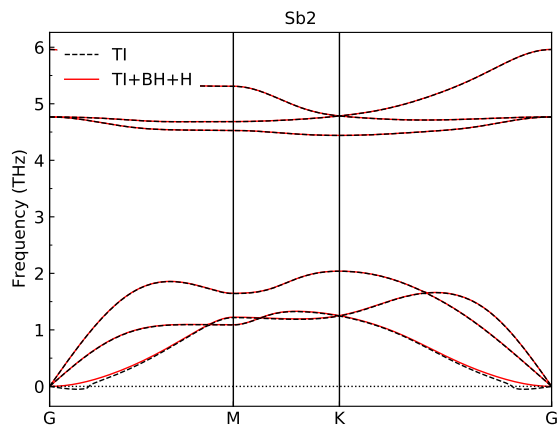

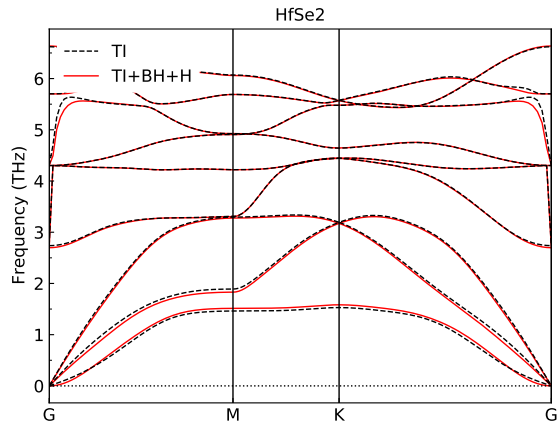

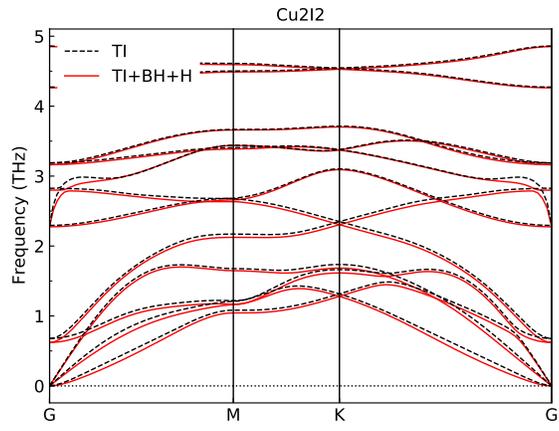

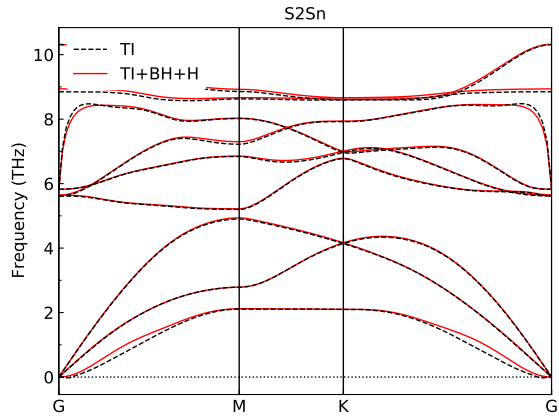

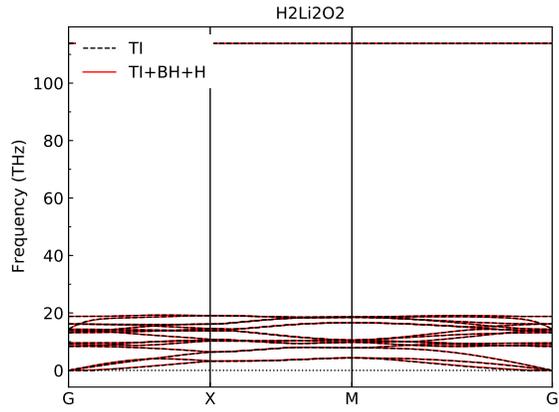

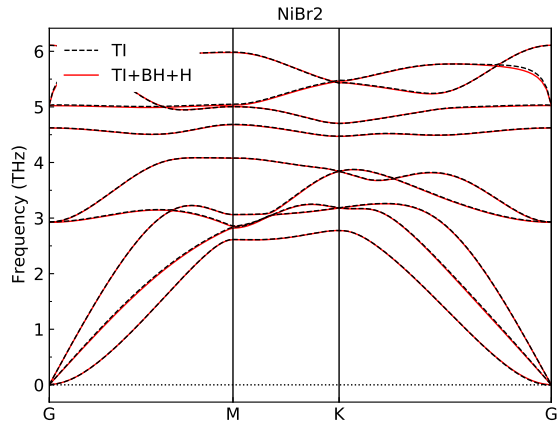

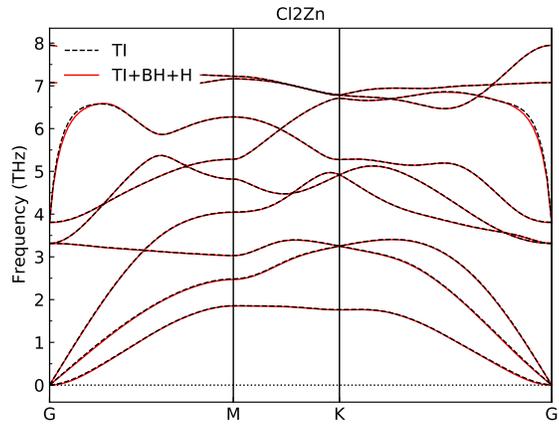

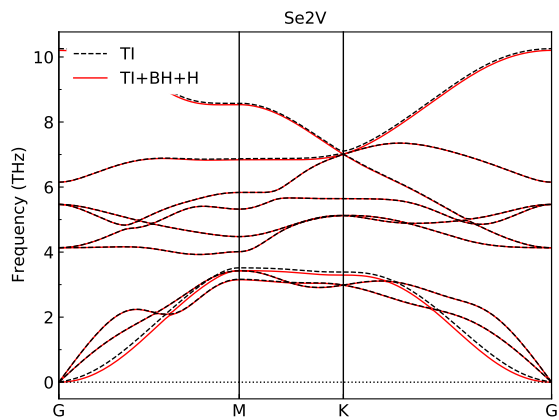

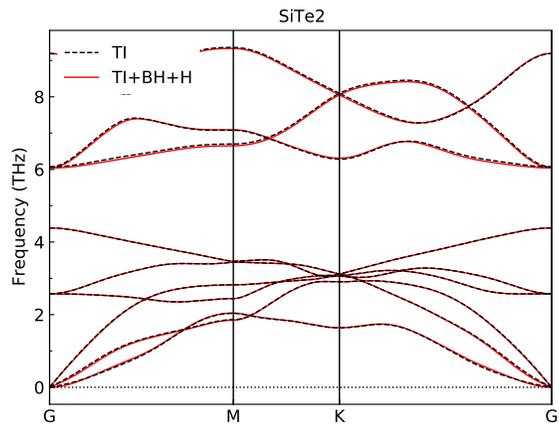

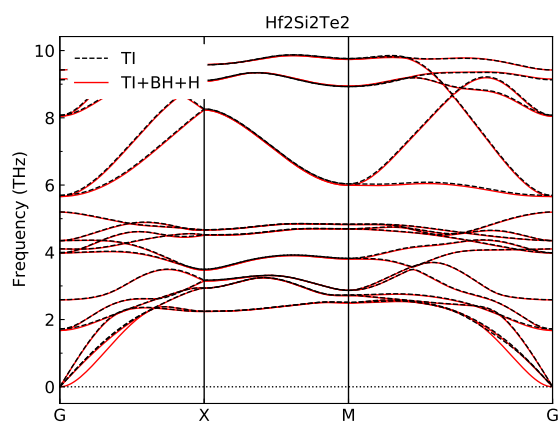
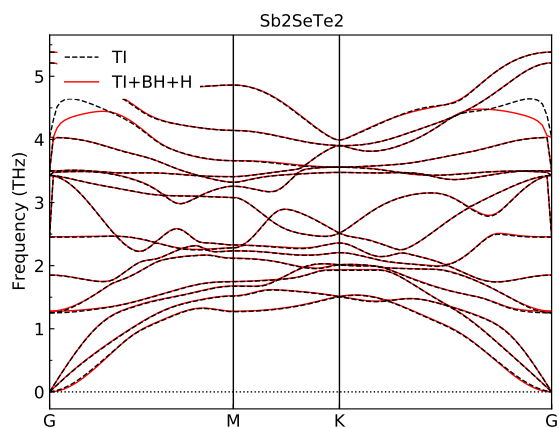
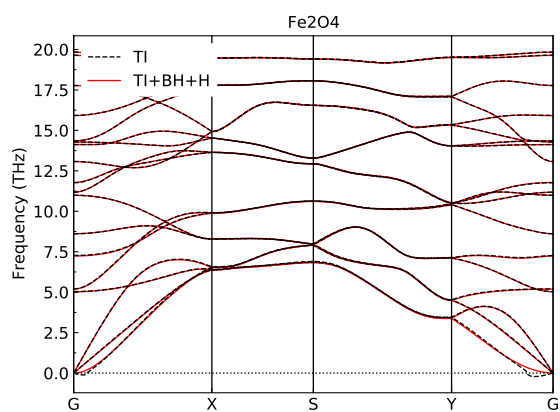
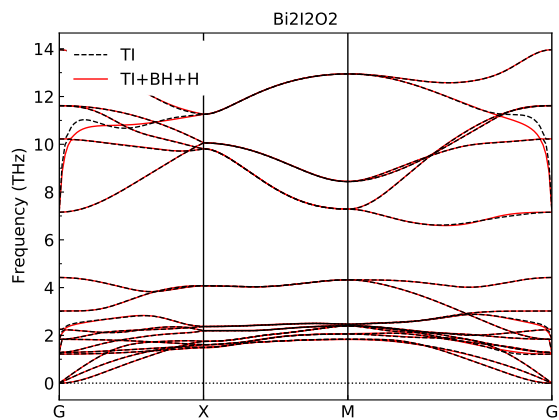
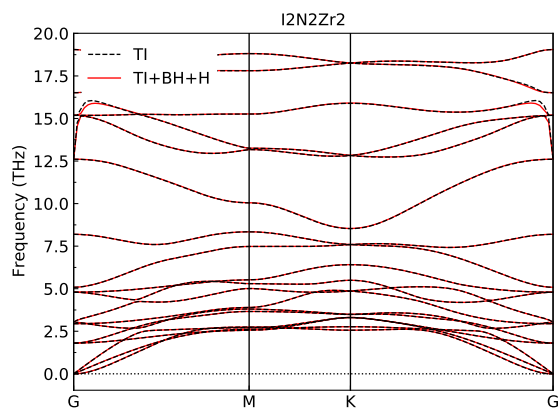
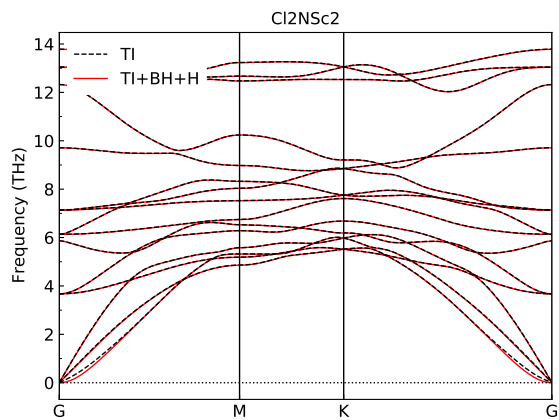
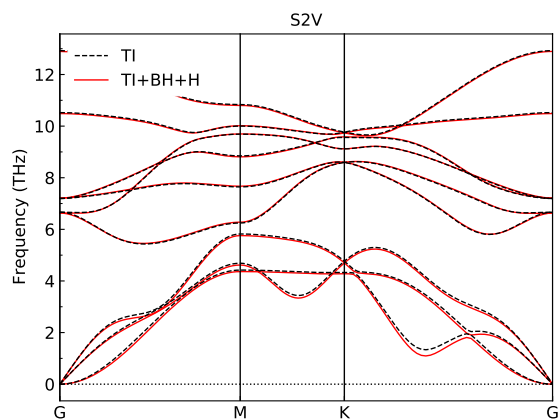
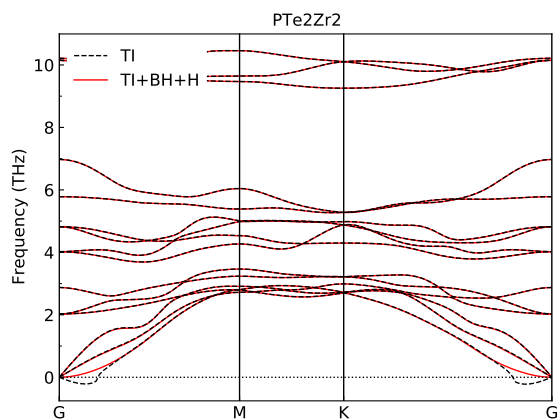

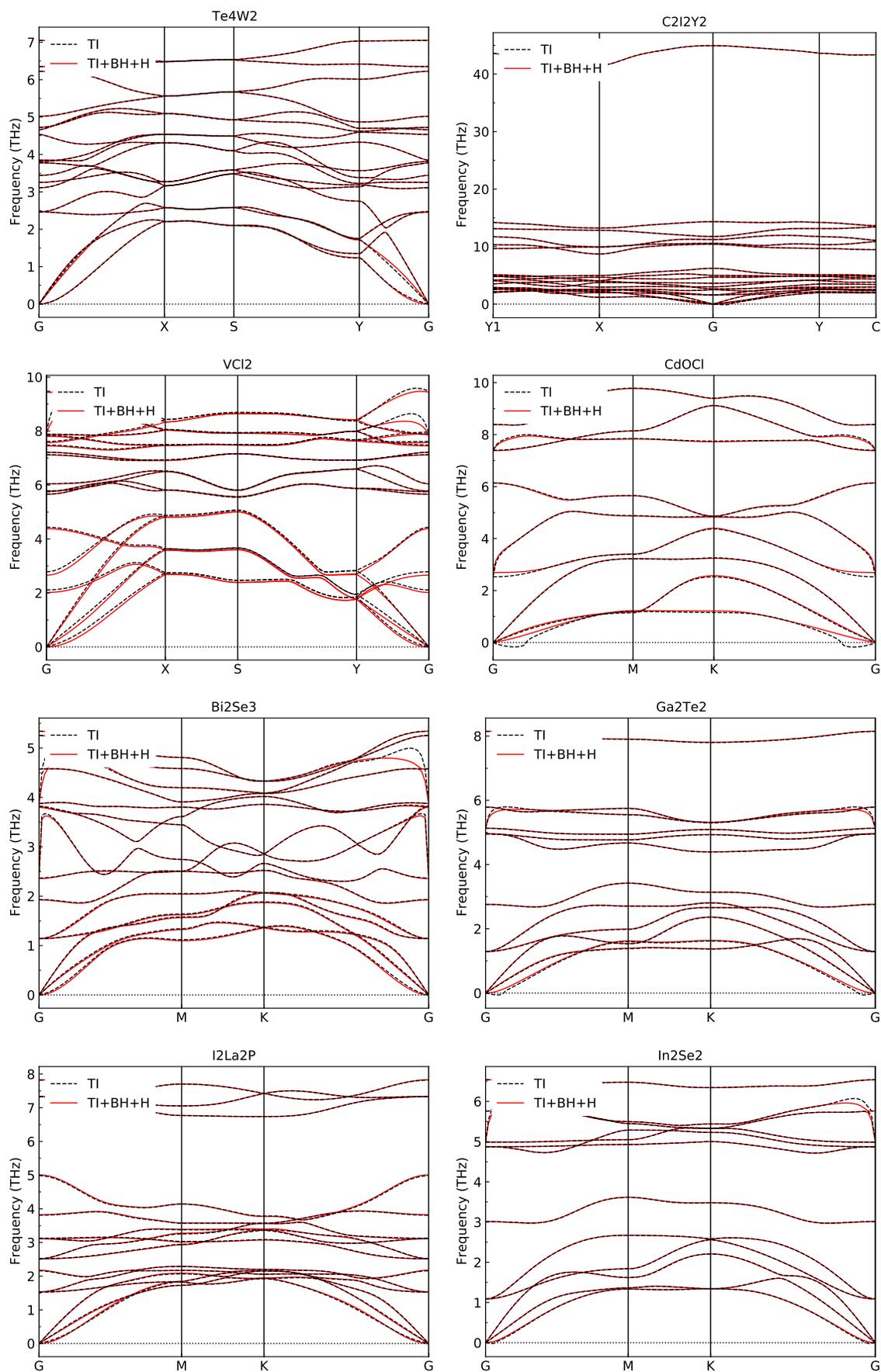

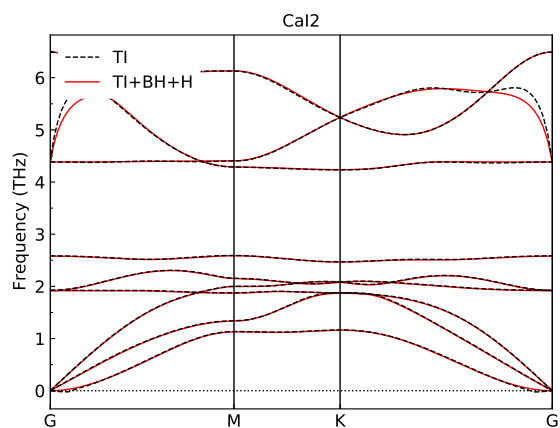

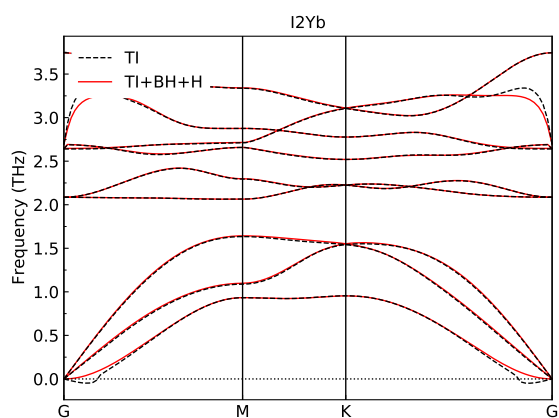

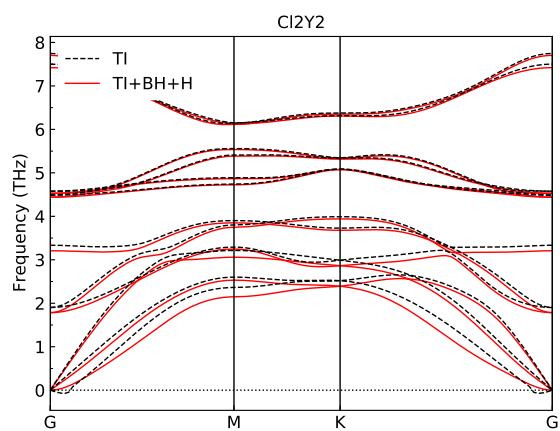

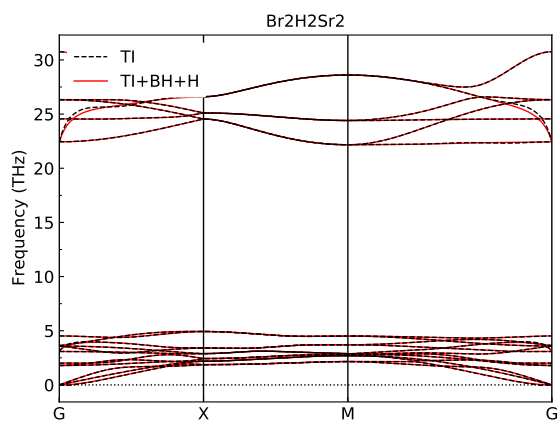

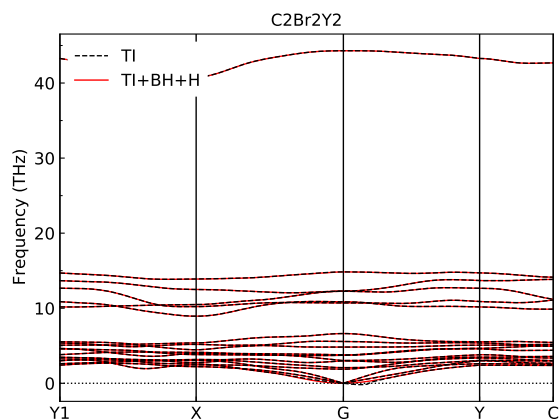

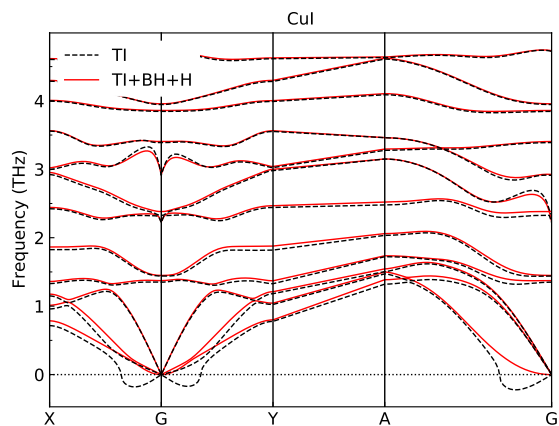

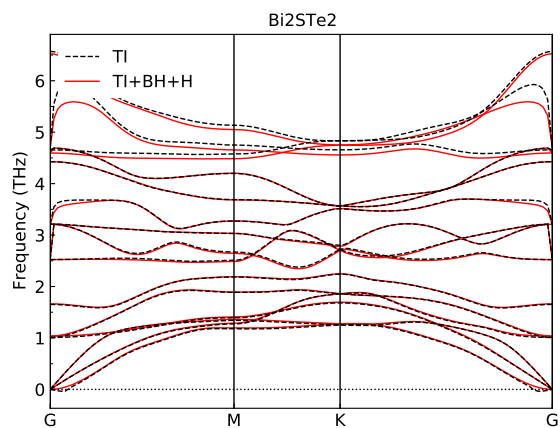

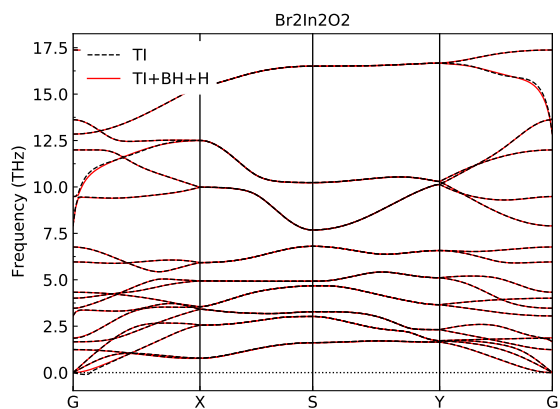

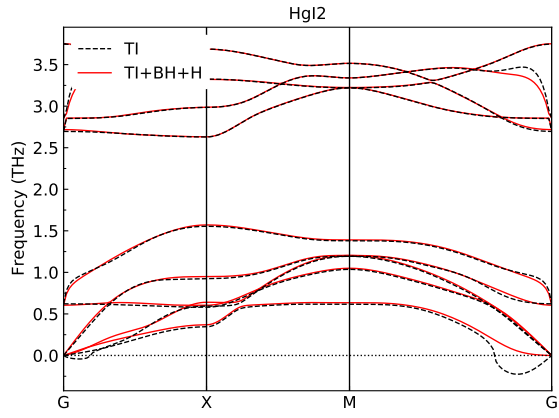

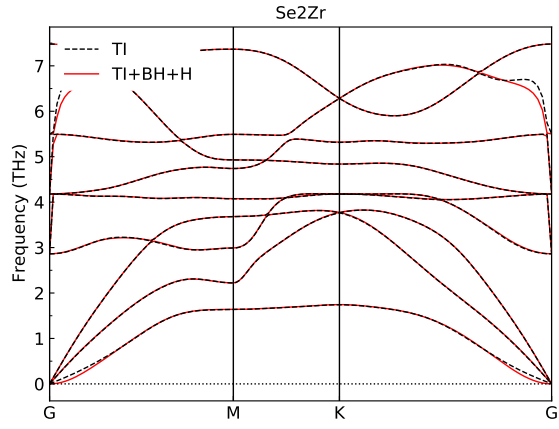

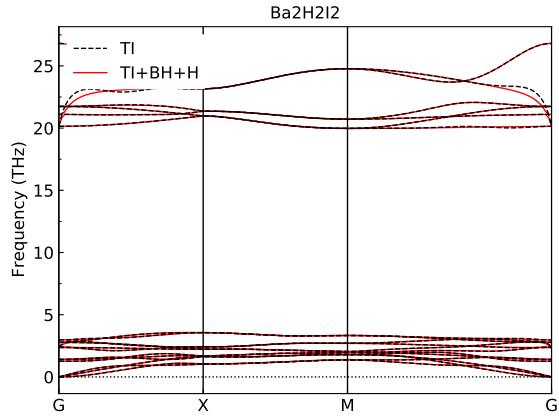

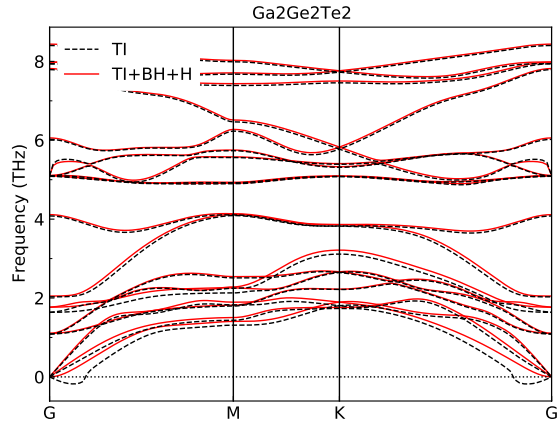

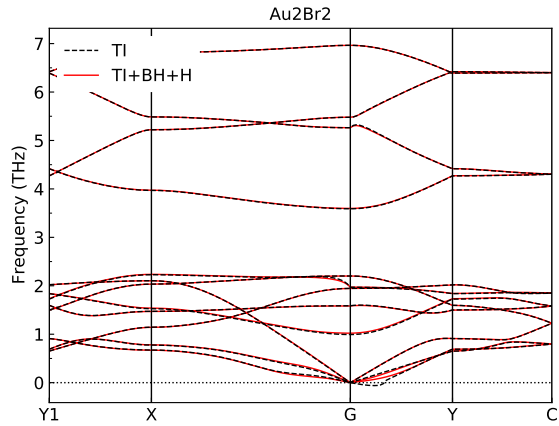

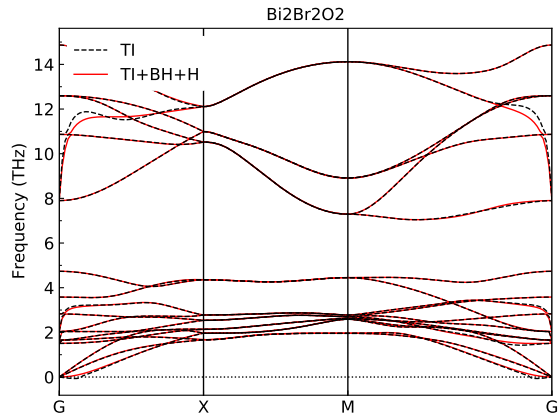

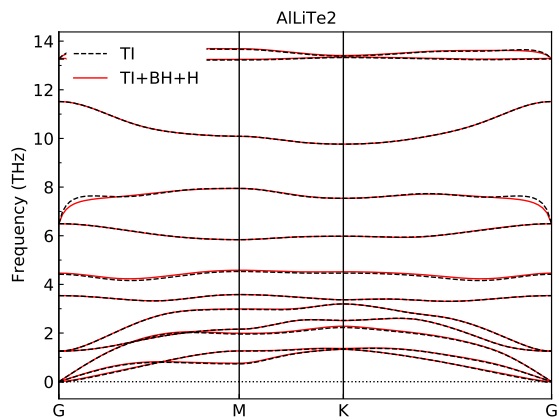

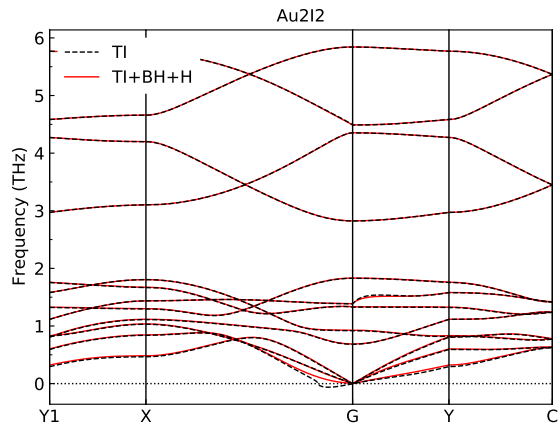

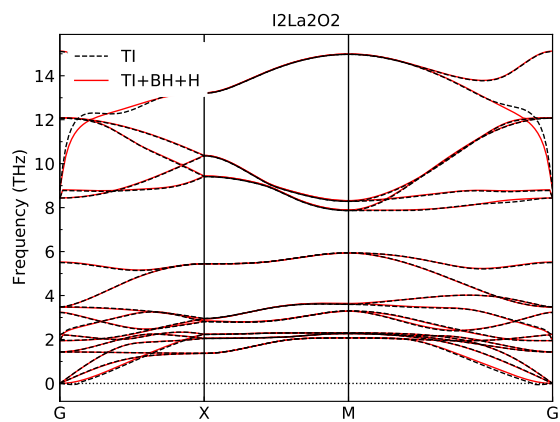

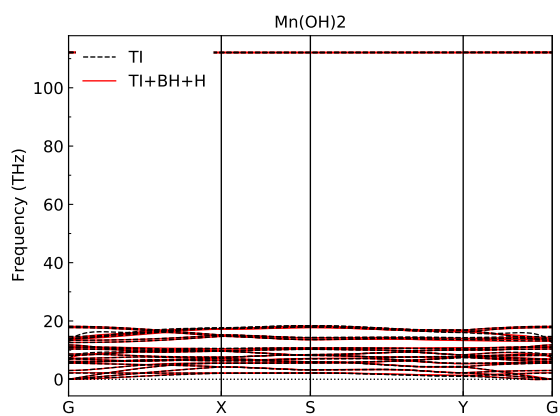

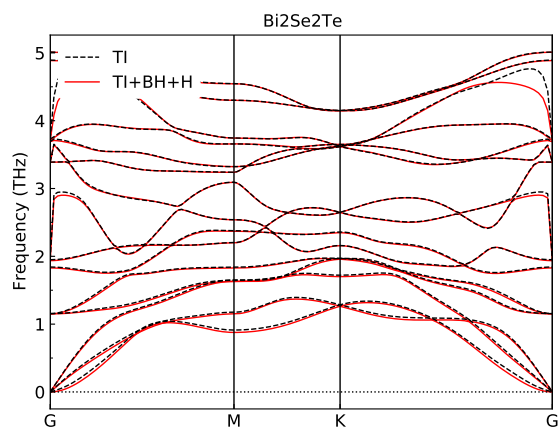

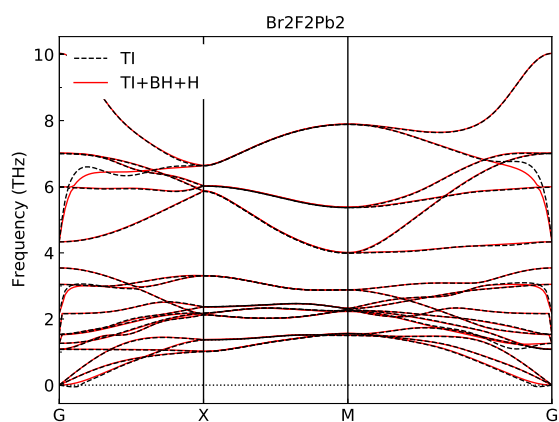

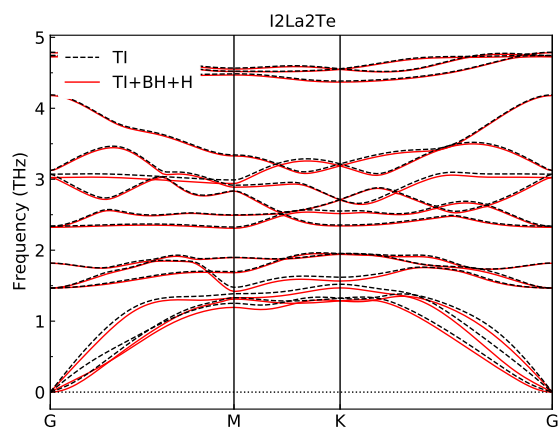

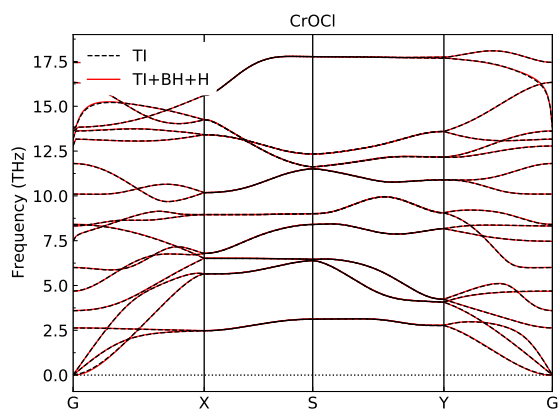

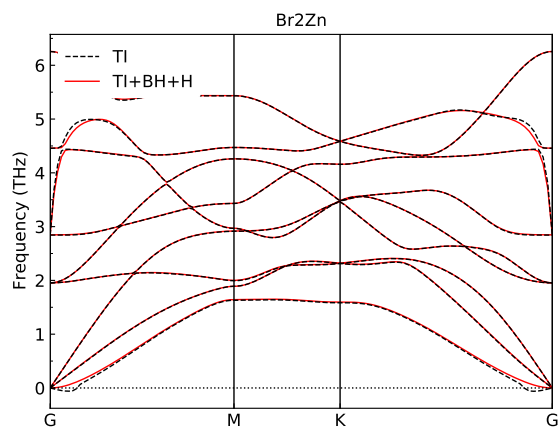

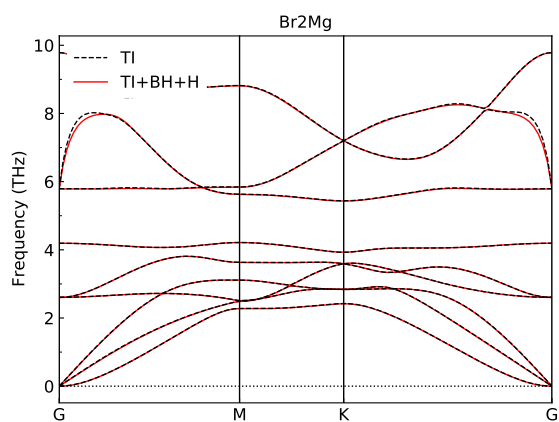

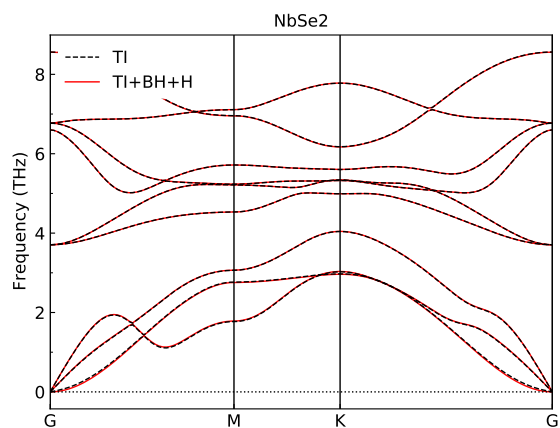

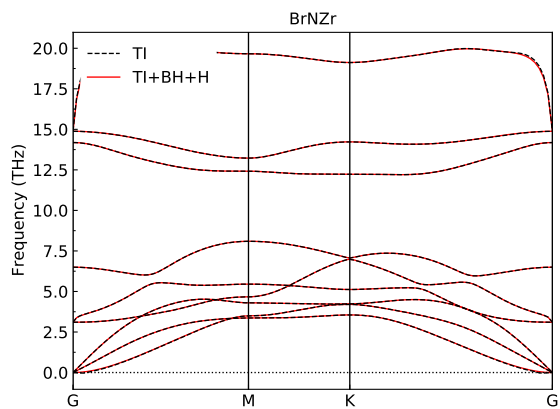

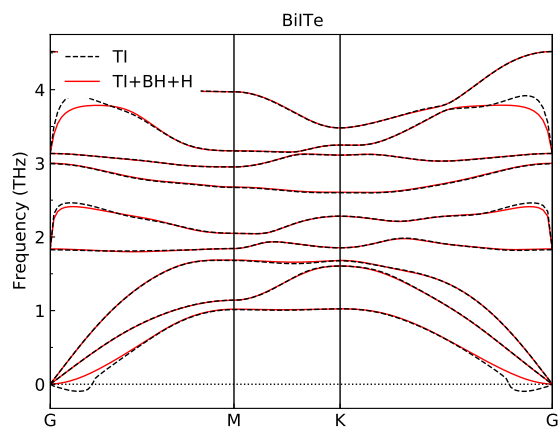

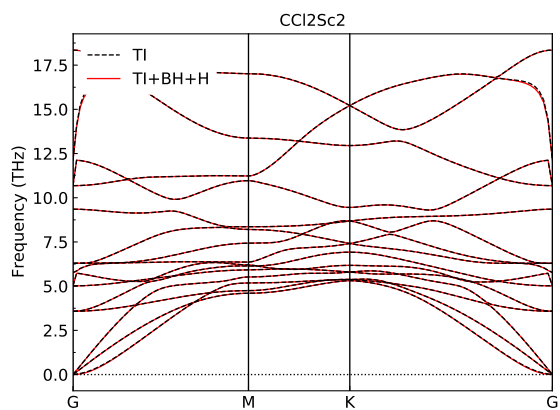

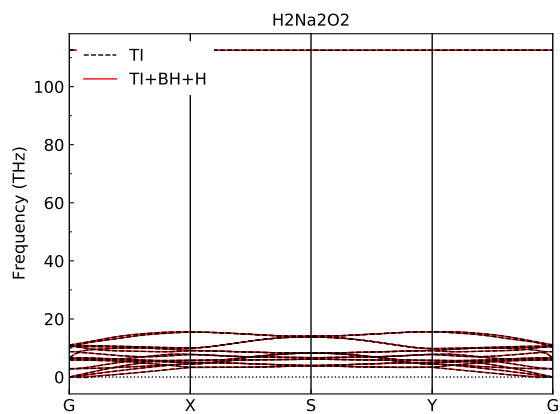

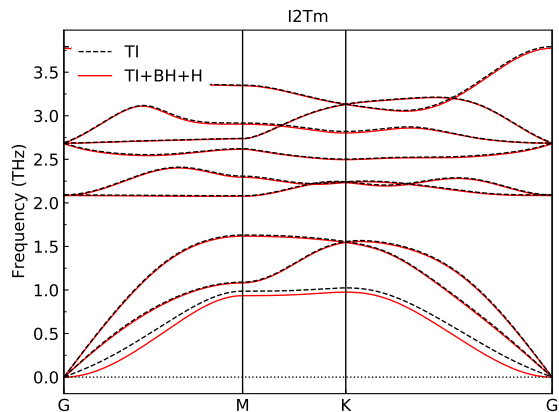

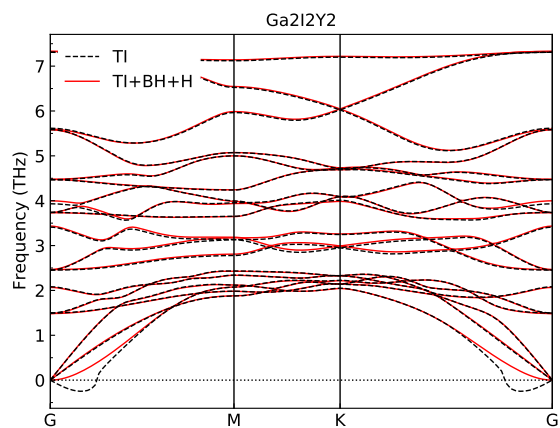

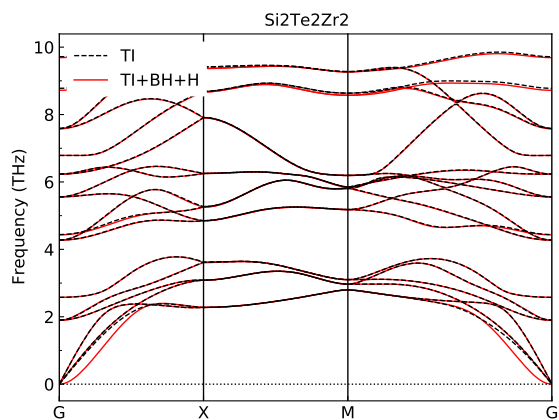

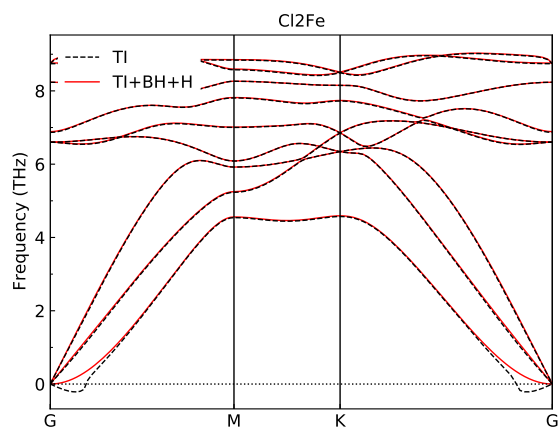

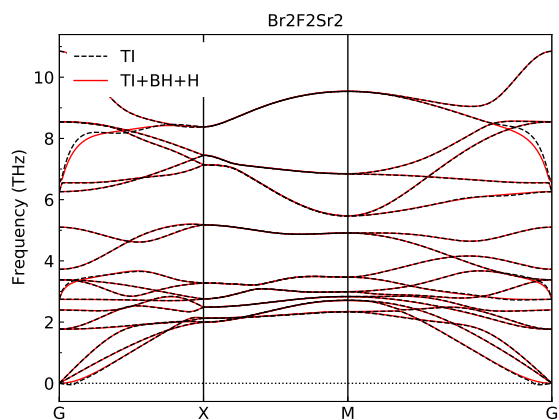

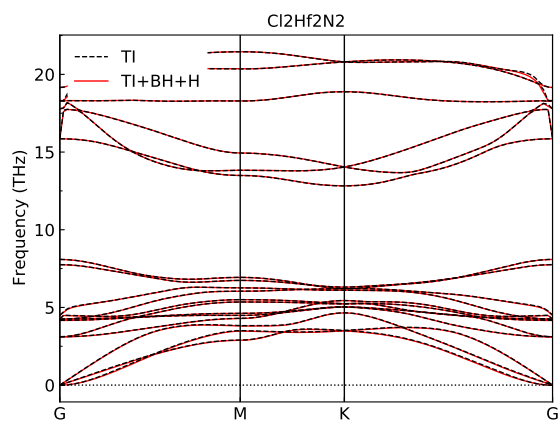

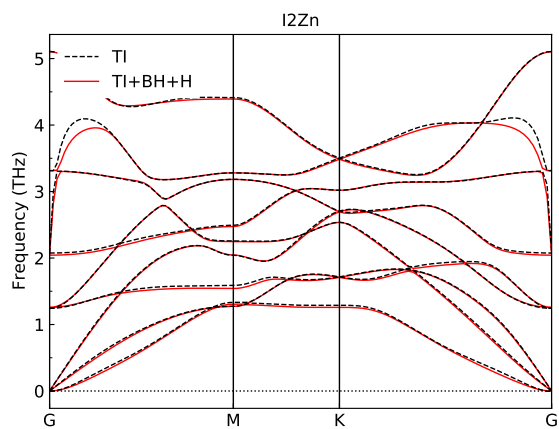

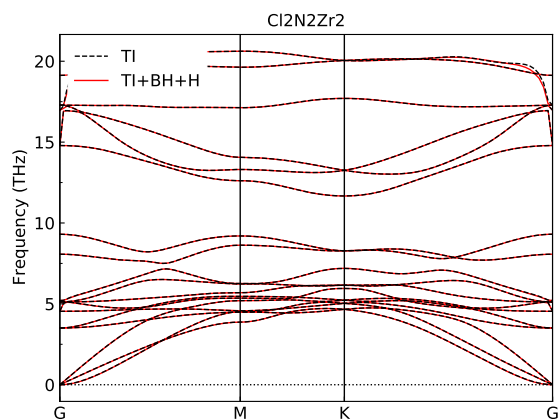

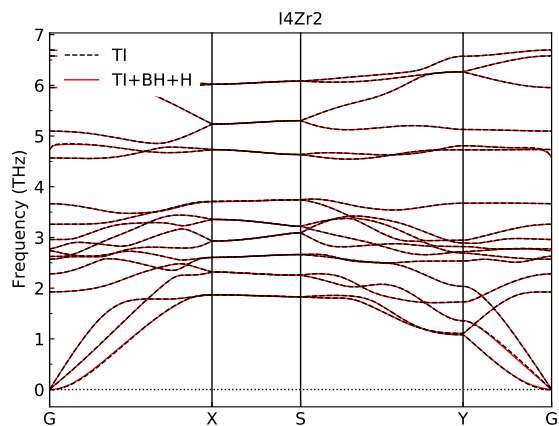

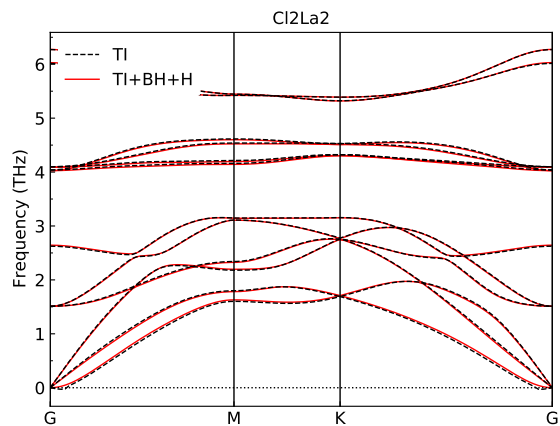

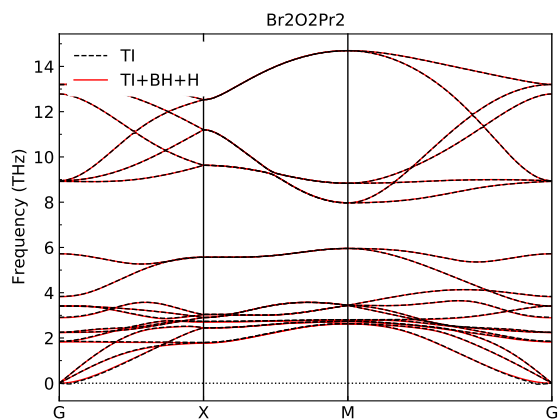

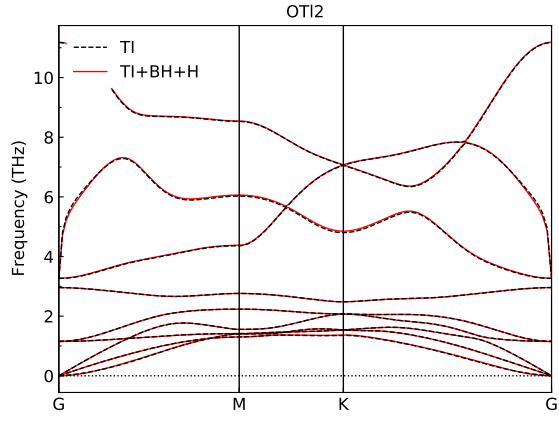

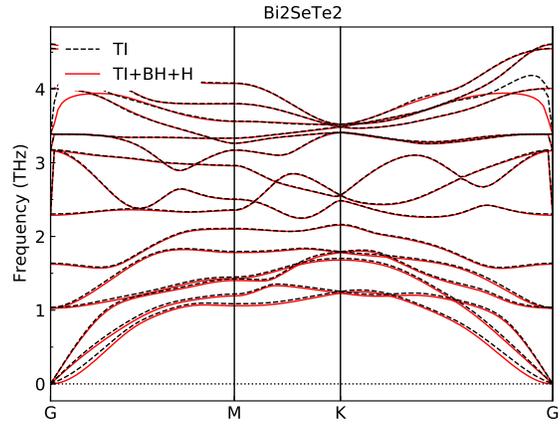

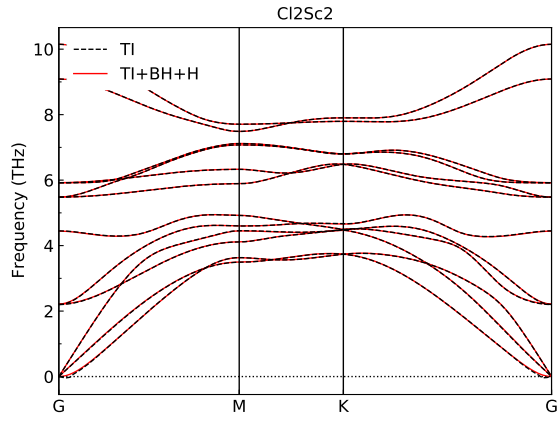

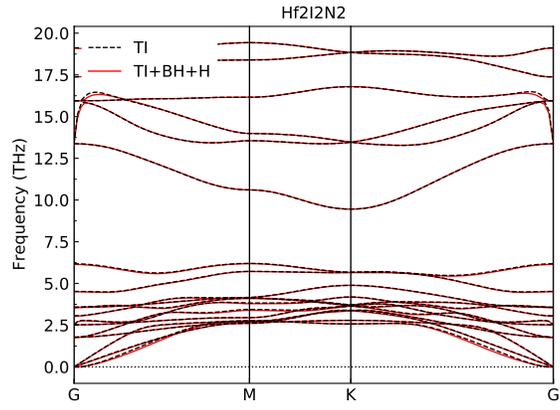

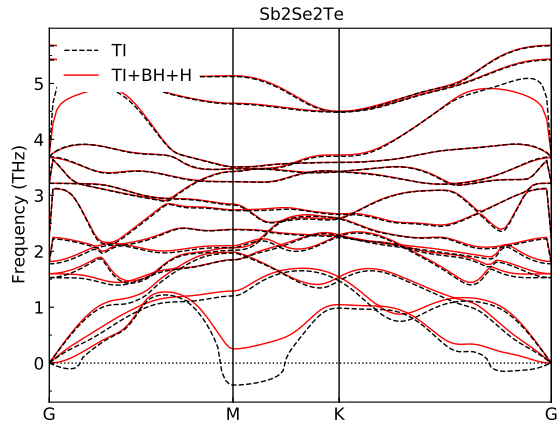

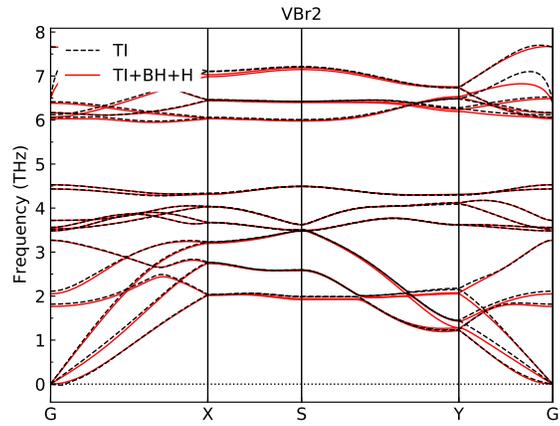

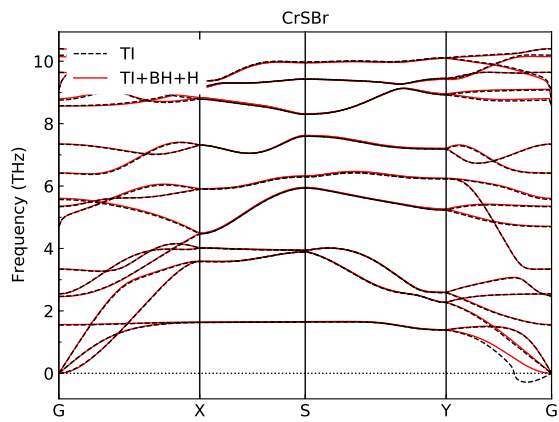

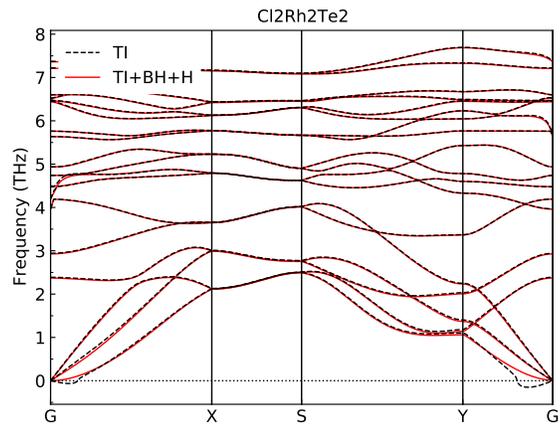

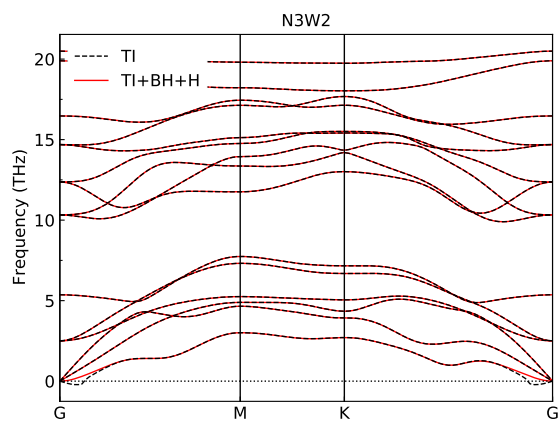

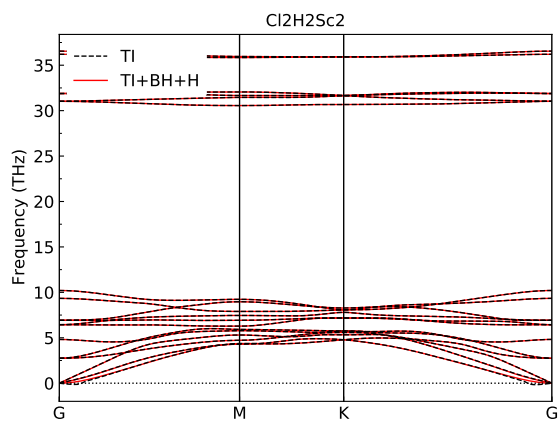

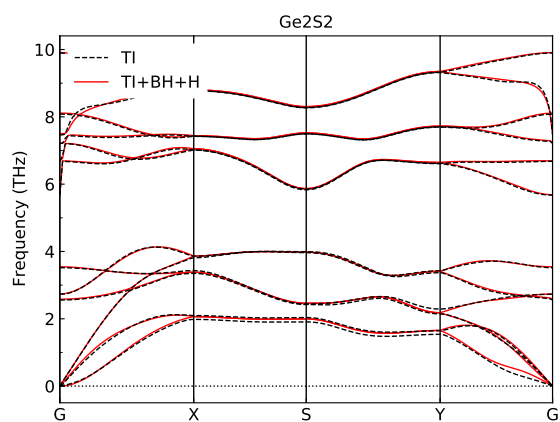

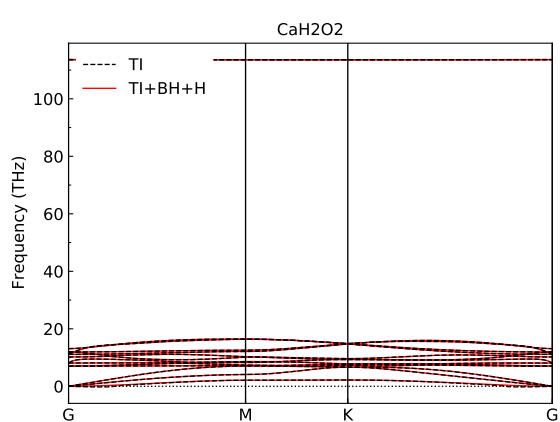

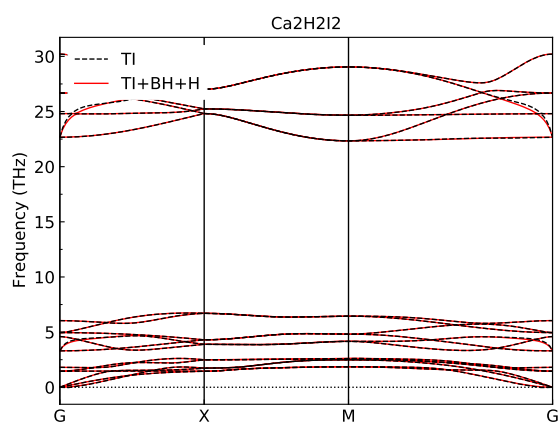

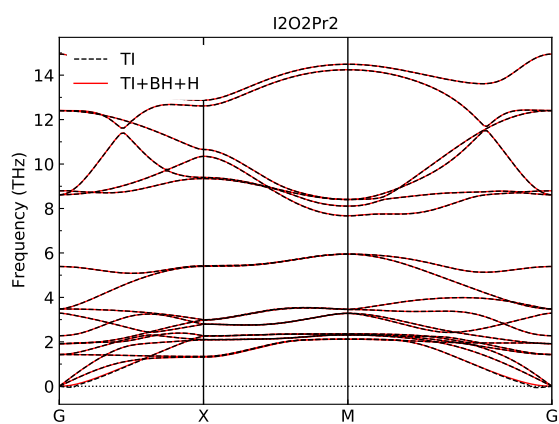

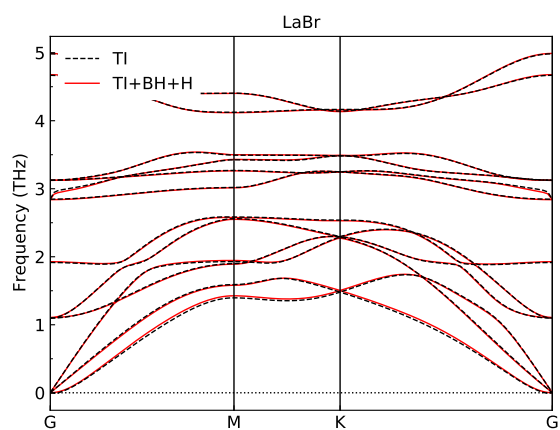

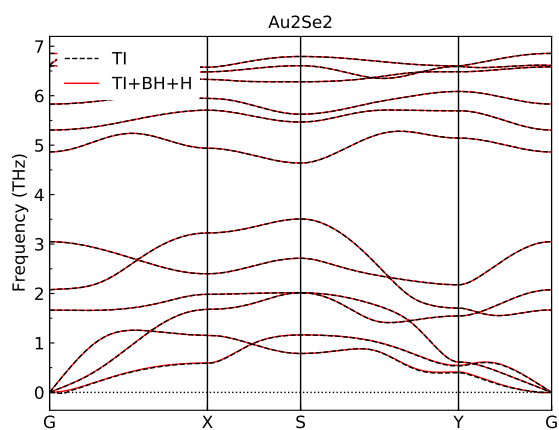

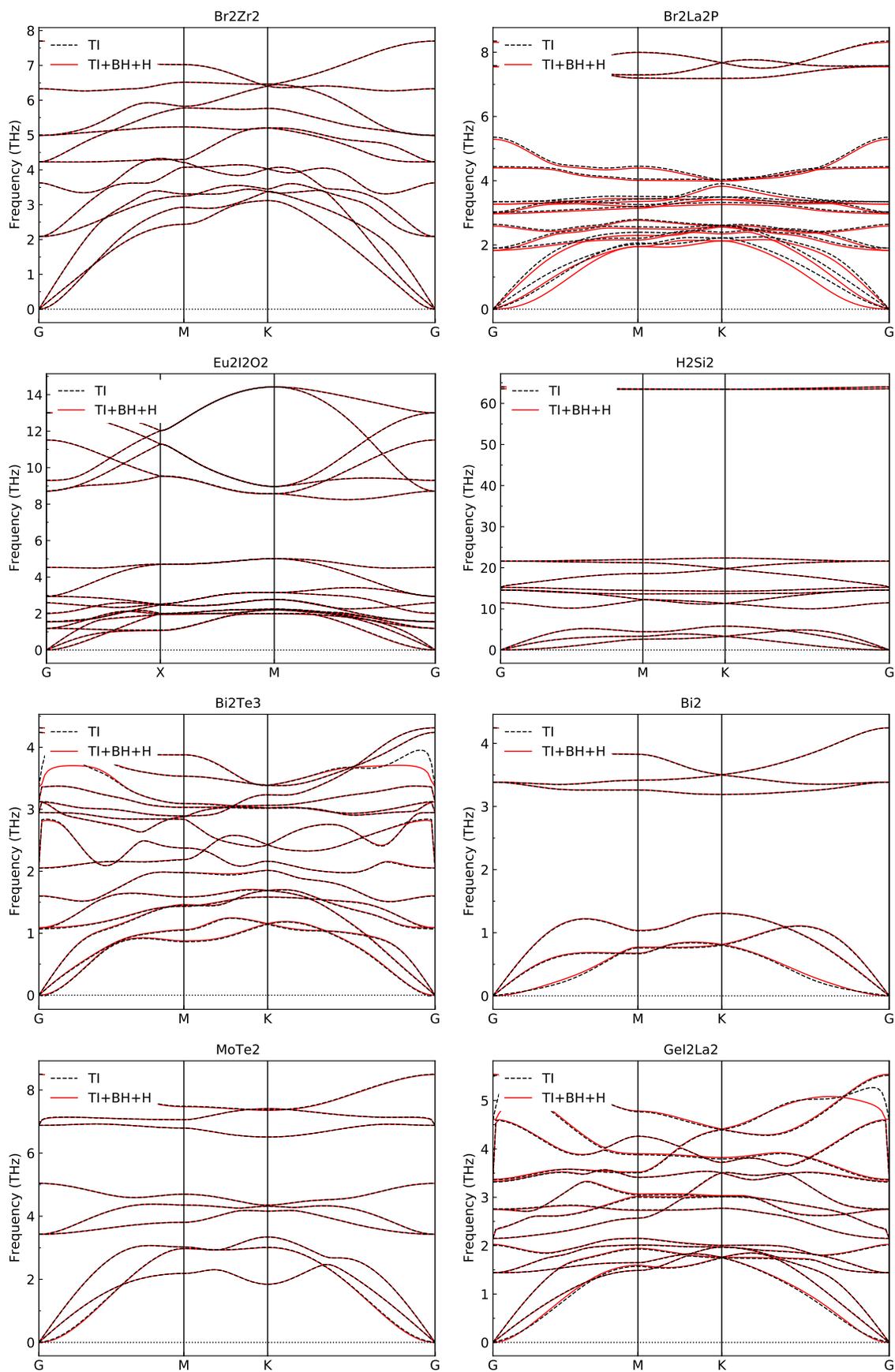

**SUPPLEMENTARY NOTE 4: 2D MATERIALS WHERE THE INVARIANCE CONDITIONS FAILED TO RECOVER A STABLE ZA MODE (33 ENTRIES)**

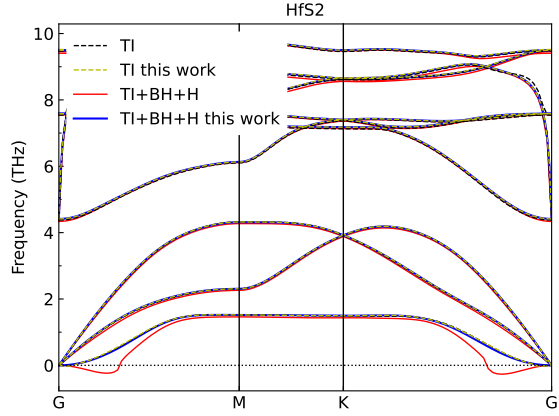

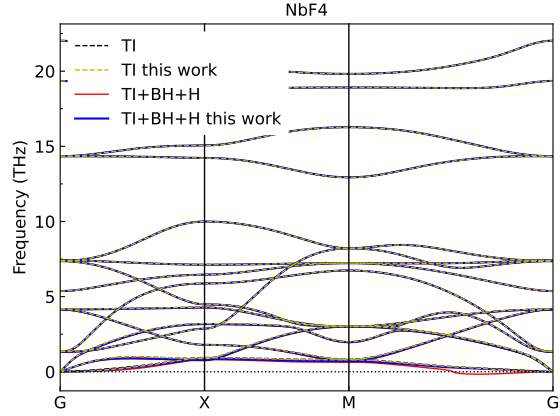

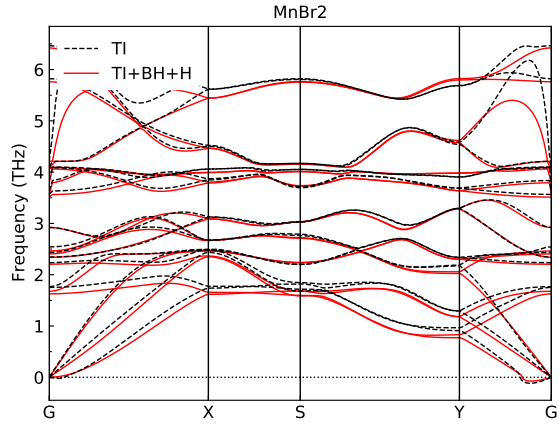

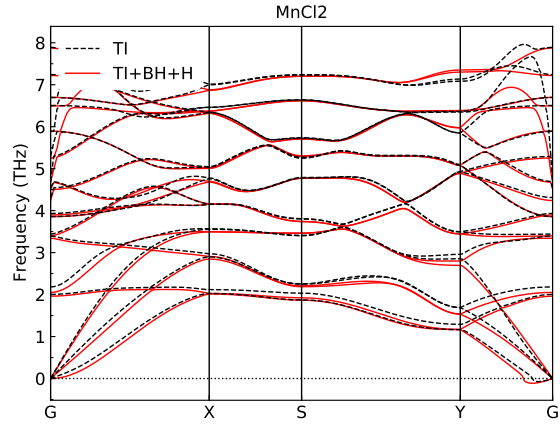

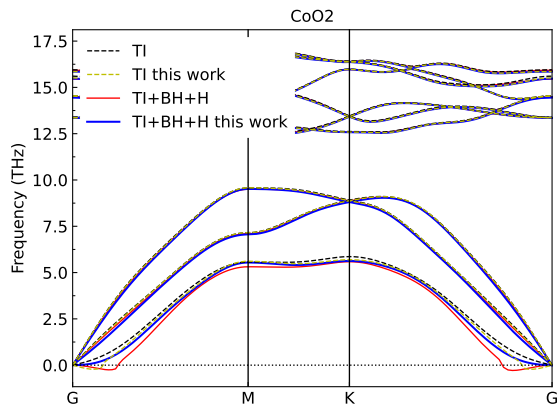

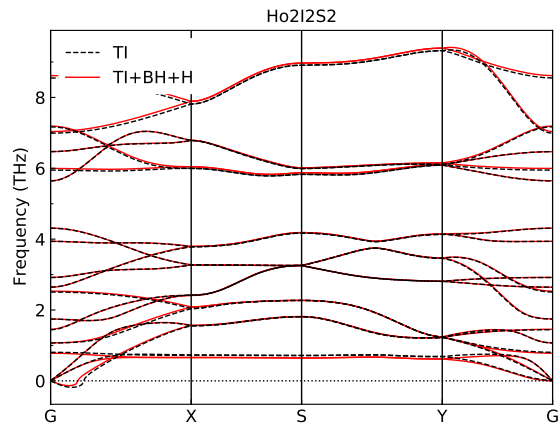

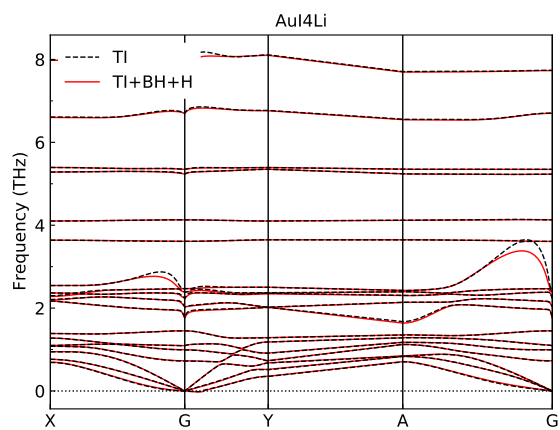
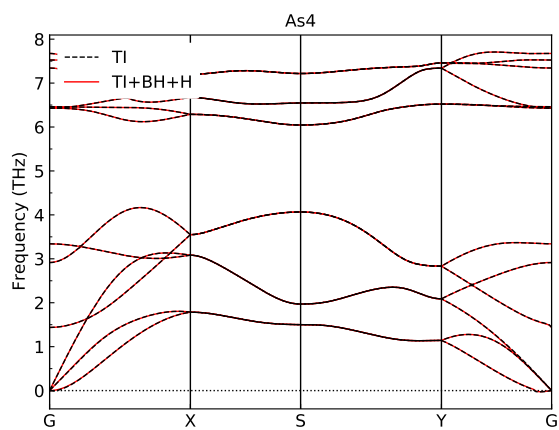
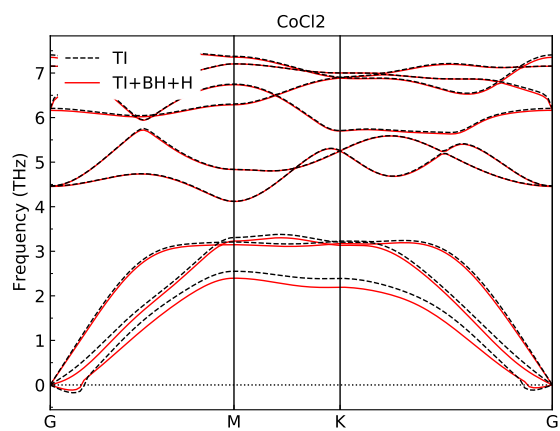
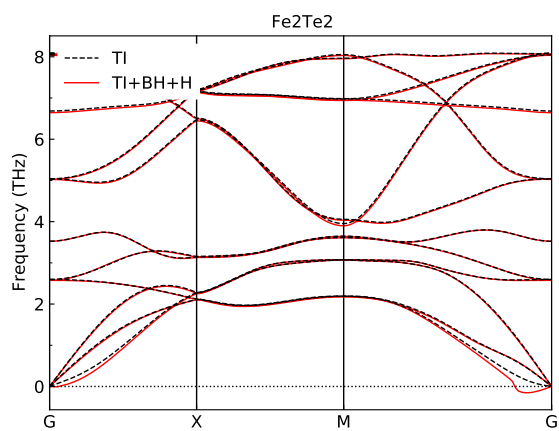
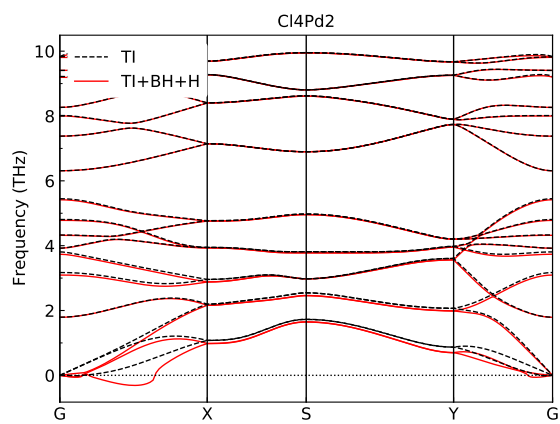
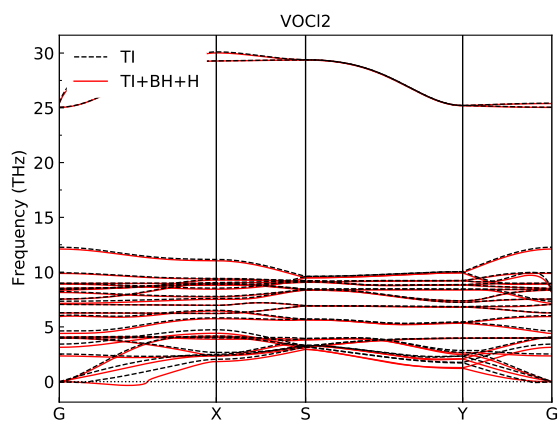
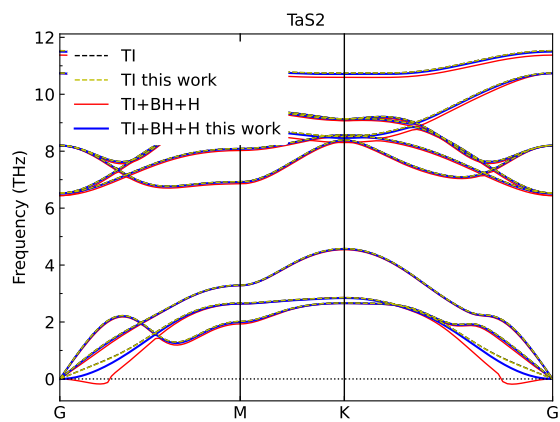
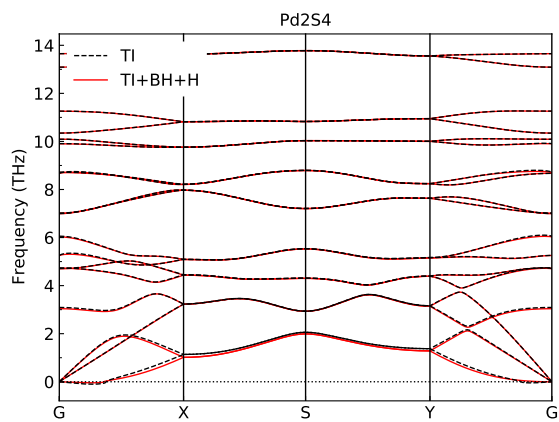

28

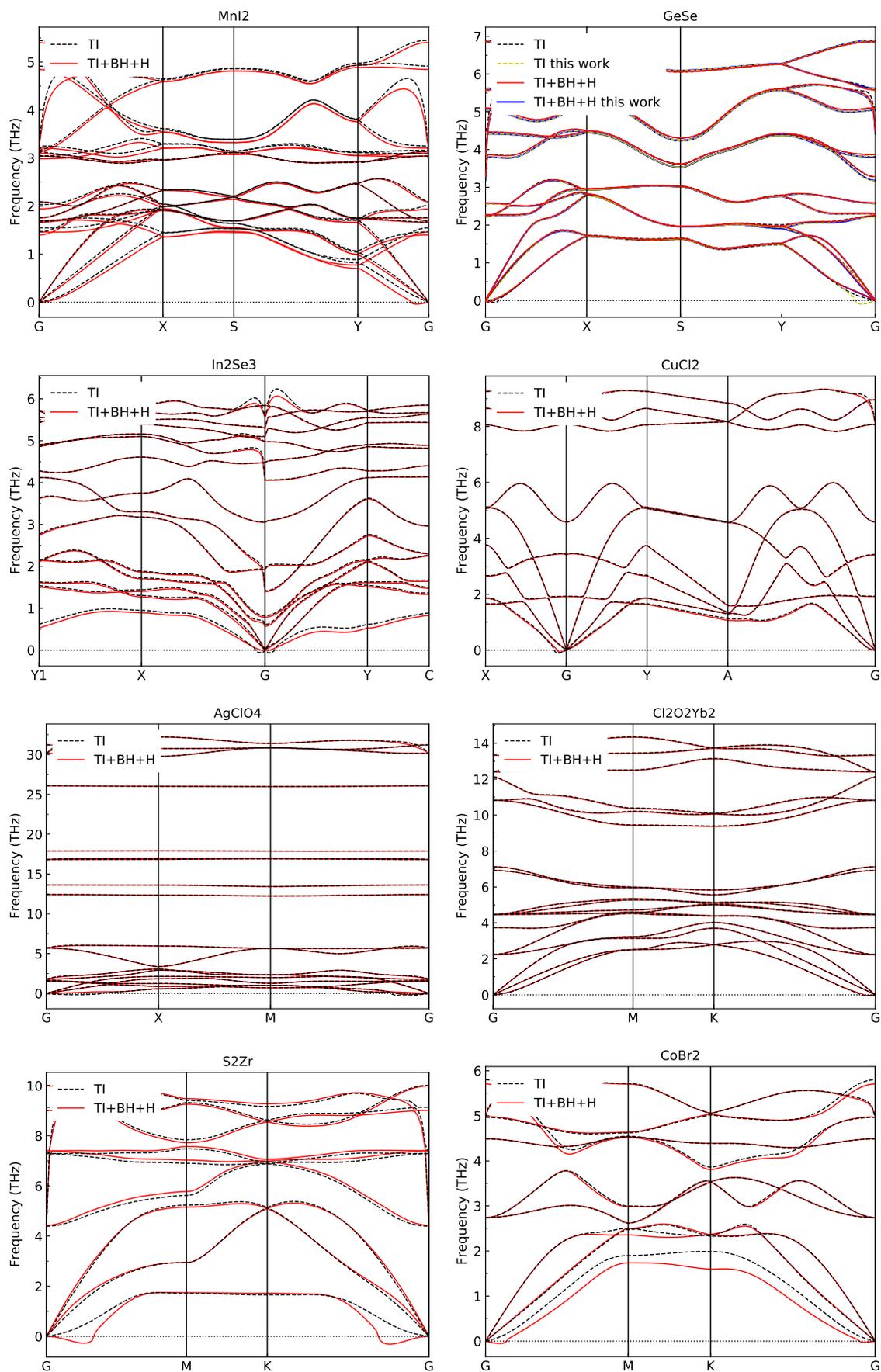

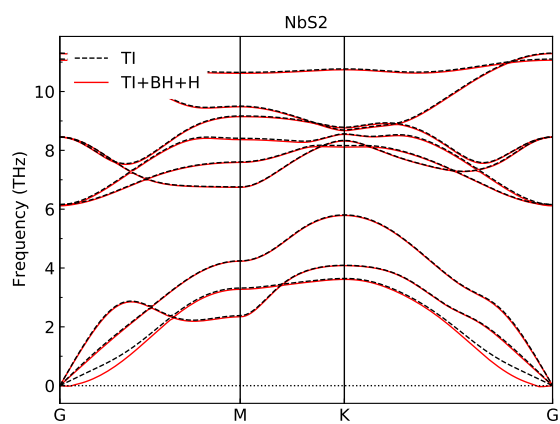
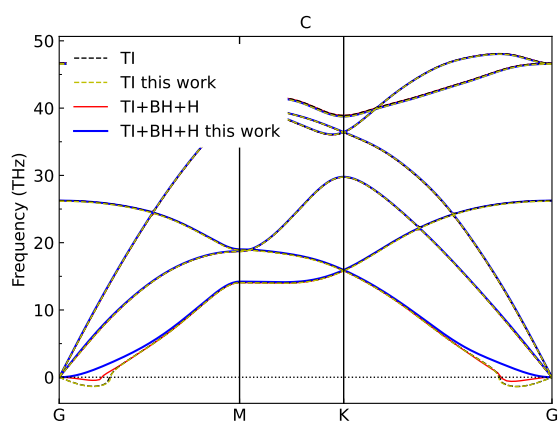
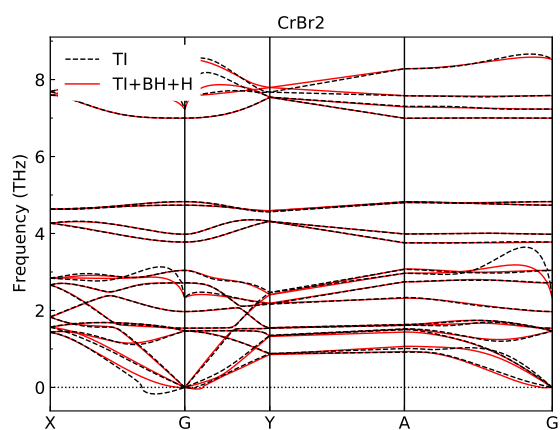
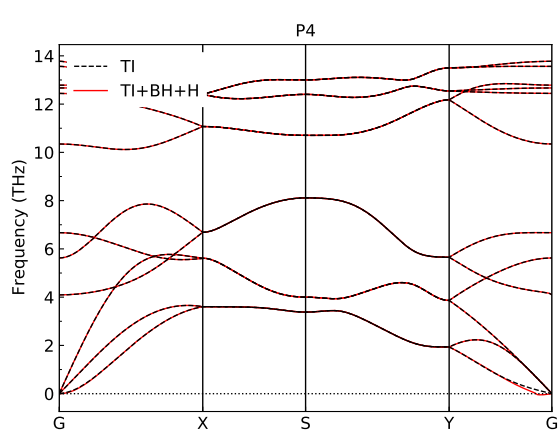
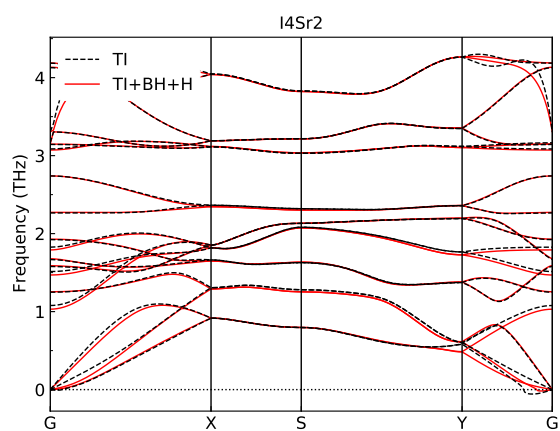
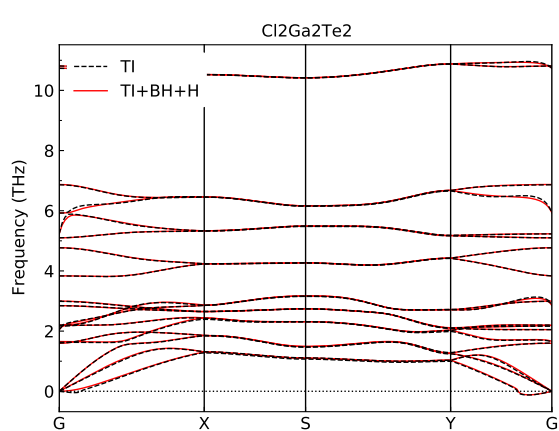
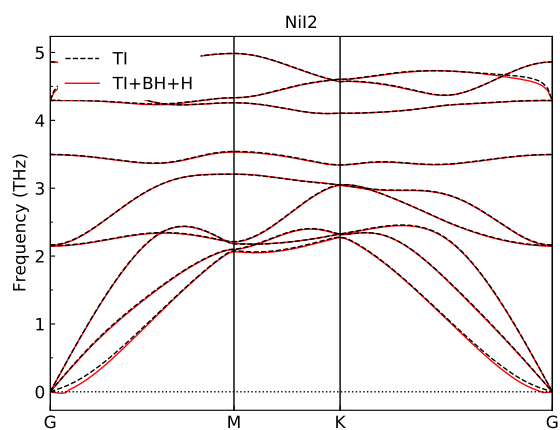
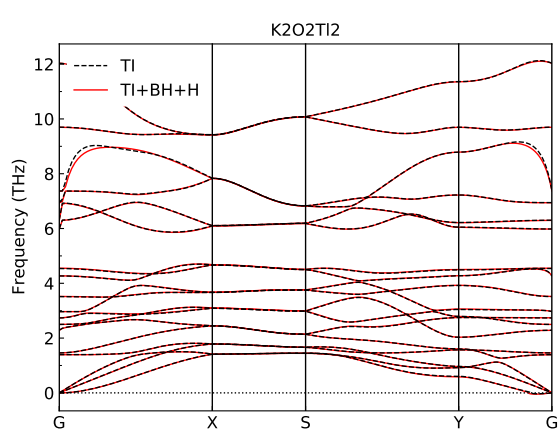

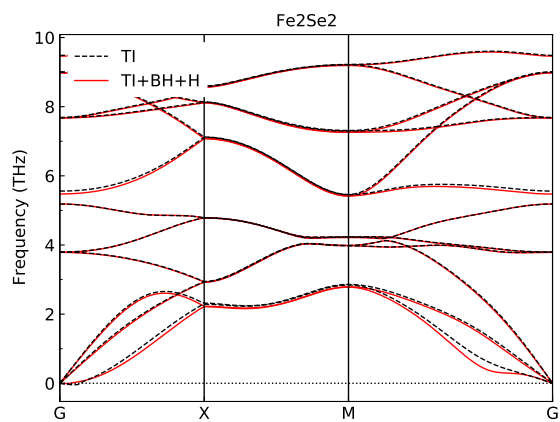
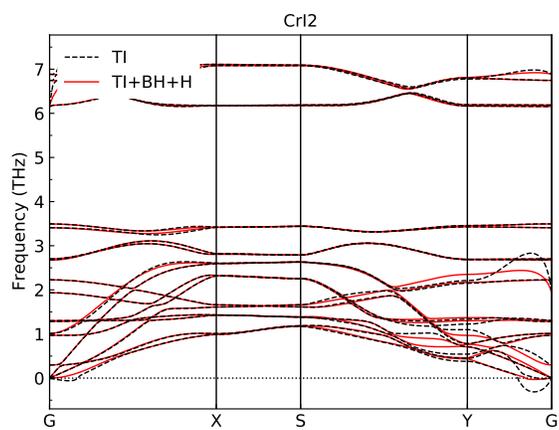
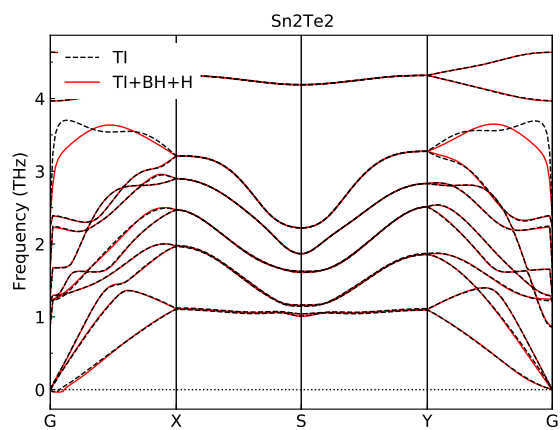

# SUPPLEMENTARY NOTE 5: UNSTABLE 2D MATERIALS (54 ENTRIES)

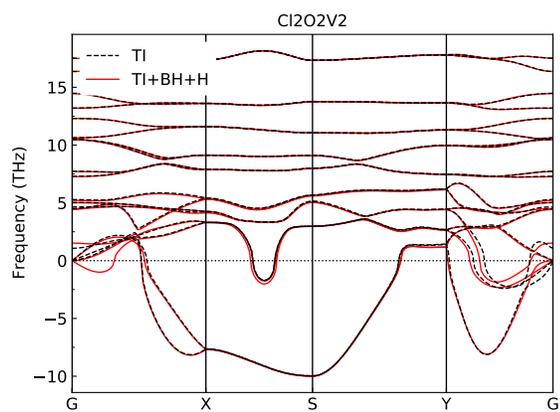

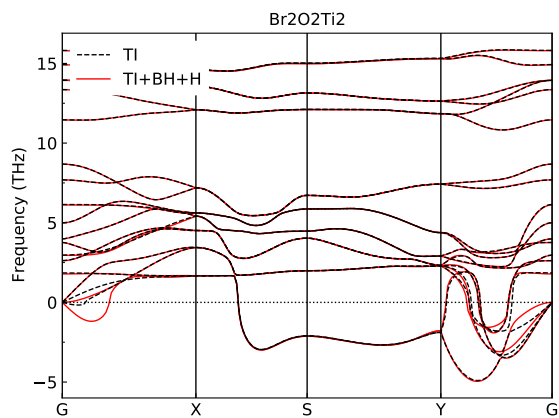

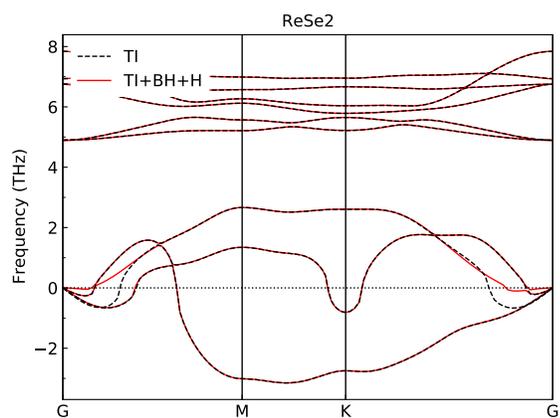

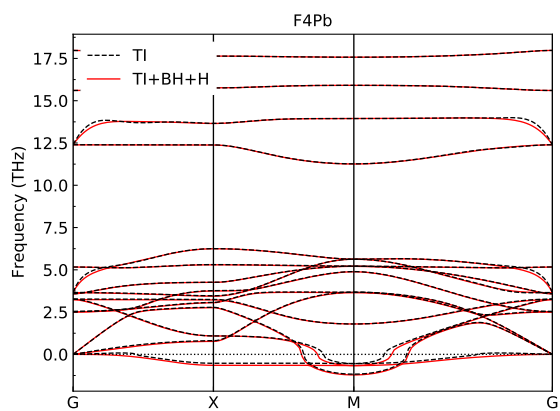

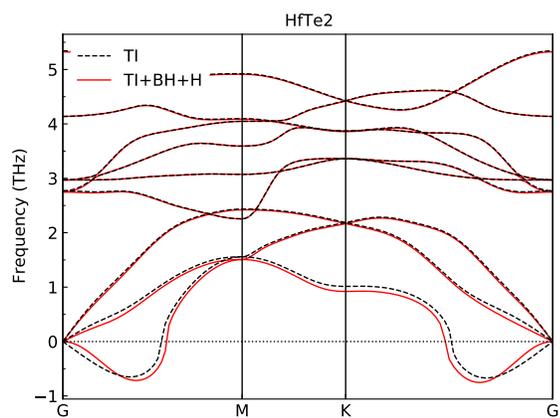

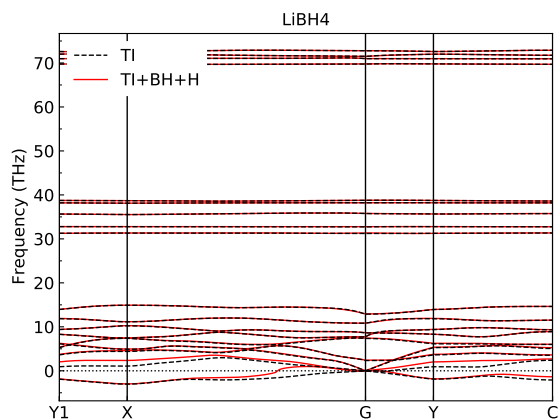

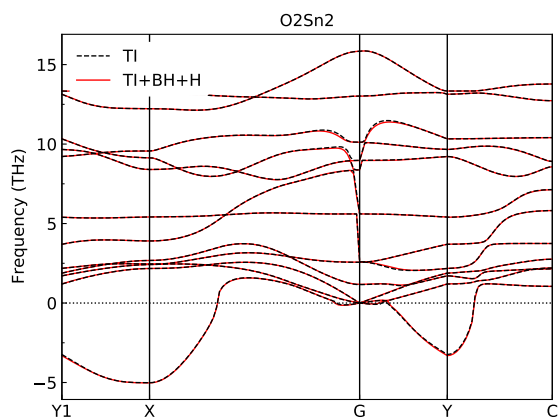

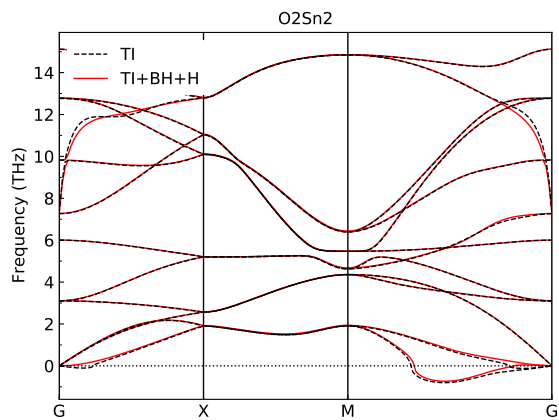

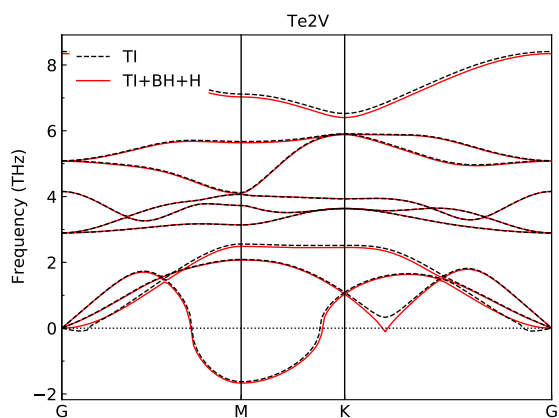

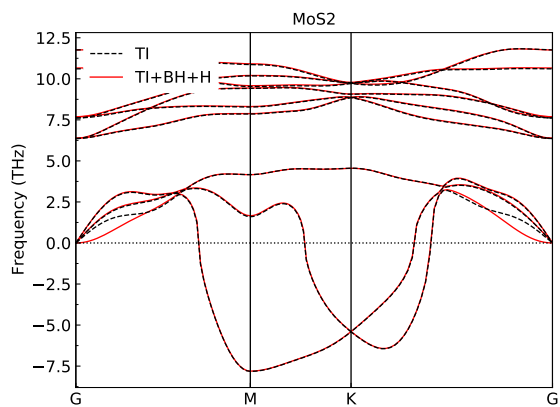

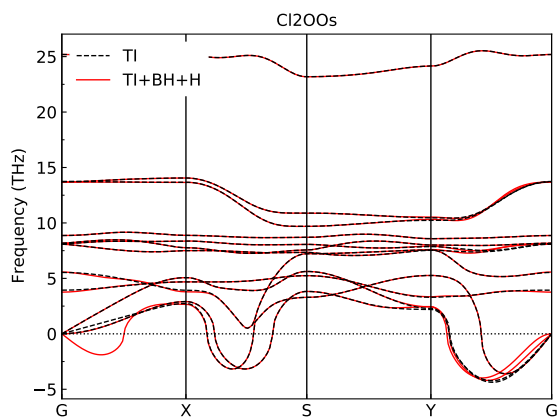

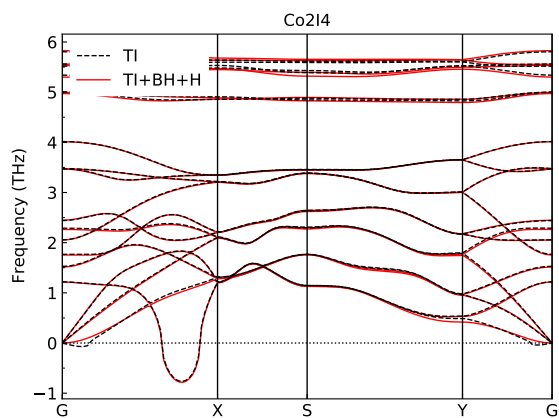

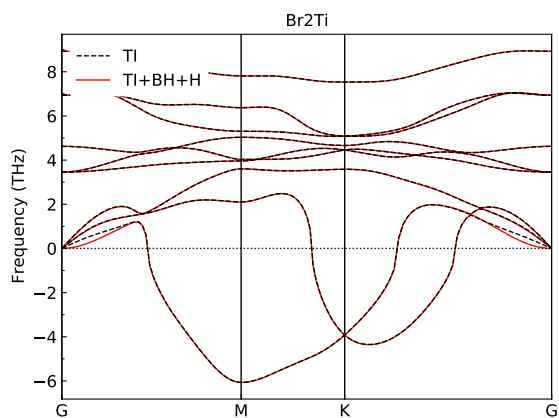

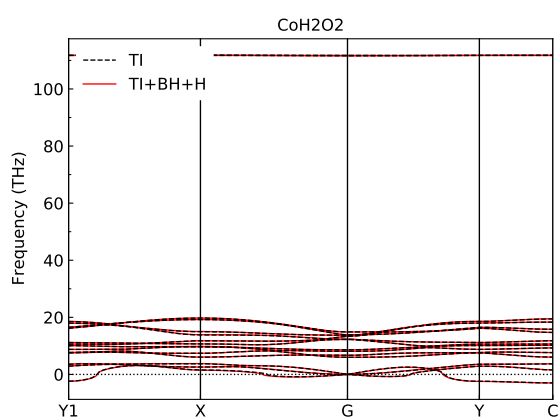

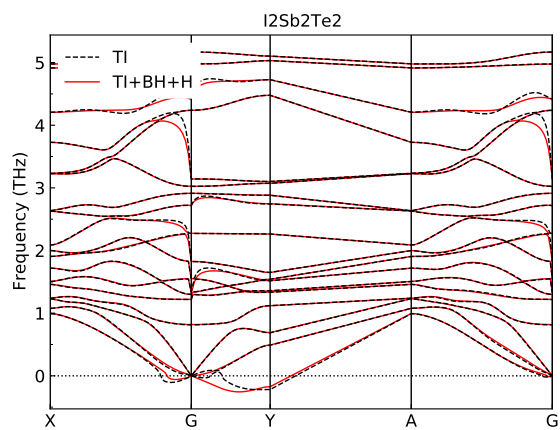

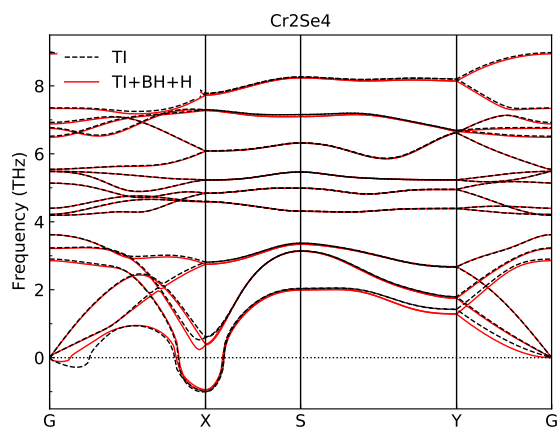

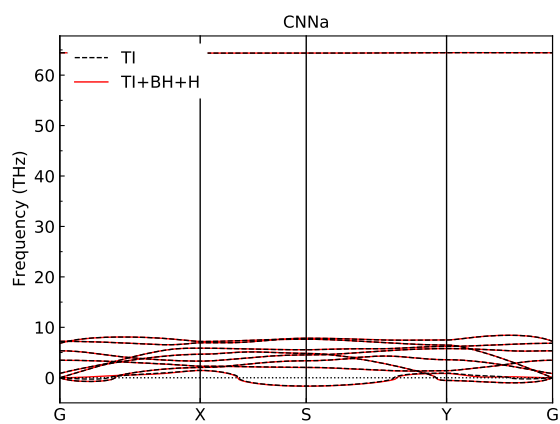

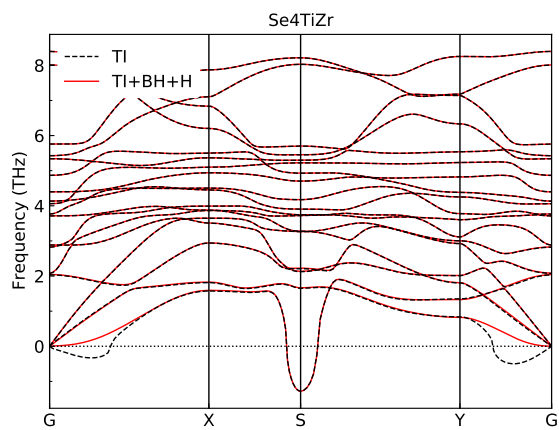

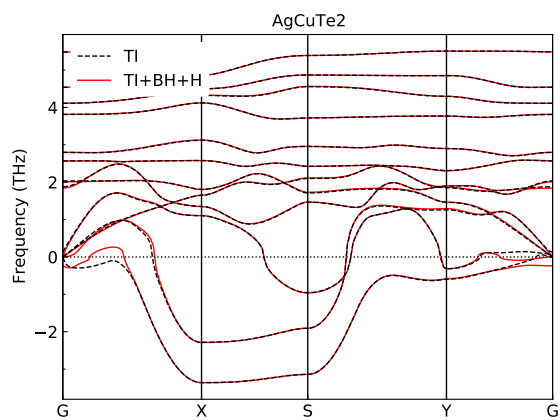

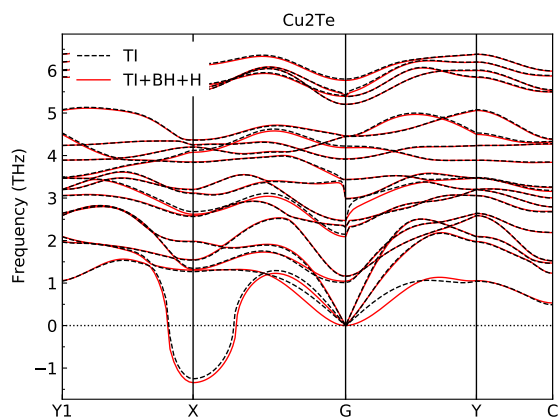

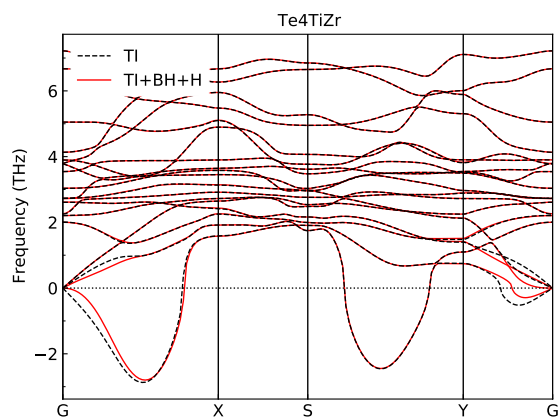

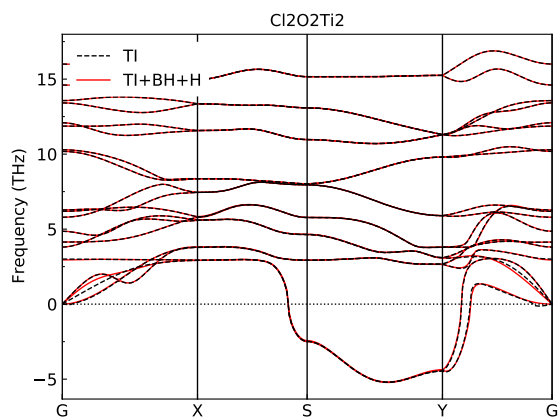

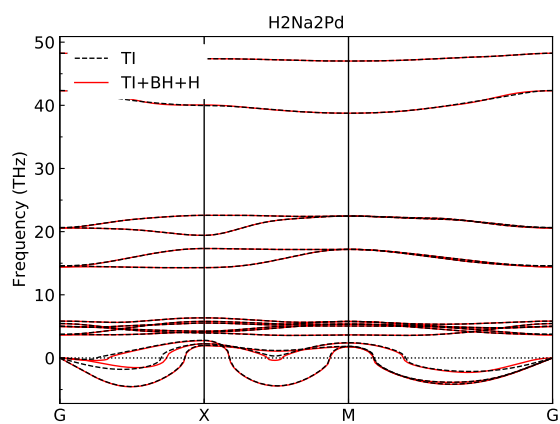
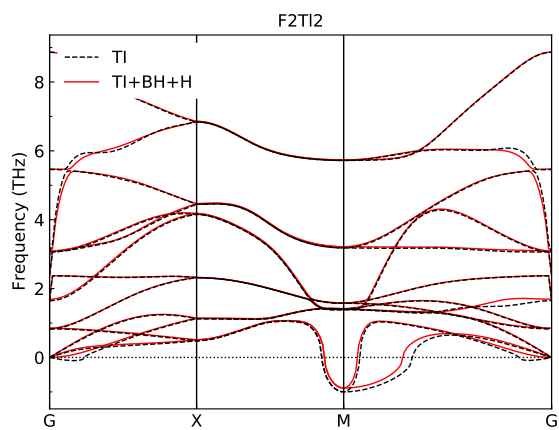
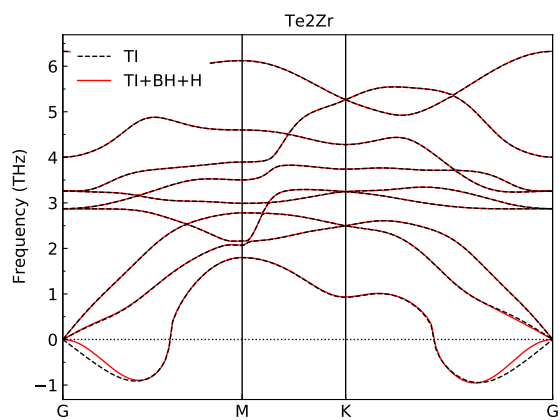
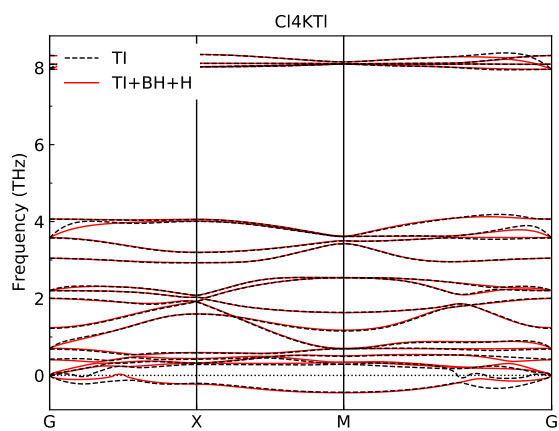
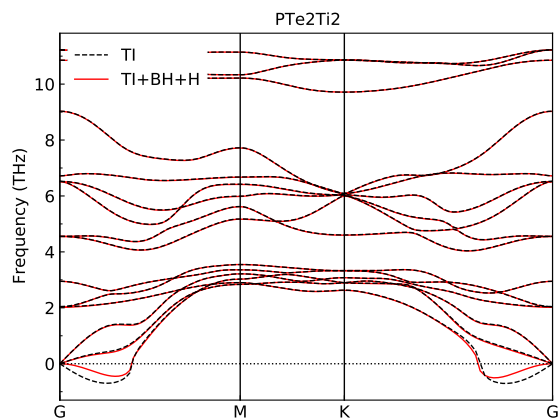
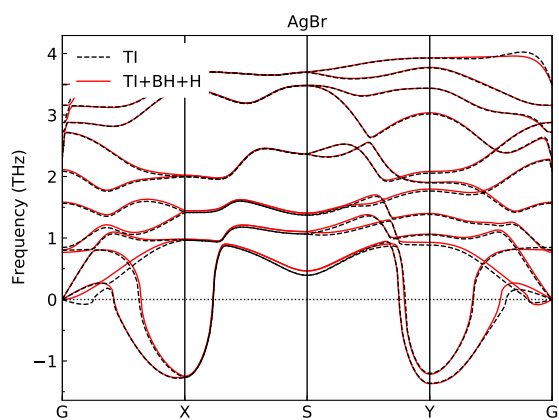
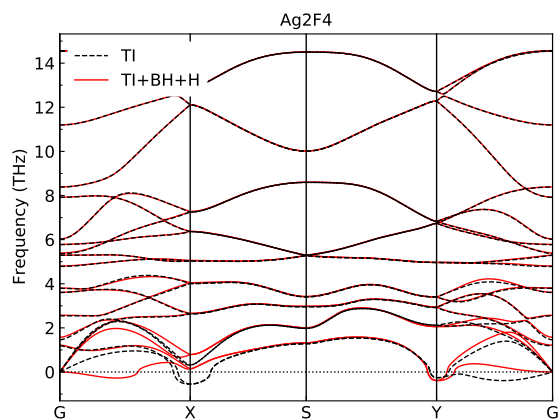
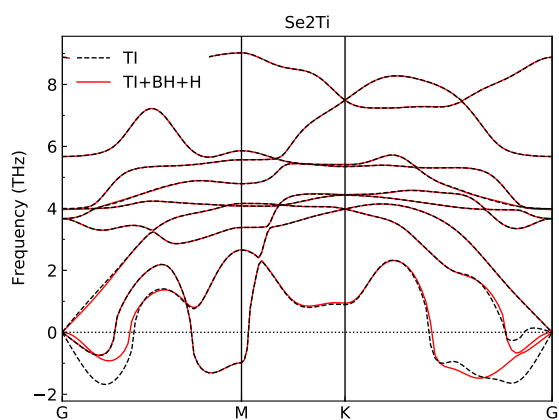

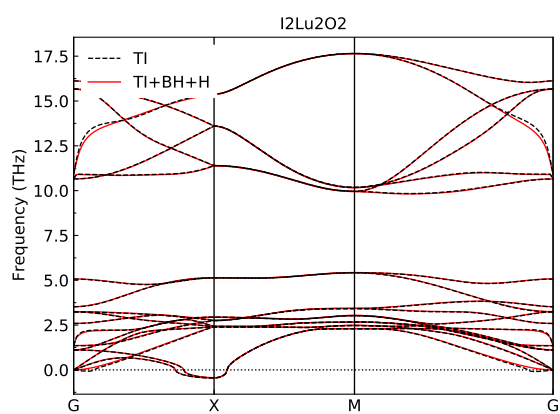

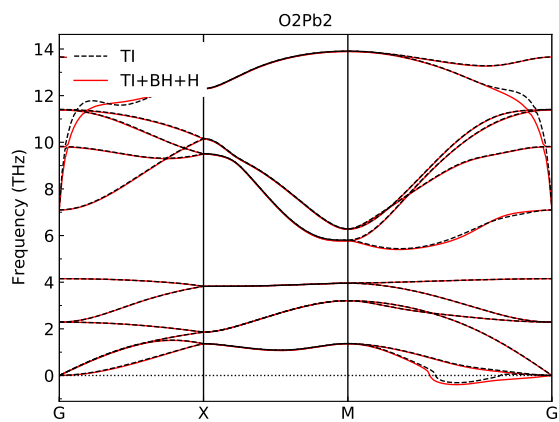

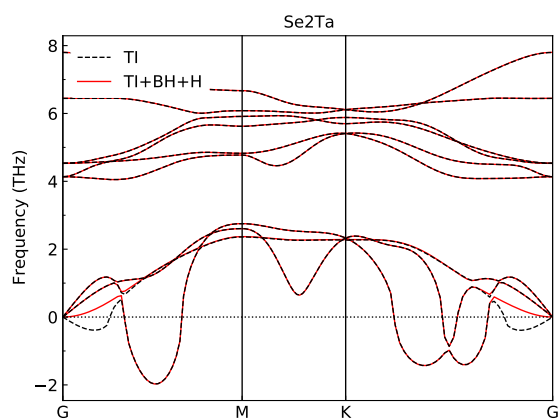

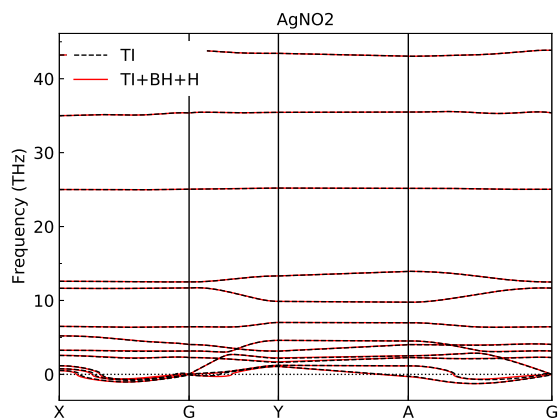

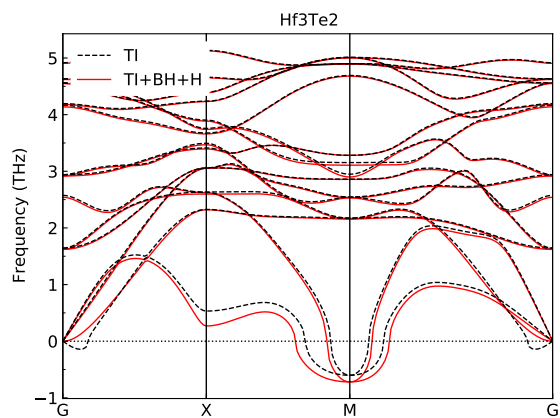

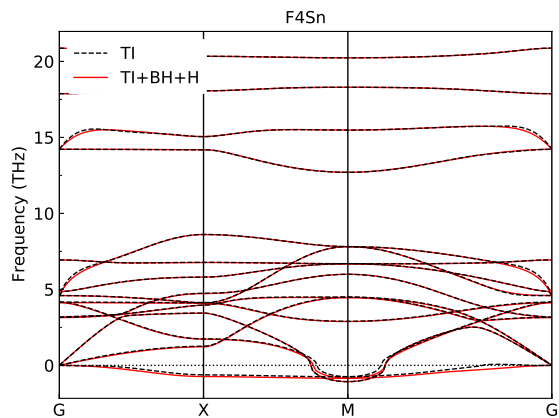

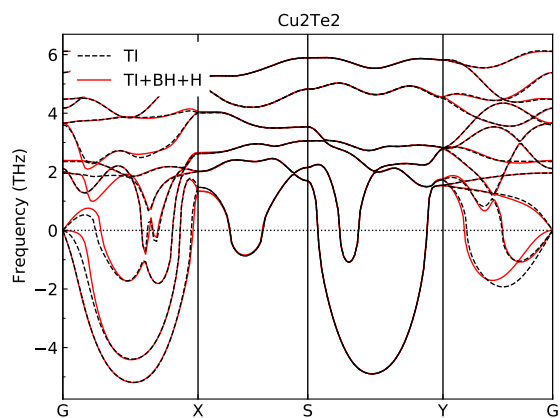

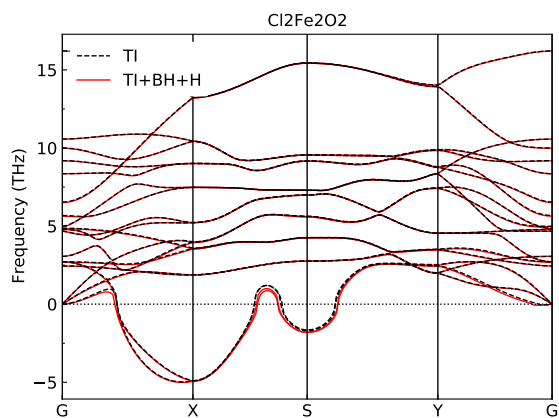

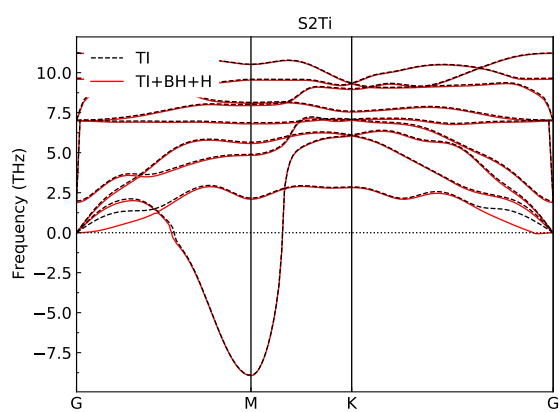

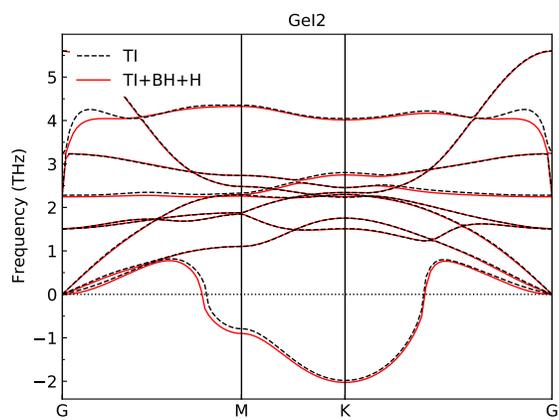

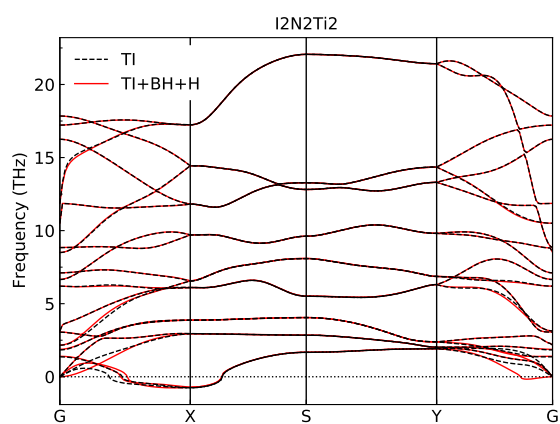

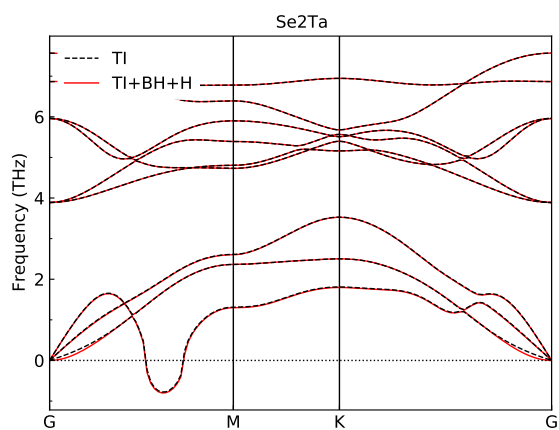

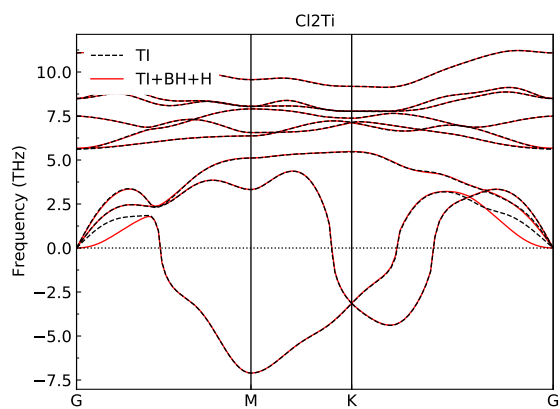

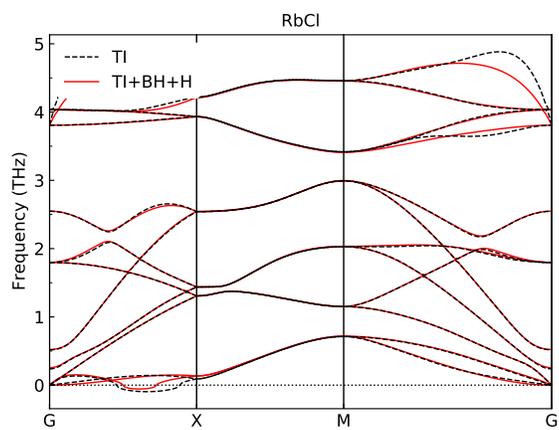

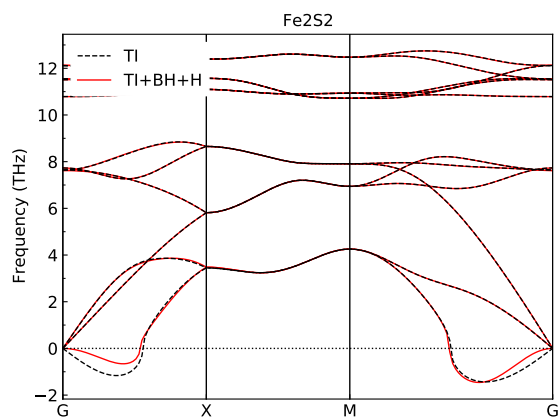

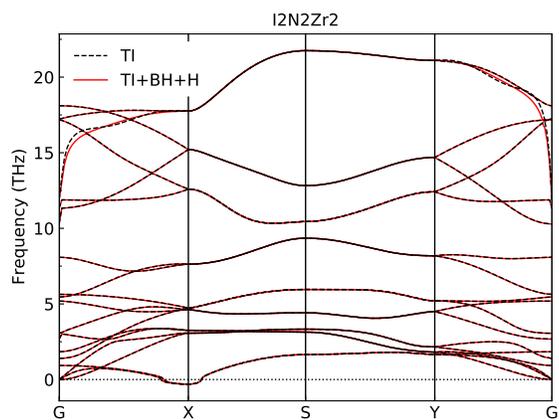

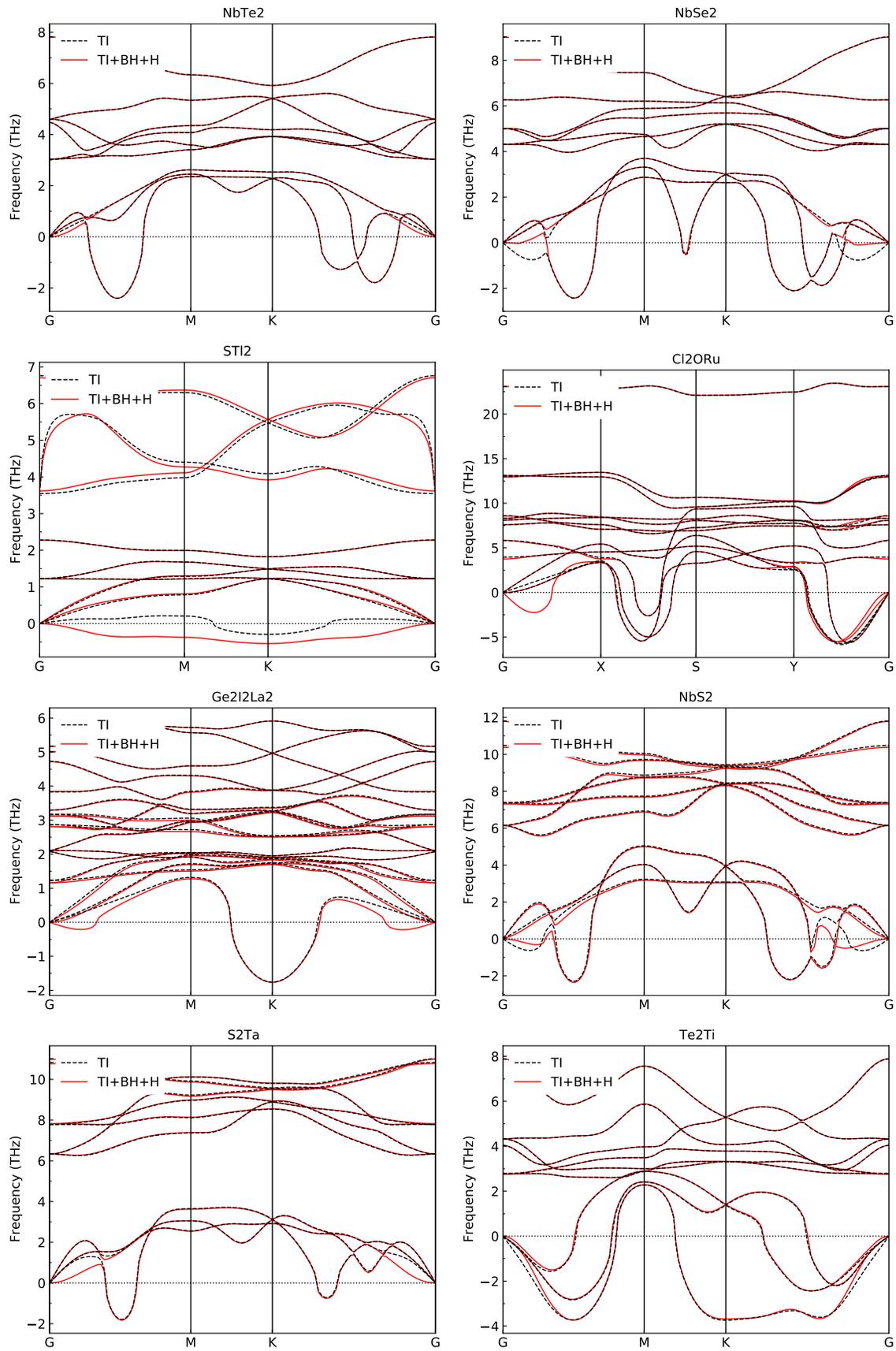

# SUPPLEMENTARY REFERENCES


\fi

\end{document}

\end{document}